\documentclass{aa}  
\usepackage{amssymb}
\usepackage{graphicx}
\usepackage{subfigure}
\usepackage[authoryear]{natbib}
\usepackage[figuresright]{rotating}

\usepackage{txfonts}
\newcommand{\teff}{\ifmmode T_{\rm eff} \else T$_{\mathrm{eff}}$\fi}
\newcommand{\logg}{\ifmmode \log g \else $\log g$\fi}
\newcommand{\lL}{\ifmmode \log \frac{L}{L_{\odot}} \else $\log \frac{L}{L_{\odot}}$\fi}
\newcommand{\mdot}{$\dot{M}$}
\newcommand{\myr}{M$_{\odot}$ yr$^{-1}$}

\newcommand{\vinf}{$v_{\infty}$}

\newcommand{\kms}{km s$^{-1}$}
\newcommand{\msun}{\ifmmode M_{\odot} \else M$_{\odot}$\fi}
\newcommand{\zsun}{\ifmmode Z_{\odot} \else Z$_{\odot}$\fi}
\newcommand{\lsun}{\ifmmode L_{\odot} \else L$_{\odot}$\fi}
\newcommand{\rsun}{\ifmmode R_{\odot} \else R$_{\odot}$\fi}
\newcommand{\qh}{\ifmmode Q_{\rm H} \else $Q_{\rm H}$\fi}
\newcommand{\qhei}{\ifmmode Q_{\ion{He}{i}} \else $Q_{\ion{He}{i}}$\fi}
\newcommand{\mum}{$\mu$m}

\newcommand{\brg}{Br$\gamma$}
\newcommand{\heii}{\ion{He}{ii} 2.189 $\mu$m}
\newcommand{\hei}{\ion{He}{i} 2.112 $\mu$m}
\newcommand{\heia}{\ion{He}{i} 2.056 $\mu$m}
\newcommand{\civ}{\ion{C}{iv} 2.070--2.083 $\mu$m}

\newcommand{\HII}{H~{\sc ii}}
\newcommand{\htwo}{H$_{2}$}

\begin{document}
   \title{Near--IR integral field spectroscopy of ionizing stars and young stellar objects on the borders of H~{\sc ii} regions \thanks{Based on observations collected at the ESO Very Large Telescope (program 081.C-0057)}}

   \subtitle{}

   \author{F. Martins\inst{1}
          \and
          M. Pomar\`es\inst{2}
          \and
          L. Deharveng\inst{2}
          \and
          A. Zavagno\inst{2}
	  \and
          J.~C. Bouret\inst{2}
          }

   \offprints{F. Martins}

   \institute{GRAAL--UMR 5024, CNRS \& Universit\'e Montpellier II, Place Eug\`ene Bataillon, F-34095, Montpellier Cedex 05, France\\
              \email{martins AT graal.univ-montp2.fr}
         \and 
             LAM--UMR 6110, CNRS \& Universit\'e de Provence, rue Fr\'ed\'eric Joliot-Curie, F-13388, Marseille Cedex 13, France\\
             }

   \date{Received ...; accepted ...}

 
  \abstract
   {}
   {We study three Galactic \HII\ regions -- RCW~79, RCW~82 and RCW~120 -- where triggered star formation is taking place. Two stellar population are observed: the ionizing stars of each \HII\ region and young stellar objects on their borders. Our goal is to show that they represent two distinct populations, as expected from successive star forming events. }
   {We use near--infrared integral field spectroscopy obtained with SINFONI on the VLT to make a spectral classification. We derive the stellar and wind properties of the ionizing stars using atmosphere models computed with the code CMFGEN. The young stellar objects are classified according to their $K$--band spectra. In combination with published near and mid infrared photometry, we constrain their nature. Linemaps are constructed to study the geometry of their close environment. }
   {We identify the ionizing stars of each region. RCW~79 is dominated by a cluster of a dozen O stars, identified for the first time by our observations. RCW~82 and RCW~120 are ionized by two and one O star, respectively. All ionizing stars are early to late O stars, close to the main sequence. The cluster ionizing RCW~79 formed 2.3$\pm$0.5 Myr ago. Similar ages are estimated, albeit with a larger uncertainty, for the ionizing stars of the other two regions. The total mass loss rate and ionizing flux is derived for each regions. In RCW~79, where the richest cluster of ionizing stars is found, the mechanical wind luminosity represents only 0.1\% of the ionizing luminosity, questioning the influence of stellar winds on the dynamics of these three \HII\ regions. The young stellar objects show four main types of spectral features: H$_{2}$ emission, \brg\ emission, CO bandheads emission and CO bandheads absorption. These features are typical of young stellar objects surrounded by disks and/or envelopes, confirming that star formation is taking place on the borders of the three \HII\ regions. The radial velocities of most YSOs are consistent with that of the ionized gas, firmly establishing that their association with the \HII\ regions. Exceptions are found in RCW~120 where differences up to 50 \kms\ are observed. Outflows are detected in a few YSOs. All YSOs have moderate to strong near--IR excess. In the [24] versus $K-$[24] diagram, the majority of the sources dominated by H$_{2}$ emission lines stand out as redder and brighter than the rest of the YSOs. The quantitative analysis of their spectra indicates that for most of them the H$_{2}$ emission is essentially thermal and likely produced by shocks. We tentatively propose that they represent an earlier phase of evolution compared to sources dominated by \brg\ and CO bandheads. We suggest that they still possess a dense envelope in which jets or winds create shocks. The other YSOs have partly lost their envelopes and show signatures of accretion disks. Overall, the YSOs show distinct spectroscopic signatures compared to the ionizing sources, confirming the presence of two stellar populations. }
    {}


   \maketitle


\section{Introduction}
\label{s_intro}

Massive stars play a significant role in several fields of astrophysics. They produce the majority of heavy elements and spread them in the interstellar medium, taking an active part in the chemical evolution of galaxies. But they also end their life as supernovae and gamma--ray bursts. Through their strong winds and ionizing fluxes they power \HII\ regions and bubbles which are often used to trace metallicity gradients in galaxies. The energy they release in the interstellar medium is thought to trigger second--generation star formation events. Observations of young stellar objects (YSO) in molecular clouds surrounding (clusters of) massive stars lend support to this mechanism \citep[e.g.][]{walborn02,hatano06}. 

A particular case concerns star formation on the borders of \HII\ regions. According to the collect and collapse model \citep{el77}, a dense shell of material is trapped between the shock and ionization front of an expanding \HII\ regions. When the amount of collected material is large enough, global shell fragmentation occurs and new stars are formed. The observation of molecular condensations on the borders of several \HII\ regions and the subsequent identification of YSOs within these clumps \citep{de03,de05,za06,za07,de08a,de09,po09} confirms that this mechanism is at work at least in some \HII\ regions. 

Other mechanisms leading to triggered star formation exist. Some work qualitatively as the collect and collapse model in the sense that the clumps are formed during the \HII\ region expansion. For instance, dynamical instabilities of the ionization front \citep{vishniac83,garcia96} create molecular condensations separated by zones of lower densities. The newly formed clumps grow until they become Jeans unstable and collapse. Alternatively, second generation star formation can happen in pre-existing clumps.
If the neutral gas in which the \HII\ region expands is not homogeneous, the outer layers of the molecular overdensities are ionized like the borders of a classical \HII\ region. A shock front precedes the ionization front inside these clumps, leading to their collapse \citep{duvert90,ll94}. 

A number of questions regarding triggered star formation remain unanswered. The properties of the observed YSOs are poorly known besides a crude classification in class I or class II objects by analogy with low--mass stars. YSOs usually display near--infrared spectra with CO, \brg\ and/or H$_{2}$ emission lines \citep{bik06}. The relation, if any, between objects with different spectroscopic appearance is not clear. Besides, in the regions where the collect and collapse process is at work, the quantitative properties of the ionizing sources of the \HII\ regions are not known. In particular, the relative role of ionizing radiation and stellar winds on the dynamics of such regions is debated. The timescales under which material accumulates and fragments, and the properties of the resulting clumps depend on the strength of those two factors \citep{whit94}. Hence, one might wonder whether the nature of the newly formed objects depends on the properties of the stars powering the \HII\ regions.

In the present study, we tackle these questions by investigating the properties of the ionizing stars and YSOs of three Galactic \HII\ regions: RCW~79, RCW~82 and RCW~120 \citep{rod60}. Those regions are known to be the sites of triggered star formation \citep{za06,za07,po09,kang09}. We have used SINFONI on the VLT to obtain near--infrared spectra of both the ionizing stars and a selection of YSOs in each region. Our main goals were: 

\begin{itemize}

\item Identify the ionizing stars of each region, and derive their stellar properties using atmosphere models. In particular, we want to determine their ionizing fluxes and mass loss rates in order to better understand the dynamics of the \HII\ regions. Equally important is the determination of the age of those stars, since it can be related to the presence of YSOs to quantitatively confirm the existence of triggered star formation. 

\item Constrain the nature of the YSOs on the borders of the \HII\ regions. In combination with infrared photometry, spectroscopy can reveal the presence of disks or envelopes. The evolutionary status of those objects can thus be better understood. In particular, it can be clearly shown whether they are stars still in their formation process or objects already on the main sequence. 

\end{itemize}

The paper is organized as follows. In Sect.\ \ref{pres_hii} we present the three \HII\ regions targeted in this study. Sect.\ \ref{s_obs} describes our observations. The analysis of the ionizing stars is presented in Sect.\ \ref{s_ex}, while the YSOs are discussed in Sect.\ \ref{s_yso}. We discuss our results in Sect.\ \ref{s_disc} and summarize our conclusions in Sect.\ \ref{s_conc}.


\section{Presentation of the observed H~{\sc ii} regions}
\label{pres_hii}

\subsection{RCW~79}

\begin{figure}[ht]
\centering
\includegraphics[width=9cm]{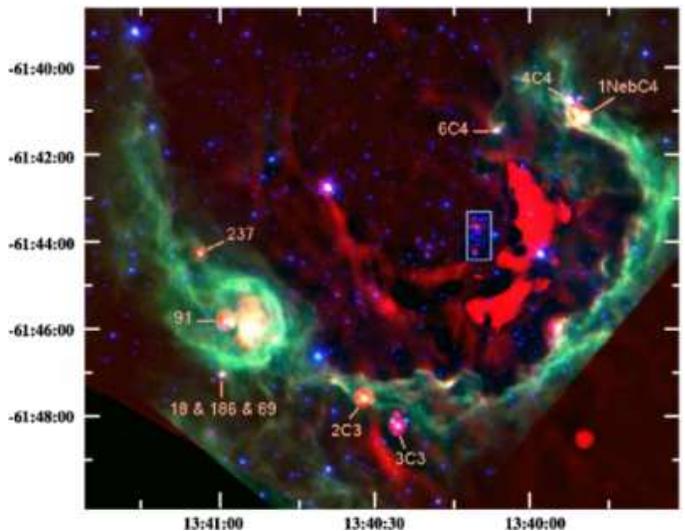}
\caption{Identification of the RCW~79 YSOs targeted in the present study on a composite infrared color image. Blue and green are the Siptzer--GLIMPSE images at 3.6 and 8.0 \mum\ respectively. Red is an unsharped masked image obtained from the Spitzer--MIPS frame at 24~\mum\ (as explained in ZA07). The bright extended emission has been subtracted to emphasize the YSOs. The candidate ionizing sources are located inside the blue rectangle (see also Fig.\ \ref{im_rcw79_ex}).}\label{yso_79}
\end{figure}

RCW~79 is a Southern Galactic \HII\ region located at 4.2$\pm$1 kpc
\citep{russeil98}. Its diameter is $\sim$6.4 pc. \citet{za06}
(hereafter ZA06) used Spitzer GLIMPSE and SEST-SIMBA 1.2-mm continuum
data to study the star formation on the borders of this region. A
layer of warm dust is clearly seen at 8\,$\mu$m surrounding RCW~79
(Fig.\ \ref{yso_79}). A compact \HII\ region is observed in this layer
as well as five cold dust condensations (masses between 100 and 1000
\msun) detected by 1.2-mm continuum emission. YSOs have also been
revealed by Spitzer-GLIMPSE observations, leading ZA06 to conclude
that triggered massive star formation was at work possibly through the
collect and collapse process \citep{el77}. Eight YSOs detected in the
near IR have been observed with SINFONI at the VLT. They are
identified on Fig.\ \ref{yso_79}. In several cases, multiple sources
have been uncovered by the SINFONI observations reported in this
paper. They are identified in Sect.\ \ref{s_yso} and Appendix
\ref{ap_yso}.

The ionizing stars of RCW~79 were unknown before the present
study. Using photometry and choosing objects inside the \HII\ region
contours, we have identified and observed a cluster of possible OB
stars candidates (blue square in Fig.\ \ref{yso_79}). They are
described in Sect.\ \ref{s_ex}.

\subsection{RCW~82}

\begin{figure}[ht]
\centering
\includegraphics[width=9cm]{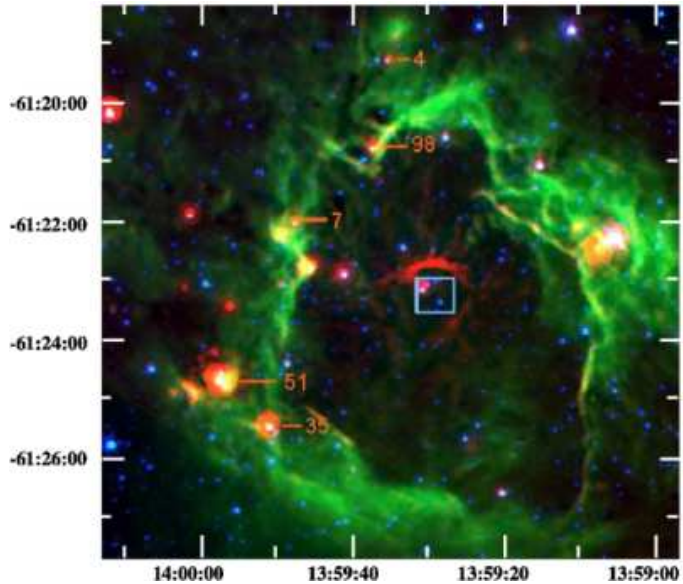}
\caption{Same as Fig.\ \ref{yso_79} for RCW~82.}\label{yso_82}
\end{figure}

RCW~82 is a Southern \HII\ region located at a distance of
3.4$\pm$1 kpc. Its radius is about 5 pc.  \citet{po09} (hereafter PO09)
detected molecular material in $^{12}$CO(1-0) and $^{13}$CO(1-0)
surrounding the \HII\ region. They showed that some of the structures
correspond to dense material collected between the shock front and the
ionization front during the expansion of the \HII\ region. Masses of
these clumps range from 200 to 2500 \msun. Star formation is observed
on the borders of RCW~82, with a total of 63 candidate YSOs. Among
these, we have selected five YSO candidates visible in the $K$--band for
our study. These objects are shown on Fig.~\ref{yso_82}.

PO09 identified four candidate ionizing sources, among which two are
likely O stars (see their Fig.\ 5). These sources are located just
south of the 24~\mum\ emission ridge in Fig\ \ref{yso_82}, and better
displayed in the upper panel of Fig.\ \ref{im_rcw82_120_ex}.

\subsection{RCW~120}
\label{desc_rcw120}

A detailed study of RCW~120 can be found in \citet{za07} (hereafter
ZA07), \citet{de08} (DE08) and \citet{de09} (DE09). RCW~120 is the
nearest of the three \HII\ regions observed in the present study. Its
distance, D=1.35$\pm$0.3 kpc, is well determined since both the
photometric and kinematic distances are in good agreement (DE09).

\begin{figure}[h]
\centering
\includegraphics[width=9cm]{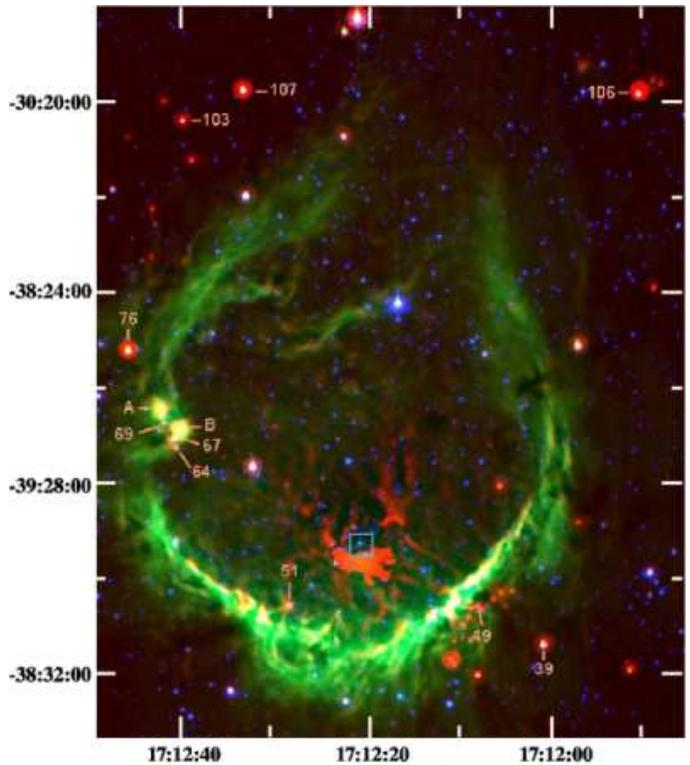}
\caption{Same as Fig.\ \ref{yso_79} for RCW~120. On this image, the extended high brightness emission has been subtracted to enhance the small scale structures and the point sources.}\label{yso_120}
\end{figure}

RCW~120 has a circular geometry (diameter $\sim$3.5~pc) and is
surrounded by a shell of dense material collected during the expansion
of the \HII\ region. The cold dust emission of this shell has been
observed with the ESO SEST at 1.2-mm (ZA07) and with the APEX-LABOCA
camera at 870~$\mu$m (DE09). Its mass is in the range 1100--1900
\msun; it is fragmented with massive fragments elongated along the
ionization front. Star formation is at work in these condensations, as
discussed by ZA07 and DE09. Twelve candidate YSOs have been observed
with SINFONI and are identified in Fig.\ \ref{yso_120}.

The central ionizing star of RCW~120 is
CD$-$38\hbox{$^\circ$}11636. Its spectral type (estimated from
spectrograms) is O8 \citep{gg70} or O9 \citep{crampton71} - see also
the discussion in DE08. Its extinction, determined by \citet{ak84} is
A$_V$=4.65~mag. It is located just north of the bright
24~\mum\ emission ridge in Fig.\ \ref{yso_120}. Its characteristics
are discussed in Sect.\ \ref{s_ex}.

\section{Observations and data reduction}
\label{s_obs}

For the current project, we selected two types of sources: the candidate ionizing stars of the \HII\ regions and a few candidate YSOs. For the latter, we selected the brightest objects in the $K$--band. 

Data have been collected at the ESO-VLT on April 25$^{th}$ and 26$^{th}$ 2008. The near infrared integral field spectrograph SINFONI \citep{spiffi} was used in seeing--limited mode to obtain medium resolution $K$--band spectra of our selected sources. We selected the 250 mas scale which provided a field of view of 8\arcsec$\times$8\arcsec. Sequences of source and sky exposures were conducted in order to ensure optimal sky subtraction. For the faintest sources, two exposures were made, with a 1\arcsec\ offset between them to minimize pixels artifacts. The exposure times ranged from 1 minute for $K$=9 sources to 10 minutes for $K$=13 objects. Telluric stars were observed regularly during the night to allow proper atmospheric correction. Standard calibration data were obtained by the ESO staff. The observations were conducted under an optical seeing ranging from 0.6\arcsec\ to 1.2\arcsec.

Data reduction was made with the SPRED software \citep{spred}. After bias subtraction, flat field and bad pixel corrections, wavelength calibration was done using a Ne-Ar spectrum. Fine tuning was subsequently performed using sky lines. Telluric correction was done using standard stars from which \brg\ and, when present, \ion{He}{i} 2.112 \mum\ were removed. The resulting spectra have a signal to noise of 10-100 depending on the source brightness and wavelength. Their resolution is $\sim$ 4000 (see also Sect.\ \ref{dyn_yso}). They were extracted with the software QFitsView \footnote{http://www.mpe.mpg.de/$\sim$ott/QFitsView/}, carefully selecting individual ``source'' pixels one by one to avoid contamination by neighboring objects. Background pixels were selected close to the ``source'' pixels in order to correct for the underlying nebular emission. In practice, the average spectrum of these background pixels is subtracted to the each individual source pixels. Then all corrected source pixels are added together to ensure optimum extraction of the pure stellar sprectrum. This method is possible with integral field spectroscopy and was one of the main drivers for the choice of SINFONI for the present observations.


\section{Ionizing stars}
\label{s_ex}

\begin{figure}[]
\centering
\includegraphics[width=9cm]{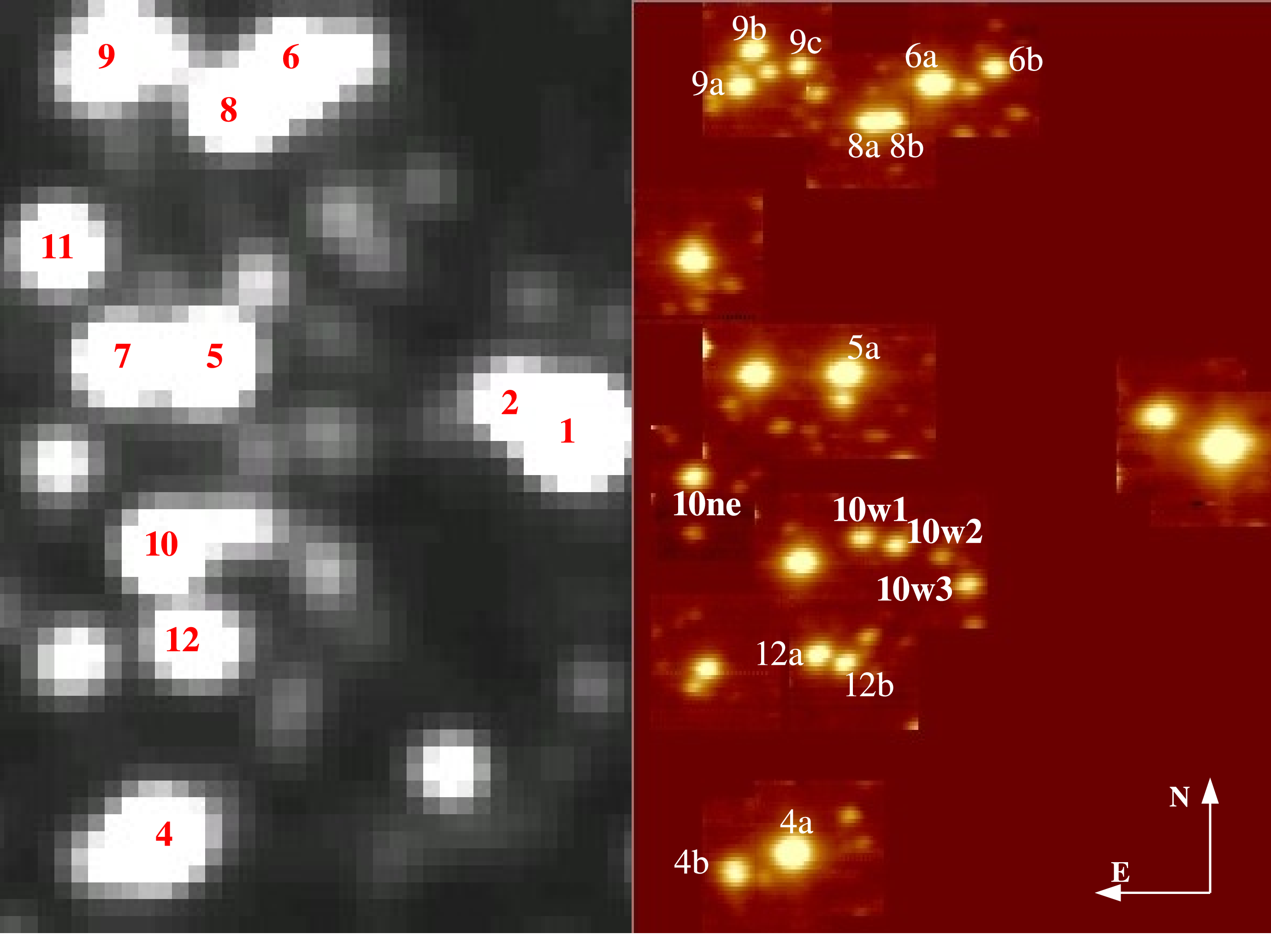}
\caption{$K$--band images of the ionizing star cluster of RCW79 from 2MASS (left) and SINFONI (right). The intensity of the SINFONI image corresponds to the median of the spectral dimension of the datacube.}\label{im_rcw79_ex}
\end{figure}

\begin{figure}[]
\centering
\includegraphics[width=9cm]{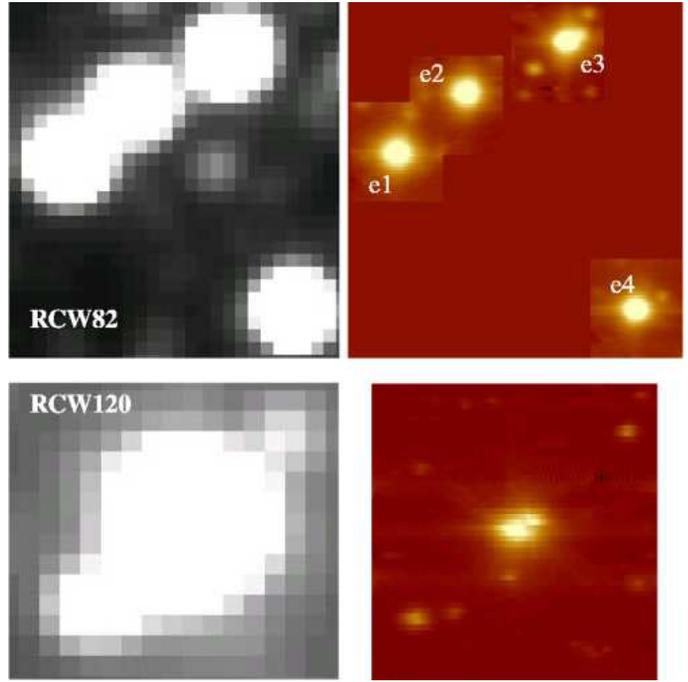}
\caption{Same as Fig.\ \ref{im_rcw79_ex} for RCW~82 (top) and RCW~120 (bottom). North is up and East is left.}\label{im_rcw82_120_ex}
\end{figure}

In this section we describe the spectral morphology of the ionizing sources of each region. We determine their stellar and wind properties using atmosphere models computed with the code CMFGEN \citep{hm98}. Fig.\ \ref{im_rcw79_ex} shows the cluster of stars responsible for the ionization of RCW~79. The stars are labeled on the 2MASS $K$--band image presented on the left part of the figure. Numbers correspond to decreasing 2MASS $K$--band magnitudes. Note that star number 3 is not indicated: it corresponds to a source out of the field which turned out to be a foreground star (its spectrum displays CO bandheads in absorption typical of cool, evolved stars). The right panel of Fig.\ \ref{im_rcw79_ex} shows a mosaic of the SINFONI fields observed. In many cases, the higher spatial resolution reveals several components for the same 2MASS source (e.g. source 9). Consequently, we have assigned new names to the resolved components. The identification is shown on the right side of Fig.\ \ref{im_rcw79_ex}. Similarly, the ionizing sources of RCW~82 and RCW~120 are displayed in Fig.\ \ref{im_rcw82_120_ex}.

\subsection{Photometry and spectral classification}
\label{s_phst}

Table \ref{tab_stars} summarizes the observational properties of the ionizing stars of the three regions. Photometry is from 2MASS. When a given 2MASS point source was resolved in several components by our observations, the relative SINFONI fluxes were used to recompute the individual magnitudes, using the 2MASS magnitude as a measure of the total flux from all components. In addition, the $K$--band magnitude of star 2 was revised because its magnitude is clearly different from that of star 1, in contradiction to the the 2MASS values (see right panel of Fig.\ \ref{im_rcw79_ex}). Inspection of the 2MASS catalog revealed that the aperture used to compute the $K$--band magnitude was wider than the separation of sources 1 and 2. Hence, the flux used to estimate the $K$--band magnitude of star 2 was contaminated by flux from star 1. We used the SINFONI flux and 2MASS magnitude of star 1 to calibrate the $K$--band magnitude of star 2 from the observed SINFONI flux. The $H$--band magnitudes are from 2MASS. Since there is no SINFONI $H$--band data, one cannot recalculate the magnitude of the components of unresolved 2MASS sources. In that case, we give the 2MASS $H$--band magnitude as a lower limit. 

For each region, we calculated the absolute $K$--band magnitude using information on the distance and extinction: 

\begin{itemize}
\item \textit{RCW~79}: a distance of 4.2$\pm$1 kpc was derived by \citet{russeil98}. We estimated the $K$--band extinction from the $H-K$ color of the early type stars of the region (see below for the spectral classification). Since ($H-K$)$_{0}$ is basically independent of the spectral type for O stars \citep[see][]{mp06}, one can use the 2MASS magnitudes even in the cases were SINFONI revealed several sub-components. In practice, we used stars 4, 5, 7, 8, 9 and 10 to obtain A$_{\rm K}$ = 0.7 $\pm$ 0.1. 

\item \textit{RCW~82}: We proceeded as for RCW~79 to calculate MK assuming a distance of 3.4$\pm$1 kpc \citep{russeil98}. The extinction estimate was based on stars RCW82 e2 and RCW82 e3. Both of them are O stars. A mean value of A$_{\rm K}$=0.36$\pm$0.2 was derived. 

\item \textit{RCW~120}: With a distance of 1.35$\pm$0.3 kpc (ZA07), RCW~120 is the closest region of our sample. The region is ionized by a single O star. The estimated extinction is A$_{\rm K}$=0.50 $\pm$ 0.1. This is in good agreement with previous estimates (see Sect.\ \ref{desc_rcw120}).

\end{itemize}

Figs \ref{fig_rcw79_ex} and \ref{fig_ex_2} show the spectra of the candidate ionizing stars of RCW~79, RCW~82 and RCW~120. Most of the stars show \brg\ and \hei\ absorption and, depending on the sources, \heii\ absorption, \civ\ emission and the \ion{N}{iii}/\ion{C}{iii}/\ion{O}{iii} emission complex at 2.115 \mum. These features are typical of OB stars. We relied on the atlases of $K$--band spectra of \citet{hanson96,hanson05} to assign spectral types to the ionizing stars of RCW~79, RCW~82 and RCW~120. In practice, we have used the following scheme:

\begin{itemize}

\item[$\bullet$] the presence of \heii\ indicates a spectral type earlier than O8

\item[$\bullet$] \civ\ emission is observed in O4-6 stars and is the strongest at O5

\item[$\bullet$] the \ion{N}{iii}--\ion{C}{iii}--\ion{O}{iii} emission complex is observed in emission together with \hei\ in absorption in O4-O7 stars

\item[$\bullet$] \hei\ disappears at spectral types later than B2.5 

\item[$\bullet$] O3--7 supergiants have \brg\ either in emission or in absorption weaker than \heii. Later type supergiants have a narrow \brg\ absorption profile from which \ion{He}{i} 2.164 $\mu$m is clearly separated. Dwarfs and giants have broader \brg\ absorption (\ion{He}{i} 2.164 $\mu$m is blended with \brg). 

\end{itemize}

\noindent Based on these features, we classified the ionizing stars as OB dwarfs or giants. The results are listed in Table \ref{tab_stars}. The earliest spectral types are O4--6 and are observed only in the richest ionizing cluster (RCW~79). Most stars are late O -- early B dwarfs/giants. We do not observe any supergiant. The distinction between dwarfs and giants is not easily feasible with our spectra. Hence we adopt the conservative approach to give the luminosity class V/III to most of our stars. Some of the stars did not show any of the above features. Instead, strong CO bandheads were clearly visible. Such spectra are typical of cool, evolved low mass stars \footnote{CO absorption is also observed in red supergiants. However, those stars are much brighter than O stars in the $K$--band. This is not the case of the stars with CO absorption of our sample.}. They do not contribute any ionizing flux and are most likely foreground or background stars. We classify them as ``late'' in Table \ref{tab_stars}. Three stars also show strong \heia\ and \brg\ emission, without \hei\ emission. In one case (RCW82 e1), the \ion{H}{i} Pfund serie and \ion{Mg}{ii} 2.138--2.144 \mum\ were also detected in emission. According to \citet{hanson96} \citep[see also][]{cs00}, these features are seen in Oe--Be stars. Such objects are thought to host a circumstellar disk contributing a significant fraction of the near-IR emission (both line and continuum). Note that the ionizing source of RCW~120 seems to be double (Fig.\ \ref{im_rcw82_120_ex}). We tried to extract the spectra of both components, but found no difference between them. This indicates either that both components have the same spectral type, or that the spatial resolution of our observations is not sufficient to separate the spectra. Hence, we have treated the ionizing source of RCW~120 as a single source.

PO09 estimated a spectral type 06.5V and 07.5V for stars RCW82 e2 and RCW82 e3 respectively from $J$ vs $J-K$ and $J-H$ vs $H-K$ diagrams. Our spectroscopic classification indicates later types: O9-B2V/III for both stars. Given the difficulty to make a spectral classification from near-infrared color--color and color--magnitude diagrams, the difference is acceptable. Similarly, star RCW82 e4 is identified as a late--type giant which is confirmed by our spectroscopy. PO09 showed that star RCW82 e1 presented a near-IR excess, which is consistent with the presence of a disk in a star of spectral type Oe or Be. ZA07 reported a spectral type O8V for the ionizing star of RCW~120. We prefer a spectral type O6--8V/III, in rather good agreement. Finally, the ionizing sources of RCW~79 were not previously identified. ZA06 estimated that a single O4V star could power the \HII\ region. This is roughly consistent with our finding that a cluster of about ten O4 to O9 stars is responsible for the ionization.

\begin{table*}
\begin{center}
\caption{Position and photometry of ionizing stars. } \label{tab_stars}
\begin{tabular}{lcccccl}
\hline
Source & RA & DEC & $H$ & $K$ & MK & ST \\
       & h m s & $^{\circ} \arcmin\ \arcsec$ & & & \\
\hline                                                              
RCW79 1      & 13:40:09.11 & -61:43:51.9 & 9.10$\pm$0.04     & 8.29$\pm$0.03  & -5.53$\pm$0.53 & OBe \\
RCW79 2      & 13:40:09.64 & -61:43:50.1 & $>$10.59          & 10.41$\pm$0.06 & -3.46$\pm$0.53 & O7.5--8V/III \\
RCW79 4a     & 13:40:12.63 & -61:44:16.0 & $>$9.41           & 9.23$\pm$0.02  & -4.64$\pm$0.53 & O4--6V/III \\
RCW79 4b     & 13:40:13.10 & -61:44:17.1 & $>$9.41           & 11.29$\pm$0.02 & -2.54$\pm$0.53 & ? \\
RCW79 5a     & 13:40:12.19 & -61:43:47.6 & $>$9.40           & 9.24$\pm$0.05  & -4.62$\pm$0.53 & O4--6V/III \\
RCW79 6a     & 13:40:11.47 & -61:43:30.6 & $>$9.24           & 9.65$\pm$0.07  & -4.22$\pm$0.53 & O6--8V/III \\
RCW79 6b     & 13:40:10.98 & -61:43:29.6 & $>$9.24           & 11.65$\pm$0.07 & -2.22$\pm$0.53 & O9--B2V \\
RCW79 6c     & 13:40:11.17 & -61:43:30.9 & $>$9.24           & 12.97$\pm$0.07 & -0.86$\pm$0.53 & $>$B2.5V \\
RCW79 7      & 13:40:12.93 & -61:43:47.8 & 9.93$\pm$0.04     & 9.64$\pm$0.03  & -4.23$\pm$0.53 & O6--8V/III \\
RCW79 8a     & 13:40:11.98 & -61:43:32.9 & $>$9.93           & 10.33$\pm$0.04 & -3.54$\pm$0.53 & O9--B2V/III \\
RCW79 8b     & 13:40:11.82 & -61:43:32.7 & $>$9.93           & 10.55$\pm$0.04 & -3.32$\pm$0.53 & O9--B2V/III \\
RCW79 9a     & 13:40:13.06 & -61:43:30.6 & $>$10.12          & 10.79$\pm$0.03 & -3.07$\pm$0.53 & O9--B2V/III \\
RCW79 9b     & 13:40:12.95 & -61:43:28.6 & $>$10.12          & 10.91$\pm$0.03 & -2.96$\pm$0.53 & O9--B2V/III \\
RCW79 9c     & 13:40:12.56 & -61:43:29.5 & $>$10.12          & 11.87$\pm$0.03 & -1.95$\pm$0.53 & B \\
RCW79 10     & 13:40:12.56 & -61:43:58.9 & 10.09$\pm$0.07    & 9.84$\pm$0.05  & -4.05$\pm$0.53 & O6.5--8V/III \\
RCW79 10ne   & 13:40:13.44 & -61:43:53.9 & 12.71$\pm$0.09    & 11.25$\pm$0.03 & -2.57$\pm$0.53 & Be \\
RCW79 10w1   & 13:40:12.06 & -61:43:57.4 & $>$11.69          & 12.11$\pm$0.05 & -1.71$\pm$0.53 & B0-2.5V \\
RCW79 10w2   & 13:40:11.79 & -61:43:57.8 & $>$11.69          & 12.04$\pm$0.05 & -1.78$\pm$0.53 & B0-2.5V \\
RCW79 10w3   & 13:40:11.19 & -61:44:00.1 & 12.31$\pm$0.10    & 11.99$\pm$0.07 & -1.83$\pm$0.53 & B1-2 \\
RCW79 11     & 13:40:13.44 & -61:43:41.0 & 10.39$\pm$0.04    & 9.93$\pm$0.03  & --             & late \\
RCW79 12a    & 13:40:12.42 & -61:44:04.2 & $>$10.67          & 10.75$\pm$0.04 & --             & late \\
RCW79 12b    & 13:40:12.19 & -61:44:04.9 & $>$10.67          & 11.49$\pm$0.04 & -2.38$\pm$0.53 & O9--B2V/III \\
RCW79 CHII 6 & 13:40:53.71 & -61:45:46.8 & 11.55$\pm$0.07    & 10.72$\pm$0.05 & -4.03$\pm$0.53 & O7.5--9.5V/III \\
 & & & & & & \\
RCW82 e1     & 13:59:29.86 & -61:23:08.9 & 9.55$\pm$0.03   & 9.09$\pm$0.02  & --             & Be \\
RCW82 e2     & 13:59:29.13 & -61:23:04.2 & 9.59$\pm$0.08   & 9.40$\pm$0.03  & -3.62$\pm$0.67 & O9--B2V/III \\
RCW82 e3     & 13:59:28.09 & -61:23:00.4 & 9.41$\pm$0.03   & 9.19$\pm$0.03  & -3.83$\pm$0.67 & O9--B2V/III \\
RCW82 e4     & 13:59:27.37 & -61:23:20.3 & 10.44$\pm$0.02  & 9.35$\pm$0.02  & --             & late \\
 & & & & & & \\
RCW120 e    & 17:12:20.82 & -38:29:25.5 & 7.71$\pm$0.04    & 7.52$\pm$0.02  & -3.62$\pm$0.49 & O6--8V/III \\
\hline
\end{tabular}
\end{center}
\end{table*}

\subsection{Spectroscopic analysis}
\label{s_spec_ana}

In this section we derive the stellar and wind properties of the
ionizing stars of the \ion{H}{ii} regions by means of spectroscopic 
analysis with atmosphere models. We subsequently use the derived
properties to constrain the age of the ionizing populations.

\subsubsection{Stellar and wind properties}
\label{s_prop}

The stellar and wind properties have been derived through spectroscopic analysis. Atmosphere models were computed with the code CMFGEN \citep{hm98}. This code solves the radiative transfer and statistical equations in the co-moving frame of the expanding atmosphere, for light elements as well as for metals. It thus produces non-LTE, line blanketed atmosphere models with winds. An exhaustive description of the code and its approximations is given in \citet{hm98}. Summaries of the main characteristics can also be found in \citet{hil03,arches}. The resulting synthetic spectra are compared to the SINFONI $K$--band spectra to constrain the main physical parameters. In practice, we have proceeded as follows:

\begin{itemize}

\item[$\bullet$] \textit{Effective temperature}: the determination of \teff\ usually relies on the ratio of \ion{He}{i} to \ion{He}{ii} lines. We have used the following lines: \hei\ and \heii\ when present. \ion{C}{iv} 2.07--2.08 \mum\ lines are used as secondary indicators since they appear in emission in the hottest O stars \citep{hanson96}. An uncertainty of 2000 to 3000K is achieved depending on the star. For stars cooler than $\sim$32000 K \heii\ vanishes. The uncertainty on \teff\ is large (5000 K) and is set by the presence and strength of \ion{He}{i} lines.

\item[$\bullet$] \textit{Mass loss rate}: The main indicator in the $K$--band spectrum of O stars is \brg. It is filled by emission as the wind strength increases. We have used this line to constrain the mass loss rates of the ionizing stars. However, as demonstrated below, it becomes almost insensitive to \mdot\ below $\sim 3 \times 10^{-8} M_{\odot} yr^{-1}$ so that we could derive only upper limits on \mdot. We stress once more that our correction for nebular contamination ensures the best extraction possible of the stellar line profiles. Thus, our mass loss rate determination does not suffer from major uncertainties due to non stellar emission.

\item[$\bullet$] \textit{Luminosity}: \lL\ was derived from the absolute magnitude and the $K$--band bolometric correction. The latter was computed from \teff\ and the calibration of \citet{mp06}. The uncertainty on \lL\ was calculated from a full error propagation and takes into account the uncertainties on the distance, \teff\ and A$_{\rm K}$. It is of the order 0.2 dex for stars in the three regions.

\end{itemize}

Several parameters could not be derived due to the lack of spectral diagnostics. In particular, the surface gravity was adopted. Since most of our stars are dwarfs/giants, we chose $\log g=4.0$ which is typical of O dwarfs \citep{msh05}. In principle \brg\ can be used to constrain \logg\ \citep[e.g.][]{repolust05}. However, the S/N of a few tens combined to the medium resolution of our spectra prevents an accurate determination. Besides, since we use this line to determine mass loss rates and since the line core depends not only on \mdot\ but also on \logg\ (although in a less sensitive way), we decided to fix \logg. 

Modern atmosphere models for massive stars also include clumping since evidence for wind inhomogeneities exist \citep[e.g.][]{hil91,hil03,lm99}. Clumping is usually quantified by a volume filling factor $f$. Currently, no diagnostics of clumping have been identified in the K--band spectra of O stars with moderate winds. Such diagnostics are traditionally found in the UV, optical and submm range \citep{hil91,eversb98,blomme02}. Hence, we have simply adopted the canonical value of 0.1 for $f$ \citep{hk98,hil01}. 

Another important parameter is the wind terminal velocity (\vinf). It is determined from the blueward extension of UV P--Cygni profiles or from the width of strong emission lines. In our case, none of these diagnostics are available. Consequently, we decided to adopt \vinf. We chose a value of 2000 \kms\ as representative of early and mid O stars, and 1000 \kms\ as typical of late O and B stars \citep[e.g.][]{prinja90}. We also adopted the so-called $\beta$ parameter. Our model atmospheres require an input velocity structure which is constructed from a quasi-static photospheric structure to which a $\beta$ velocity law is connected. This law is of the form: $v=v_{\infty}(1-R/r)^{\beta}$ where $R$ is the stellar radius. The value of 0.8 we adopted for $\beta$ is typical of O dwarfs/giants \citep[e.g.][]{repolust04}. For the input photospheric structure, we used the OSTAR2002 TLUSTY models \citep{lh03}. Finally, we used the solar abundances of \citet{gre07} for the elements included in our models, namely H, He, C, N, O, Ne, Mg, Si, S, Fe and Ni. 

Fig.\ \ref{fit_79} and \ref{fit_82} shows the best--fit we obtained for the brightest ionizing sources of the three \HII\ regions. In general, those fits are of good quality. The only important discrepancy in a few stars is the \heia\ line which is predicted too strong. However, this line has been shown to be extremely sensitive to line-blanketing \citep{paco94,paco06}. Since our models include only a limited number of elements and since some atomic data for metals remain uncertain \citep[see][]{paco06}, the observed discrepancy is not surprising. The derived parameters for each star are gathered in Table \ref{tab_param}.

\begin{figure}[]
\centering
\includegraphics[width=9cm]{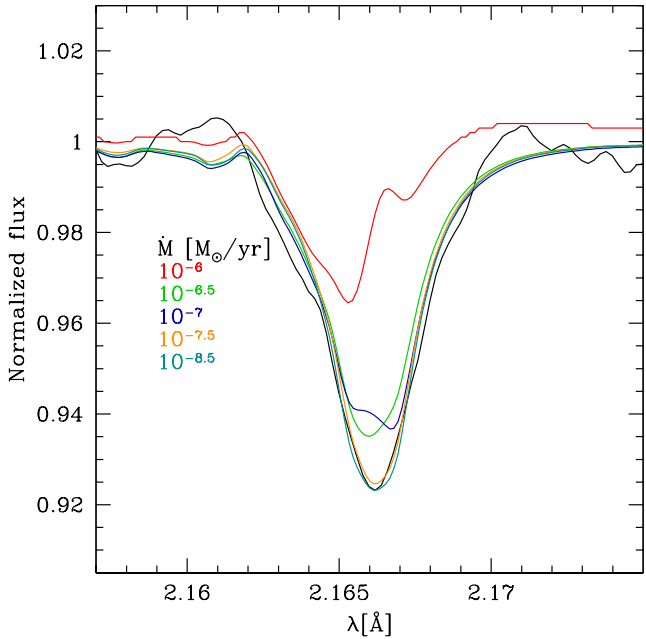}
\caption{Mass loss estimate for star RCW79 5a. The black solid line is the observed spectrum. The colored lines refer to models with different mass loss rates. A conservative upper limit of $10^{-7}$ \myr\ is derived.}\label{mdot_79_5a}
\end{figure}

Fig.\ \ref{mdot_79_5a} shows an example of mass--loss rate determination based on \brg\ in the case of star RCW79 5a. Reducing \mdot\ produces a deeper absorption line. However, below a value of $3 \times 10^{-8}$ \myr\ the line becomes little sensitive to any change of mass--loss rate. In practice, it is the wind density which is derived in the fitting process. Since it is proportional to \mdot/\vinf\, an error on \vinf\ translates into an error on \mdot. However, it is very unlikely that our estimates of \vinf\ are off by more than a factor of 2 (which is already a rather extreme case). Hence, a rather conservative upper limit of $10^{-7}$ \myr\ (including the uncertainty on \vinf\ and on the normalization and S/N of the observed spectrum) is determined. Since star RCW79 5a is one of the earliest of our sample, it is expected to show the strongest mass loss rate. Consequently, the upper limit on \mdot\ derived for this star also applies to all other ionizing stars. The upper limits we derive are fully consistent with recent determination of mass loss rates of Galactic O5--9 dwarfs based on optical and UV diagnostics \citep{jc03,repolust04}. Actually, mass loss rates up to two orders of magnitudes smaller are routinely found in late O dwarfs \citep{n81,ww05,marcolino09}. 

We have fitted only the $K$--band part of the observed spectra, but our best fit models yield the complete spectral energy distribution from which one can compute the number of ionizing photons and the ionizing luminosity. The values we derived are listed in columns 10 and 11 of Table \ref{tab_param}. Such quantities are important to understand the physics and dynamics of \HII\ regions (see Sect.\ \ref{disc_wind}).

\subsubsection{Age determination}
\label{s_age}

\begin{figure*}[!ht]
\begin{center}
\begin{minipage}[b]{0.4\linewidth} 
\centering
\includegraphics[width=7cm]{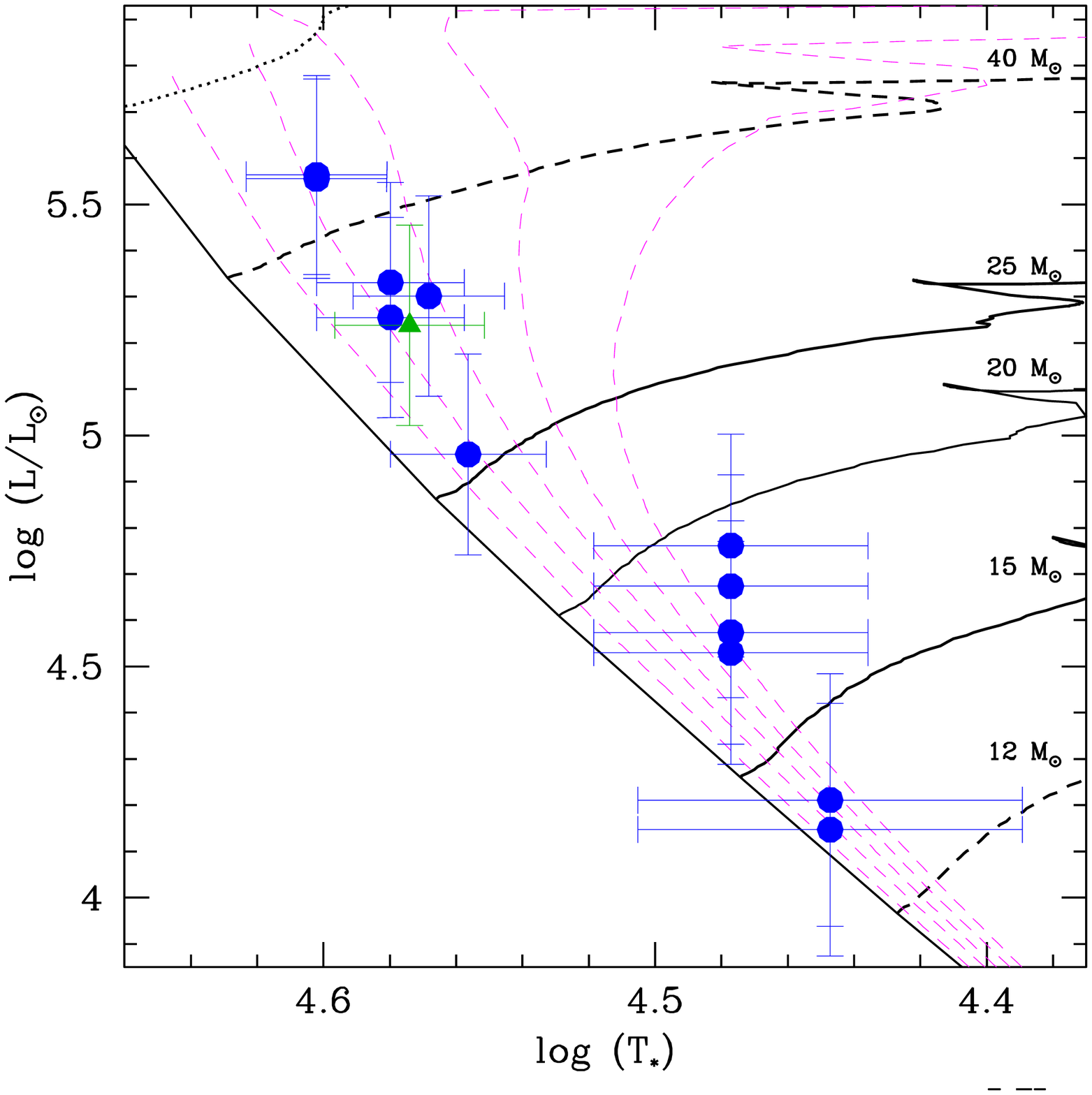}
\end{minipage}
\hspace{0.5cm} 
\begin{minipage}[b]{0.4\linewidth}
\centering
\includegraphics[width=7cm]{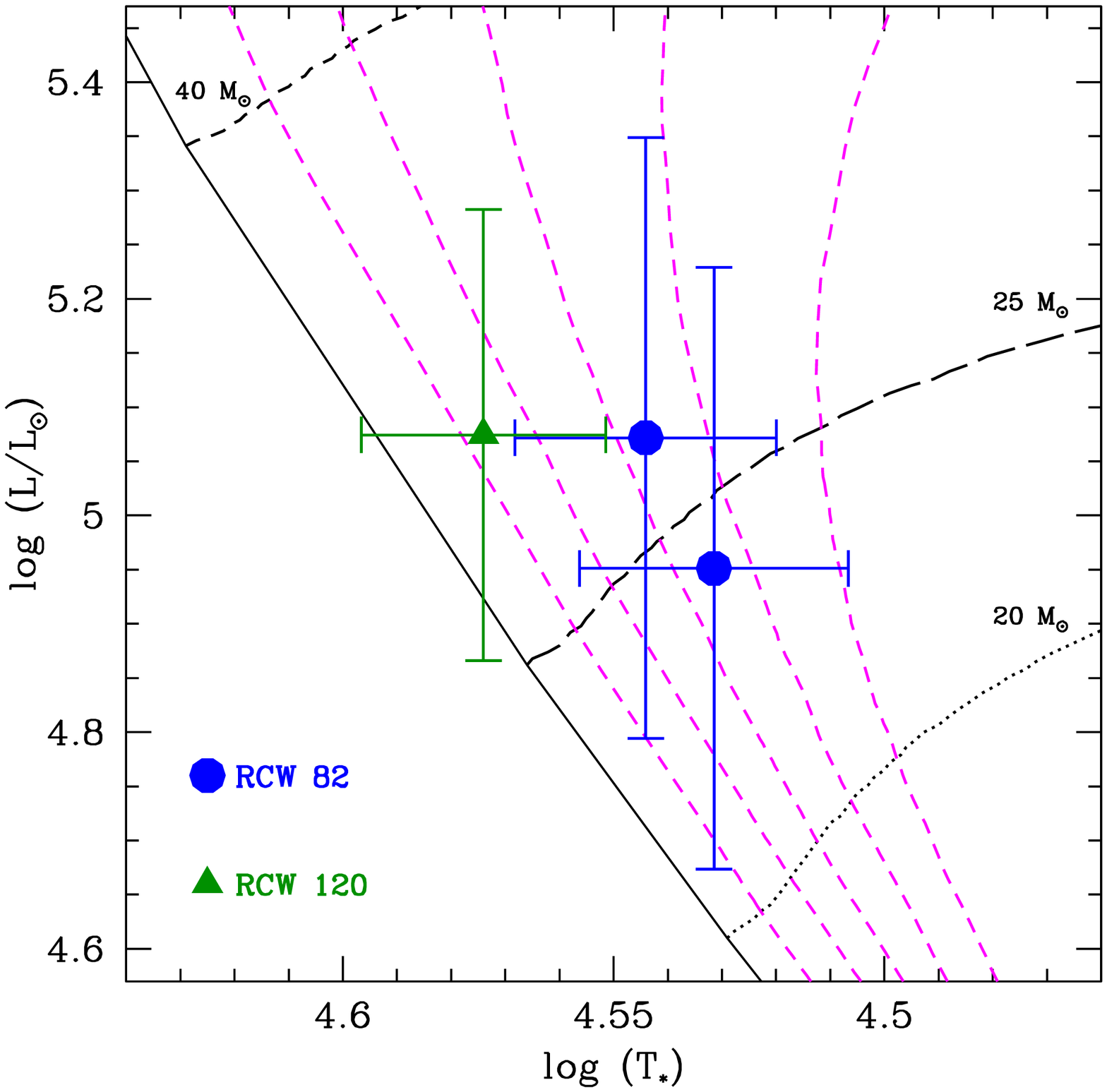}
\end{minipage}
\caption{HR diagram of the ionizing sources of RCW~79 (left) and RCW~82 and RCW~120 (right). Evolutionary tracks are from \citet{mm03}. Isochrones corresponding to ages of 1, 2, 3, 4, 5 Myr are shown by magenta dashed lines. \textit{Left}: Blue circles are the ionizing stars of RCW~79 and the green triangle is star number 6 in the compact \HII\ region on the border of RCW~79. \textit{Right}: the ionizing source of RCW~120 is shown by the green triangle, while the two O stars ionizing RCW~82 (sources e2 and e3) are shown by the blue circles. The Be star (e1) is not shown because its properties have not been derived in our study.} \label{hr_rcw}
\end{center}
\end{figure*}

\begin{figure}[]
\centering
\includegraphics[width=9cm]{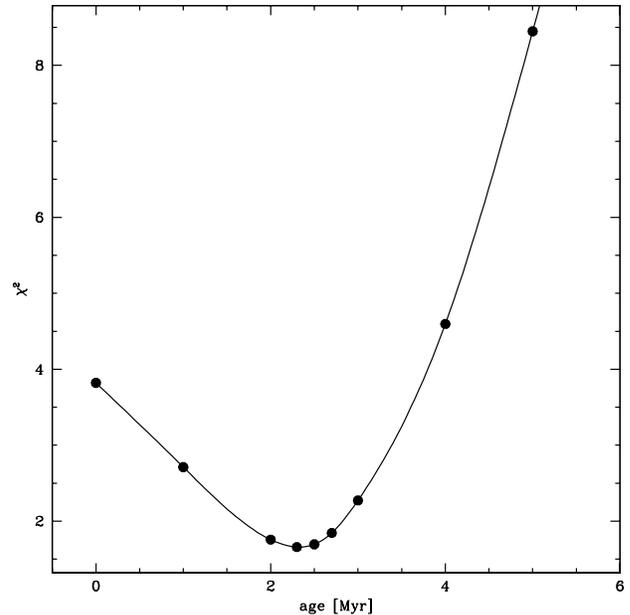}
\caption{Age estimate of the cluster of the ionizing sources of RCW~79. A value of 2.3$\pm$0.5 Myr is obtained. The solid line is a spline fit curve. }\label{age_rcw79}
\end{figure}

Fig.\ \ref{hr_rcw} (left) shows the HR diagram of the ionizing cluster of RCW~79. Evolutionary tracks are from \citet{mm03} and include the effects of rotation. All stars lie rather close to the main sequence, on a relatively narrow band, indicating that they most likely formed in a single star formation event. By finding the theoretical isochrone best representing the distribution of stars, the age of the population can be derived. For this, we use the following function:

\begin{equation}
\label{eq_age}
\chi_{iso}^{2}=\sum_{i=1}^{n} \frac{(log(T_{eff}^{i})-log(T_{eff}^{iso}))^{2}}{\sigma^{2}(log(T_{eff}^{i}))} + \frac{(log(L^{i})-log(L^{iso}))^{2}}{\sigma^{2}(log(L^{i}))}
\end{equation}

\noindent where ``iso'' stands for ``isochrone'' and $n$ is the number of stars. $T_{eff}^{iso}$ and $L^{iso}$ are the coordinates of the point of the isochrone closest to the values (\teff, $L$) representing a star's position. Hence, $\chi_{iso}^{2}$ is an estimate of the sum over all stars of the distance of a star's position to the selected isochrone. The isochrone for which this distance is minimized indicates the age of the population. In practice, $\chi_{iso}^{2}$ is evaluated for several isochrones corresponding to ages from 0 to 5 Myr. The results are displayed in Fig.\ \ref{age_rcw79}. The minimum is found for an age of 2.0--2.5 Myr and a typical uncertainty of $\sim$0.5 Myr. Note that this estimate takes all stars into account. If we use only the five brightest stars, we find a very similar result. This is because the lower part of the HRD is almost insensitive to age due to larger error bars and small separation between isochrones.

The position of the brightest source of the compact \HII\ region (CHII) on the border of RCW~79 (see Figs.\ 4 of ZA06 and Fig.\ \ref{source_id_uchii}) is shown by the green triangle in Fig.\ \ref{hr_rcw} (left). It is essentially indistinguishable from the position of the ionizing sources. An age estimate for that star would give the same result as for the ionizing sources, within the error bars. The other stars of the CHII region do not help to refine this estimate since they are much fainter and cooler and fall in the region of the HR diagram where the isochrones are tightly packed. Given the errors on the effective temperature and luminosity of star number 6, one can exclude an age younger than 0.5 Myr. This means that if there is any age difference between the CHII region and the ionizing stars of RCW~79 (as would be expected in case of triggered star formation) it is not larger than 2 Myr (see also ZA06). 

The right panel of Fig.\ \ref{hr_rcw} shows the position of the ionizing stars of RCW~82 and RCW~120. Only the two O stars (e2 and e3) are shown for RCW~82 since we have not analyzed the properties of the Be star e1 (our atmosphere code does not allow a treatment of circumstellar material). Unlike RCW~79, the small number of stars prevents an accurate age determination. Besides, the large uncertainty on \teff\ and the relatively low luminosity of the objects (compared to the brightest stars in RCW~79) complicates any attempt to constrain the stars' age. The only safe conclusion one can draw is that the ionizing stars are younger than 5 Myr. Other than that, any age smaller than 5 Myr is possible.

From interpolation between evolutionary tracks in Fig.\ \ref{hr_rcw}, one can estimate the masses of the stars of RCW~79. They are summarized in Table \ref{tab_param}. Five stars have masses larger than 30\msun. If we assume a Salpeter IMF between 0.8 and 100 \msun\ for the entire population formed with the ionizing sources, one finds a total mass of about 2000\msun. Hence, if the star formation event which produced the ionizing stars of RCW~79 also gave birth to lower mass stars, a rather massive cluster was born 2.0--2.5 Myr ago. In that case, the faint sources seen in Fig.\ \ref{im_rcw79_ex} are probably intermediate mass main sequence stars born at that time.

\begin{table*}
\begin{center}
\caption{Derived stellar properties of the ionizing stars of RCW~79, RCW~82 and RCW~120. }
\begin{tabular}{llcccccc}
\hline
Source          & ST & \teff\ & \lL\ & M$_{evol}$ & log(Q$_{0}$) & log(L$_{0}$) \\
                &    & [kK]    &     & [\msun]   & &  \\
\hline                                                              
RCW79 2    & O7.5--8V/III &  36$\pm$2 & 4.95$\pm$0.22 & $26.2^{+6.1}_{-4.8}$   & 48.45$\pm$0.23 & 37.90$\pm$0.23 \\
RCW79 4a   & O4--6V/III   &  40$\pm$2 & 5.56$\pm$0.22 & $46.1^{+13.3}_{-12.7}$ & 49.25$\pm$0.23 & 38.74$\pm$0.23 \\
RCW79 5a   & O4--6V/III   &  40$\pm$2 & 5.56$\pm$0.22 & $46.1^{+13.3}_{-12.7}$ & 49.25$\pm$0.23 & 38.74$\pm$0.23 \\
RCW79 6a   & O6--8V/III   &  38$\pm$2 & 5.33$\pm$0.22 & $36.2^{+7.2}_{-6.2}$   & 48.91$\pm$0.23 & 38.40$\pm$0.23 \\
RCW79 6b   & O9--B2V      &  28$\pm$4 & 4.15$\pm$0.27 & $13.2^{+3.5}_{-3.8}$   & 47.00$\pm$0.31 & 36.42$\pm$0.31 \\
RCW79 7    & O6--8V/III   &  37$\pm$2 & 5.30$\pm$0.22 & $35.1^{+6.4}_{-7.3}$   & 48.80$\pm$0.23 & 38.28$\pm$0.23 \\
RCW79 8a   & O9--B2V/III  &  30$\pm$3 & 4.76$\pm$0.24 & $19.2^{+5.0}_{-3.7}$   & 47.33$\pm$0.26 & 36.75$\pm$0.26 \\
RCW79 8b   & O9--B2V/III  &  30$\pm$3 & 4.67$\pm$0.24 & $18.6^{+4.6}_{-4.2}$   & 46.75$\pm$0.26 & 36.16$\pm$0.26 \\
RCW79 9a   & O9--B2V/III  &  30$\pm$3 & 4.57$\pm$0.24 & $17.7^{+4.3}_{-4.4}$   & 47.95$\pm$0.26 & 37.38$\pm$0.26 \\
RCW79 9b   & O9--B2V/III  &  30$\pm$3 & 4.53$\pm$0.24 & $17.3^{+4.1}_{-4.5}$   & 48.52$\pm$0.26 & 37.99$\pm$0.26 \\
RCW79 10   & O6.5--8V/III &  38$\pm$2 & 5.26$\pm$0.21 & $34.5^{+6.8}_{-5.9}$   & 48.96$\pm$0.22 & 38.45$\pm$0.22 \\
RCW79 12b  & O9--B2V/III  &  28$\pm$4 & 4.21$\pm$0.27 & $13.6^{+3.7}_{-3.8}$   & 47.06$\pm$0.31 & 36.48$\pm$0.31 \\
 & & & & & & \\
CHII \#6  & O7.5--9.5V/III& 37.5$\pm$2& 5.25$\pm$0.22 & $34.1^{+6.6}_{-6.8}$   & 48.79$\pm$0.23 & 38.26$\pm$0.23 \\
 & & & & & & & \\
RCW82 e2   & O9--B2V/III  & 34$\pm$2  & 4.95$\pm$0.27 & $24.1^{+6.5}_{-5.0}$ & 48.37$\pm$0.28 & 37.81$\pm$0.28 \\
RCW82 e3   & O9--B2V/III  & 35$\pm$2  & 5.07$\pm$0.27 & $27.5^{+8.5}_{-5.8}$ & 48.59$\pm$0.28 & 38.05$\pm$0.28 \\
 & & & & & & \\
RCW120 e   & O6--8V/III   & 37.5$\pm$2& 5.07$\pm$0.21 & $29.9^{+5.9}_{-6.8}$ & 48.58$\pm$0.22 & 38.05$\pm$0.22 \\
\hline
\end{tabular}\\
\end{center}
{Note 1: The columns are: target name, spectral type, effective temperature, luminosity, evolutionary mass, ionizing flux, ionizing luminosity.} \\
{Note 2: The errors on the last two parameters only take into account the uncertainties on \teff\ and \lL.}\\
\label{tab_param}
\end{table*}


\section{YSOs on the borders of \HII\ regions}
\label{s_yso}

In this section we now focus on the second groups of objects: embedded sources on the borders of the \HII\ regions. They have been classified as candidate YSOs by ZA06, ZA07, DE09 and PO09. Our aims are the following: 1) confirm spectroscopically the YSO nature of these sources, 2) get more insight into their physical properties (presence of disks, envelopes, jets, outflows). For this, twenty-three YSOs have been observed: 8 around RCW~79, 5 around RCW~82, and 10 around RCW~120. They were selected mainly from their high $K$--band magnitude. Their position and photometry is given in Table \ref{TableYSOs}. Appendix \ref{ap_yso} gives further details on each individual objects. We first present a general description of their spectral properties (Sect.\ \ref{yso_spec}) before confronting them to near and mid infrared photometry (Sect.\ \ref{yso_photom}). We then focus on YSOs with strong \htwo\ emission lines in Sect.\ \ref{yso_h2}. The morphological and kinematic properties of a few objects are presented in Sect.\ \ref{dyn_yso} where we show that the YSOs are associated to the HII regions.


\subsection{Spectroscopy of embedded sources}
\label{yso_spec}

The 23 objects we have observed on the borders of the three \HII\ regions show four main type of spectral signatures: \brg\ emission, \htwo\ emission, CO bandheads emission and CO bandheads absorption. Some sources are entirely dominated by one type of spectroscopic signatures, but others can show several features, such as \brg\ and \htwo\ emission, or \htwo, \brg\ and CO emission. Table \ref{tab_specYSOs} summarizes the main spectroscopic properties of each source. Their spectra are shown in Figs.\ \ref{spec_yso} and \ref{spec_mixed}. Below, we describe the various sources, grouping them according to the main feature they display.

\begin{itemize}

\item[$\bullet$] \textit{\htwo\ emission sources}: ro--vibrational \htwo\ transitions are observed in these objects. The strongest one is \htwo\ 1--0S(1) at 2.122 \mum\ and it is sometimes the only line observed. However, in the majority of ``\htwo\ emission sources'' a number of other lines are observed. They are identified on Fig.\ \ref{spec_yso} (top right). Such lines are known to be formed either thermally in shocks \citep[e.g.][]{burton90}, or by fluorescence after excitation of \htwo\ molecules by non ionizing FUV radiation in the Lyman-Werner band \citep[e.g.][]{bvd87}. In the latter case, the lines are formed under non LTE conditions. They can thus trace either outflows or photo-dissociation regions (PDRs) in young star forming regions. In Sect.\ \ref{yso_h2}, we will analyze quantitatively these sources to track the nature of the exciting mechanism. 

\item[$\bullet$] \textit{\brg\ emission sources}: strong \brg\ emission is observed in several sources, sometimes in combination with \htwo\ emission. \brg\ may have various origins. According to Bik et al.~(2006), the large line widths (100--200 \kms) could be explained by ionized gas flowing from the surface of a circumstellar rotating disk, rather than a pure nebular origin. Other mechanisms such as accretion flows or outflows are alternative possibilities. In the former case, magnetospheric accretion \citep{vda05} or inner disk accretion \citep{muz04} theoretical models predict \brg\ emission. For outflows, classical stellar winds, X--winds \citep{shu94} or disk winds \citep[e.g.]{bp82} lead to ionized hydrogen emission. Interferometric observations of Herbig Ae/Be stars by \citet{kraus08} favour \brg\ emission in extended stellar or disk winds.

\item[$\bullet$] \textit{CO bandheads emission sources}: CO first overtone emission at 2.2935 \mum\ (CO(2-1)), 2.3227 \mum\ (CO(3-1)), 2.3535 \mum\ (CO(4-2)), 2.3829 \mum\ (CO(5-3)) is observed in eight sources. This spectral signature has been observed in a number of low and high mass YSOs \citep{sco79,chandler93,bik06}. It is commonly attributed to irradiation of an accretion disk by the central object \citep{chandler95,kraus00}. This interpretation is motivated by the temperature (2500 to 5000 K) and density ($n \sim 10^{10-11}$ cm$^{-3}$) conditions required to produce such an emission, placing the emitting zone close to the central star \citep[within 0.5 AU for the Be star 51 Oph studied by][]{berthoud07,tatulli08}. The geometry of the bandheads depends on the Keplerian rotation and viewing angle. \citet{bik04} used the formalism of \citet{kraus00} to show that CO emission in the massive YSOs observed by \citet{bik05} was produced in a keplerian disk. Our spectra do not have high enough resolution to detect line asymmetries. In the strongest CO emitters of our sample, \ion{Na}{i} 2.206--2.209 \mum\ and \ion{Ca}{i} 2.261--2.264 \mum\ emission is also detected (see Fig.\ \ref{spec_yso}).

\item[$\bullet$] \textit{CO bandheads absorption sources}: the CO first overtone lines are observed in absorption in a few sources (Fig.\ \ref{spec_yso}, bottom right). The line depth ranges from less than 10$\%$ to about 40$\%$ of the continuum.. \ion{Na}{i} 2.206--2.209 \mum, \ion{Ca}{i} 2.261--2.264 \mum\ and \ion{Mg}{i} 2.281 \mum\ lines are usually observed in absorption as well. In one object (RCW120 51b) a significant \brg\ emission is observed, while in another one (RCW82 98) \htwo\ emission lines are present. CO absorption in the $K$--band is typical of cool stars (red supergiants, AGBs, giants), but it is also detected in YSOs \citep{cm92,hoff06,aspin09}. In the latter case, it is thought that CO absorption is observed due to the absence of a strong disk and/or envelope emission which otherwise dominates the $K$--band spectrum and leads to either featureless or emission line spectra. This is consistent with the fact that CO absorption YSOs tend to be class II (and not class I) objects \citep{cm92}. 
CO absorption is also observed in the rare class of eruptive FU Orionis objects \citep[FUOr,][]{reip97}. They are thought to be active T--Tauri stars experiencing an accretion burst. \citet{aspin09L} collected spectra of V1647 Orionis, the FUOr star responsible for the appearance of the McNeil nebula \citep{mcneil04}, and show that its CO bandheads shift from absorption to emission when the star experiences a FUOr event. \citet{calvet91} explained theoretically this type of transition by an increase of the accretion rate. EXor variables which are pre main sequence objects similar to Fuor stars but with weaker outbursts also display CO absorption bandheads \citep{lorenz09}.

\end{itemize}

All these spectral signatures are typical of YSOs. In spite of this general grouping in four categories, we stress again that a few sources share several of these spectral characteristics, highlighting the complexity of the physical phenomena associated with YSOs. In the following, we attempt to better characterize the nature of the sources using infrared photometry.


\subsection{Near and mid infrared photometry}
\label{yso_photom}

\begin{figure}[]
\centering
 \includegraphics[angle=0,width=9cm]{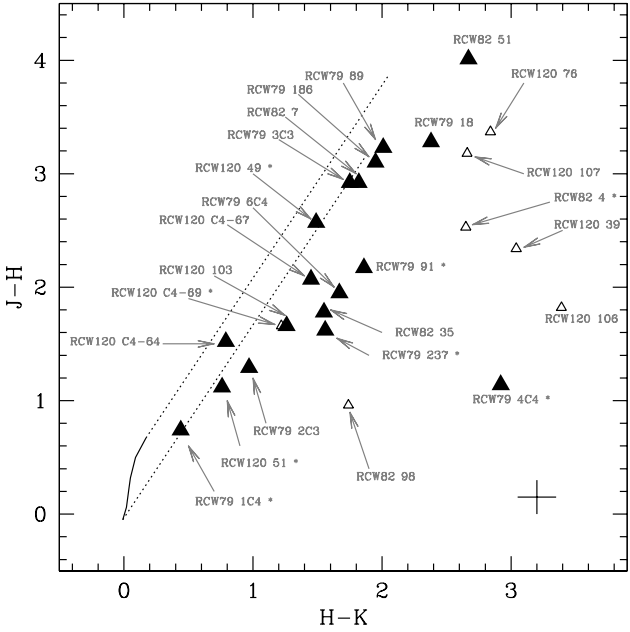}
  \caption{$JHK$ Color-color diagram (magnitudes from 2MASS
  and NTT-SofI) of the YSOs associated with RCW~79, RCW~82 and RCW~120. The location of main sequence stars is shown by the solid line.
  Reddening lines have a length corresponding to a visual extinction
  of 40~mag. Filled (open) triangles correspond to good (uncertain, from 2MASS catalog flag) magnitude measurements. Star symbols following the names indicate multiple sources. The typical error bar of good measurements is indicated. }\label{ccd_jhk}
\end{figure}

All our sources have been observed by 2MASS. We have seen that in the case of the ionizing stars of RCW~79, some 2MASS sources were actually multiple. Similarly, some YSO candidates have been resolved into several sub--components by our SINFONI observations (see Figs.\ \ref{source_id_1}, \ref{source_id_2} and \ref{source_id_120c4}). Hence, the 2MASS photometry is only indicative for those specific sources. Fig.\ \ref{ccd_jhk} shows a $JHK$ color--color diagram based on 2MASS photometry. Star symbols refers to multiple sources. When available, we have used ESO/NTT/SofI photometry from ZA06, ZA07 and PO09 since it is more accurate and does not suffer from the same uncertainties. The main sequence is also plotted together with A$_{V}$=40 mag reddening vectors. The main conclusion is that most of the objects show near--IR excess. Eight objects lie close to the reddening vectors. Out of them, three are multiple sources. Among the five remaining objects, RCW120 C4-64 shows CO absorption bandheads and is most likely a foreground giant star. The near--IR excess observed in the objects far from the reddened main sequence is usually indicative of the presence of an envelope or a disk. It thus confirms that most of our sources are YSOs.

\begin{figure}[ht]
\centering
 \includegraphics[angle=0,width=9cm]{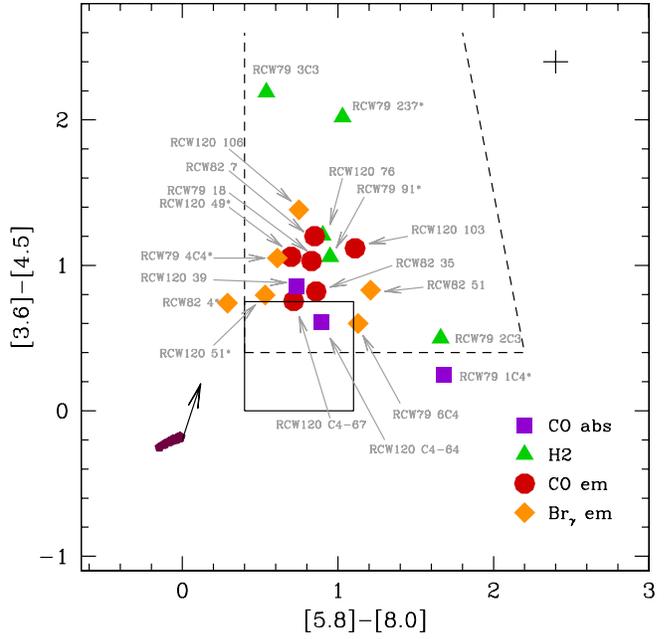}
  \caption{Color--color diagram with SPITZER/IRAC magnitudes of the YSOs associated with RCW~79, RCW~82 and RCW~120. Different symbols are used for the four spectroscopic groups defined in Sect.\ \ref{yso_spec}: triangles for \htwo\ sources, squares for CO absorption sources, circles for CO emission sources and diamonds for \brg\ dominated sources. Pentagons indicate main sequence massive stars \citep[from][]{msh05}. Colors of class II (respectively I) sources are delineated by solid (resp.\ dashed) lines according to \citet{allen04}. The reddening vector shown by the black arrow is computed for a visual extinction of 40 magnitudes with the law of \citet{indeb05}. Star symbols following source names indicate multiple sources.}\label{ccd_irac}
\end{figure}

To get more insight into the nature of the sources, we have retrieved
Spitzer/IRAC photometry (when available) from the GLIMPSE survey
\citep{ben03} to build the color--color diagram shown in
Fig.\ \ref{ccd_irac}. The different symbols correspond to the four
spectroscopic groups defined in Sect.\ \ref{yso_spec}. Class I and II
objects gather within the dashed and solid lines respectively. Stars
are shown by pentagons close to (0,0).

All our sources fall in the area corresponding to Class I and Class II
YSOs.  They gather around the color point (0.8, 0.8); some of them
may be Class II sources displaced by extinction (see
Fig.\ \ref{ccd_irac}). As discussed by \citet{robitaille08} a few of
these sources may be extreme AGB stars \citep[see also][]{sri09}.  It
is probably the case of the CO absorption sources RCW120 39 and
64. Four sources have different colors: two are pure H$_{2}$ sources
(RCW79 237 and 3C3), the two others have colors typical of PDRs
(RCW79 1C4, associated with extended 8~$\mu$m and 24~$\mu$m emission,
and RCW79 2C3). The two sources with the largest [3.6]$-$[4.5] color are the ones with the strongest \htwo\ lines (Fig.\ \ref{spec_yso}, top right). \citet{sr05} showed that numerous \htwo\ lines contributed to the IRAC fluxes, but that the integrated line intensity was the largest in the 4.5~\mum\ band. Hence, the presence of strong \htwo\ lines in the $K$--band of RCW79 3C3 and RCW79 237 is probably responsible for an intense emission in the 4.5~\mum\ band, leading to an increased flux compared to the 3.6~\mum\ channel. Consequently, the value of [3.6]$-$[4.5] is larger in those two objects, explaining their location in Fig.\ \ref{ccd_irac}. The other \htwo\ dominated sources have weaker lines, and consequently their IRAC colors are not affected.

Except for the four cases discussed above, the general conclusion is that objects with similar IRAC colors can present different spectroscopic signatures. Said differently, there is no direct correlation between IRAC colors and a spectroscopic group. Once again, the multiplicity of a few sources complicates the interpretation.

\begin{figure}[ht]
\centering
 \includegraphics[angle=0,width=9cm]{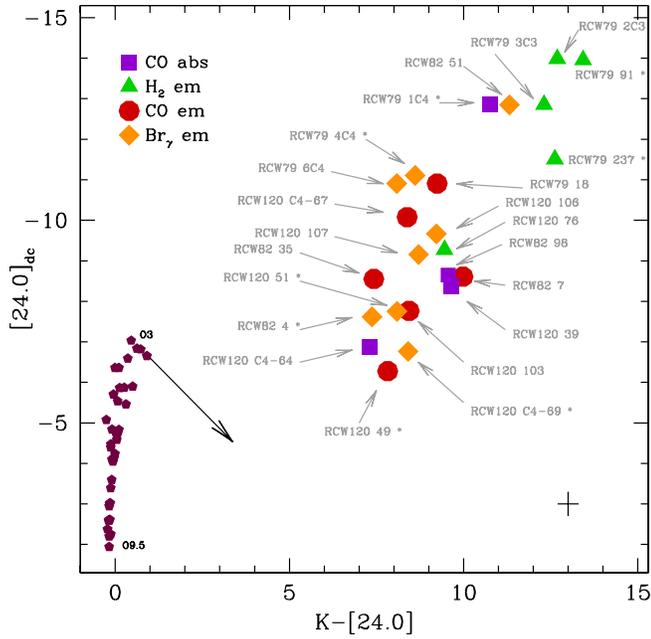}
  \caption{Color--magnitude diagram with SPITZER/MIPS 24~\mum\ magnitudes of the YSOs associated with RCW~79, RCW~82 and RCW~120. On the Y axis, the magnitudes have been corrected for distance. The symbols have the same meaning as in Fig.\ \ref{ccd_irac}. The arrow indicates an extinction of A$_{V}$=40 magnitudes \citep[law of][]{lutz99}. The pentagons are O stars from the theoretical SEDs of \citet{msh05}. Star symbols following source names indicate multiple sources.}\label{cmd_mips}
\end{figure}

Fig.\ \ref{cmd_mips} shows a color--magnitude diagram based on Spitzer/MIPS 24~\mum\ photometry \citep{carey09}, complemented by AKARI observations (Zavagno et al.\ in prep). The magnitude in this band corrected for the distance modulus of each source (noted [24]$_{dc}$), is plotted as a function of $K-$[24]. For comparison, the theoretical SEDs of \citet{msh05} have been used to calculate the positions of O stars in this diagram. They are shown by the pentagon symbols. A reddening vector corresponding to A$_{V}$=40 mag and to the extinction law of \citet{lutz99} is indicated. All YSOs are brighter at 24~\mum\ and much redder (in $K-$[24]) than O stars. This indicates once more that they have a strong infrared excess due to envelopes and/or disks which dominate the SED at those wavelengths. The H$_{2}$ sources seem to stand out in this diagram. While most sources are grouped in a region defined by -12 $<$ [24]$_{dc}$ $<$ -6 and 7 $<$ $K-$[24] $<$ 10, four out of the five \htwo\ sources for which infrared photometry is available are much brighter and redder ([24]$_{dc}$ $<$ -12 and $K-$[24] $>$ 10). This might be an indication that those objects have larger amount of circumstellar material tracing an earlier state of evolution, or that they have larger masses.

\begin{figure}[h]
\centering
 \includegraphics[angle=0,width=9cm]{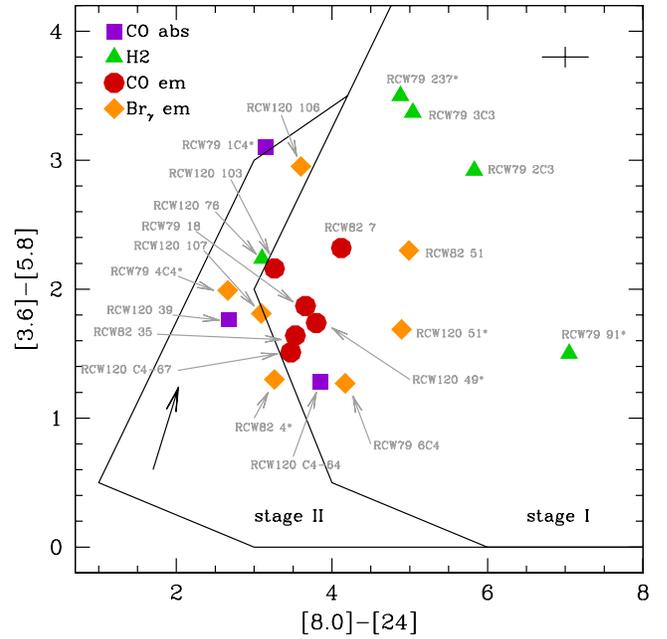}
  \caption{[3.6]$-$[5.8] vs [8.0]$-$[24] diagram. The symbols have the same meaning as in Fig.\ \ref{ccd_irac}. The arrow indicates an extinction of A$_{V}$=40 magnitudes. The solid lines delineates the areas where stage I and II objects of \citet{robitaille06} are found (see their Fig.\ 23). Star symbols following source names indicate multiple sources.}\label{cmd_iracmips}
\end{figure}

This is confirmed by Fig.\ \ref{cmd_iracmips} showing the [3.6]$-$[5.8] vs [8.0]$-$[24] diagram. Here again, the \htwo\ objects are clearly located towards redder colors (with the exception of RCW120 76). The solid lines delineates the location of stage I and II objects of \citet{robitaille06}. This classification is the theoretical analog of class I and II sources, meaning that stage I objects correspond to models with an envelope, while stage II sources have disks and possibly a tenuous envelope. \htwo\ sources are therefore consistent with being stage I objects as suspected previously. The bulk of the other objects are again located in the transition region between stage I and stage II sources. Only RCW120 51 and RCW82 51 might appear as stage I objects. We will discuss further these results in Sect.\ \ref{disc_yso} using additional findings presented below.

The main conclusion of this analysis of color--color and color--magnitude diagrams is that the objects dominated by \htwo\ emission seem to be different from the remaining of the sources we observed. They probably have an optically thick envelope. The other objects are more consistent with YSO surrounded by disks. Another conclusion regarding these sources is that they can have different spectroscopic signatures but similar colors. This can be partly explained by geometrical effects. Indeed, depending on the inclination and size of the putative disks, different regions are visible, and thus different spectral lines are expected. The corresponding photometry is also modified. \citet{whitney03a,whitney03b} showed that in the [3.6]$-$[4.5] vs [5.8]$-$[8.0], a spread of about one magnitude can be expected due to inclination effects. This corresponds to what we observe. Hence, among the CO emitters, absorbers and \brg\ emitters, one probably sees a population of class II objects at different inclinations.

\begin{table*} \tiny{ \caption{2MASS and NTT/SofI $JHK$, Spitzer-GLIMPSE, Spitzer-MIPS and AKARI magnitudes of YSOs. \label{TableYSOs}}
\begin{tabular}{l r r r r r r r r r r r}
\hline\hline
Star & RA & DEC & $J$ & $H$ & $K_{\rm{S}}$ & [3.6] & [4.5] & [5.8] & [8.0] & [24] & nature\\
       & h m s & $^{\circ} \arcmin\ \arcsec$ & & & & & & \\
 \hline
 \hline
 \multicolumn{11}{c}{\bf YSOs around RCW~79} \\
 \hline
 237    & 13:41:02.3 & -61:44:16 & 17.41  & 15.79  & 14.23 & 11.02       & 9.00 & 7.52   & 6.49 & 1.61$^{a}$ & Class~I \\
 91     & 13:40:57.6 & -61:45:44 & 16.62  & 14.45  & 12.59 & 8.66$^{b}$  & 7.60$^{b}$ & 7.16 & 6.21 & -0.84$^{a}$ & Class~I \\
 18     & 13:40:58.3 & -61:46:50 & 17.11  & 13.83  & 11.45 & 8.57        & 7.54 & 6.70 & 5.87 & 2.21 & Class~I \\
 89     & 13:40:59.0 & -61:46:51 & 18.58  & 15.35  & 13.34 & 10.83       & 10.71& --   & --   & --  & -- \\
 186    & 13:40:58.5 & -61:46:48 & 18.96  & 15.86  & 13.91 & --          & --   & --   & --   & --  & -- \\
 2C3    & 13:40:32.7 & -61:47:20 & 14.08  & 12.79  & 11.82 & 9.54        & 9.04 & 6.62 & 4.96 & -0.87$^{a}$ & Class~I \\
 3C3    & 13:40:26.6 & -61:47:56 & 17.24  & 14.32  & 12.57 & 9.21        & 7.02 & 5.84 & 5.30 & 0.26$^{a}$ & Class~I \\
 1C4    & 13:39:53.6 & -61:41:21 & 12.19  & 11.45  & 11.01 & 8.19$^{b}$  & 7.94$^{b}$ & 5.09$^{b}$ & 3.41$^{b}$ & 0.26$^{a}$ & -- \\
 4C4    & 13:39:55.9 & -61:40:58 & 14.68  & 13.54  & 10.62 & 7.27        & 6.22$^{b}$ & 5.28 & 4.67 & 2.01$^{b}$ & Class~I \\
 6C4    & 13:40:08.9 & -61:41:38 & 13.91  & 11.96  & 10.29 & 8.78        & 8.18 & 7.51 & 6.38 & 2.21$^{a}$ & Class~I \\
 \hline
 \multicolumn{11}{c}{\bf YSOs around RCW~82} \\
 \hline
 4      & 13:59:34.8 & -61:19:22 & 17.59: & 15.06 & 12.41 & 9.89 & 9.15 & 8.59 & 8.30 & 5.04 & Class~I \\
 7      & 13:59:47.3 & -61:22:02 & 18.77$^{c}$ & 15.85$^{c}$ & 14.03$^{c}$ & 11.34 & 10.14 & 9.02 & 8.17 & 4.05 & Class~I  \\
 35     & 13:59:50.9 & -61:25:22 & 14.29 & 12.51 & 10.96 & 8.61 & 7.79 & 6.97 & 6.11 & 2.58 & Class~I \\
 51     & 13:59:57.6 & -61:24:37 & 17.81$^{c}$ & 13.80$^{c}$ & 11.13$^{c}$ & 8.31 & 7.48 & 6.01 & 4.80 & -0.19 & Class~I \\
 98     & 13:59:36.7 & -61:20:48 & 16.27: & 15.31: & 13.57 & 10.09 & 9.24 & 8.47 & -- & 4.01 & -- \\
 \hline
 \multicolumn{11}{c}{\bf YSOs around RCW~120} \\
 \hline
 39     & 17:12:00.9 & -38:31:40  & 17.29: & 14.95: & 11.91  &  7.43  &  6.57  &  5.66  &  4.93  &  2.26 & Class~II \\
 49     & 17:12:08.0 & -38:30:52  & 16.24  & 13.67  & 12.18  & 10.59  &  9.53  &  8.85  &  8.16  &  4.36 & Class~I/II \\
 51     & 17:12:28.5 & -38:30:49  & 12.85  & 11.73  & 10.97  & 10.00  &  9.20  &  8.31  &  7.78  &  2.88 & Class~I/II \\
 C4-64  & 17:12:40.8 & -38:27:29  & 13.37  & 11.85  & 11.06  &  9.78  &  9.17  &  8.50  &  7.60  &  3.75 & Class~I/II \\
 C4-67  & 17:12:41.1 & -38:27:17  & 13.03  & 10.96  &  9.51  &  7.78  &  7.03  &  6.27  &  5.56  &  2.09 & Class~I/II \\
 C4-69  & 17:12:42.2 & -38:26:58  & 15.16: & 13.50  & 12.28  & 10.82  &  9.75  &  8.78  &    --  &  3.87 & Class~I/II \\
 76     & 17:12:46.0 & -38:25:25  & 17.03: & 13.66  & 10.82  &  7.61  &  6.40  &  5.37  &  4.47  &  1.37 & Class~I/II \\
 103    & 17:12:40.2 & -38:20:31  & 14.23  & 12.57  & 11.31  &  9.40  &  8.29  &  7.24  &  6.13  &  2.87 & Class~I/II \\
 106    & 17:11:50.7 & -38:19:55  & 17.62: & 15.80: & 12.41  &  8.27  &  6.89  &  5.79  &  4.95  &  0.97 & Class~I \\
 107    & 17:12:33.6 & -38:19:52  & 16.03: & 12.85  & 10.19  &  7.13  &    --  & 5.32   &  4.57  &  1.48 & Class~I/II \\
 \hline
 \label{tab_yso}
\end{tabular}\\
\\
{Note: Classification into class I/II objects is from ZA06, PO09, DE09. The notation ``:'' uncertain flux measurements due to the weakness of the source. The typical errors are 0.1 mag (respectively 0.3 mag) for IRAC (resp. MIPS) photometry.} \\
{$^{a}$ AKARI magnitude} \\
{$^{b}$ Magnitude obtained by aperture photometry (see ZAV05)} \\
{$^{c}$ NTT magnitude}
 }
\end{table*}

\begin{sidewaystable*}
\begin{center}
\caption{Characteristics of the YSO spectra. When present, lines are in emission unless stated differently.}
\begin{tiny}
\begin{tabular}{llllll}
Source & Br$\gamma$ & H$_2$ & CO & Notes\\
\hline
\multicolumn{6}{c}{\bf YSOs around RCW~79} \\
\hline
 237    & no & 2.0338, 2.1218, 2.2233, 2.2477 & no & \\
 237-b  & no & no & absorption & \ion{Na}{i}, \ion{Ca}{i} absorption \\
 91     & yes & 2.0338, 2.1218, 2.2233 & no &  \\
 91-a   & yes & 2.033, 2.1509 & emission & \\
 91-b   & yes & 1.9576, 2.0338, 2.1218, 2.2233 &  weak emission & \\
 18     & yes & no & emission & \ion{Na}{i} emission \\
 186    & yes & no & no & $+$3198 & \\
 89     & yes, weak & no & no &  \\
 2C3    & yes, on top of absorption & 2.0338, 2.1218, 2.2233 & no & \\
 3C3    & yes, weak & 2.0338, 2.1218, 2.2233, 2.2477 & no & \\
 3C3-b,c,d & no & 2.1218 & no &  \\
 1C4    & yes, absorption & no & absorption & \ion{Na}{i}, \ion{Ca}{i} absorption. PDR associated with this objet \\
 4C4    & yes, weak & no & no & faint companions \\
 6C4    & yes & no & no & \\
\hline
\multicolumn{6}{c}{\bf YSOs around RCW~82} \\
\hline
 4      & yes & no & no & \\
 7      & ? & 1.9576, 2.0338, 2.1218, 2.2233 & emission & \\
 35     & yes & 2.0338, 2.1218, 2.2233 & emission & \ion{Na}{i} emission\\
 51     & yes & 2.0338, 2.1218 & emission (weak) & \\
 98     & no & 1.9576, 2.0338, 2.1218, 2.2233, 2.4066, 2.4134 & absorption & \htwo\ lines probably not related to YSO, see Fig.\ \ref{source_id_1} \\
\hline
\multicolumn{6}{c}{\bf YSOs around RCW~120} \\
\hline
 39     & no & no & absorption & \ion{Na}{i}, \ion{Ca}{i} absorption \\
 49     & yes & 2.0338, 2.1218, 2.2233, 2.2477 & emission &  \\
 49-a   & no & 2.0338, 2.1218, 2.2233 & absorption & \\
 49-b   & no & 2.0338, 2.1218, 2.2233 & absorption? & \\
 51-a   & yes & no & no &  \\
 51-b   & no & no & absorption & \\
 51-c   & no & no & absorption & \ion{Na}{i}, \ion{Ca}{i} absorption \\
 C4-64  & absorption & no & absorption & \ion{Na}{i}, \ion{Ca}{i} absorption\\
 C4-67  & yes & 2.1218 & emission &  \\
 C4-69a & no & no & no & featureless \\
 C4-69b & yes & no & absorption & \\
 C4-69c & no & no & absorption & \ion{Na}{i}, \ion{Ca}{i} absorption\\
 C4 A   & absorption & no & absorption & PDR associated with this object \\
 C4 Ba  & absorption & no & no & PDR associated with this objet \\
 C4 Bb  & no  & no & absorption & \ion{Na}{i}, \ion{Ca}{i} absorption  \\
 76     & no & 2.0338, 2.1218, 2.2233 & no & \\
 103    & yes & 1.9576, 2.0338, 2.1218, 2.2233, 2.4066, 2.4237 & emission & \ion{Na}{i}, \ion{Ca}{i} emission\\
 106    & yes & no & no & \\
 107    & yes, weak & 2.0338, 2.1218, 2.2233 & no & \\
\hline
\label{tab_specYSOs}
\end{tabular}
\end{tiny}
\end{center}
\end{sidewaystable*}


\subsection{Nature of \htwo\ sources}
\label{yso_h2}

Several sources observed on the border of the \HII\ regions show
\htwo--dominated spectra (see top right panel of Fig.\ \ref{spec_yso}
for line identification). As described in Sect.\ \ref{yso_spec},
\htwo\ emission is due either to thermal processes or to fluorescence.
The relative intensity of individual lines is different in either
cases. Comparing \htwo\ line strength can thus lead to the underlying
excitation mechanism. For a thorough comparison, line fluxes must be
dereddened, which requires the knowledge of extinction. This piece of
information is difficult to obtain since the intrinsic SED of our
sources is not known. From Fig.\ \ref{ccd_jhk} one can see that most
objects have a visual extinction in the range 10--40 mag. We have thus
adopted a typical extinction A$_{K}$ = 2.5$\pm$1.5 for the present
study. We used the Galactic extinction law $\propto \lambda^{-1.7}$
similar to \citet{moneti01} to deredden the line fluxes.

\begin{figure*}[]
\centering
\includegraphics[width=16cm]{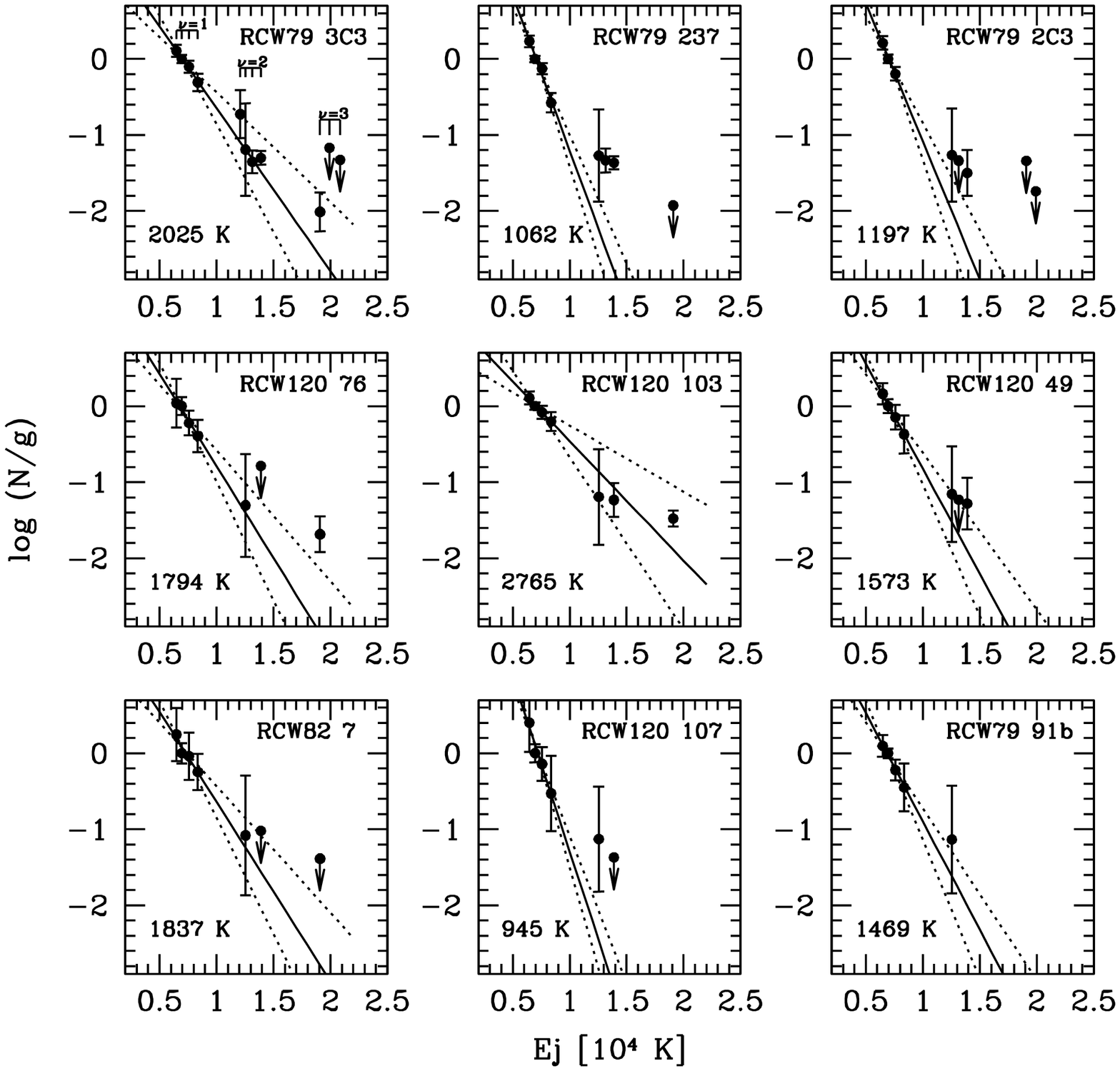}
\caption{H$_{2}$ excitation diagram of the sources showing H$_{2}$ emission spectra (either pure or mixed with other spectral features). All N/g numbers are relative to N/g for transition 1-0S(1). The solid line are fits to the $\nu=1$ level populations assuming LTE and A$_{K}$=2.5. The corresponding temperature is indicated for each YSO. The dotted lines are fits for the cases where A$_{K}$=1.0 and A$_{K}$=4.0. The error bars take into account measurement errors and an uncertainty of $\pm$1.5 in A$_{K}$. }\label{diag_h2}
\end{figure*}

We have computed the column densities $N_{i}$ of several excited levels and plotted them as a function of level energy $E_{i}$. This has been shown to be a powerful tool to distinguish between thermal and radiative emission \citep[e.g.][]{hase87}. If \htwo\ emission is thermal, the values of $N_{i}/g_{i}$ ($g_{i}$ being the statistical weight) follow a Boltzmann distribution and are thus aligned in a $log(N_{i}/g_{i}) - E_{i}$ diagram. Alternatively, radiative pumping should increase the population of high energy levels and introduce a departure from linearity in the aforementioned diagram. We have computed the column densities of several levels using the same formalism and molecular data as \citet{leticia08}. Our results are shown in Figs.\ \ref{diag_h2}. The solid and dotted lines correspond to a linear fit of the $\nu=1$ levels. The slope is directly related to the temperature of the Boltzmann distribution best representing these energy levels. Solid lines are for the case $A_{K}$=2.5, while the dotted lines are for $A_{K}$=1.0 and $A_{K}$=4.0. When extinction varies, line fluxes are affected, which explains the different slopes for different $A_{K}$. From Fig.\ \ref{diag_h2} RCW79 3C3, RCW120 103 and RCW79 91b appear consistent with pure thermal emission. For RCW 79 2C3, RCW120 76, RCW120 49 and RCW120 107, a marginal deviation from pure thermal emission is found, although the bulk of the emission seems to be thermal. Source RCW82 7 shows too few lines to draw any final conclusion. Besides, we will see in Sect.\ \ref{dyn_yso} that \htwo\ emission is much extended and patchy. Finally, RCW79 237 clearly shows a significant non thermal contribution.

Two more sources (RCW79 89 and RCW79 91) display \htwo\ lines, but not enough to build excitation diagrams. For these sources, we calculated the ratios $\frac{I_{1-0s(1)}}{I_{2-1S(1)}}$ and $\phi_{1}$ \citep[defined by][]{smith97}. Under LTE conditions, the first ratio should have values between 5 and 100 \citep[for temperatures ranging between 3000 and 1000 K respectively, see][]{gus08}, and the second ratio should be close to 3. When fluorescence dominates, lower values are expected. For RCW79 89, the observed values are 4.61$\pm$0.79 and 2.89$\pm$0.56 respectively, indicating that \htwo\ has probably multiple origins. For RCW79 91, FUV excitation is clearly favored since $\frac{I_{1-0s(1)}}{I_{2-1S(1)}}$ = 6.33$\pm$0.14 and $\phi_{1}$ = 0.99$\pm$0.14.

In conclusion, most sources show a thermal H$_{2}$ spectrum with, in some cases, a possible contribution from fluorescence. Three sources (RCW79 237, RCW79 89 and RCW79 91) are probably dominated by non thermal emission, most likely due to the presence of nearby PDRs. Thermal emission is usually attributed to shocks. \citet{davies00} summarized the various types of shocks encountered in the ISM \footnote{Two types of shocks are usually encountered: J--shocks (for Jump shocks) in which the change in density and velocity occurs on scales much shorter than the radiative cooling scale, and C--shocks (for Continuous shocks) in which the changes take place on a longer scale.} using the models of \citet{burton90}. They note in particular that fast J--shocks are usually dissociative. In that case, molecules are formed on grains in the post-shock region and are produced in excited states. Consequently, their emission spectrum is more typical of a non-thermal mechanism since the molecules will cascade down to lower energy levels. Hence, fast J--shocks seem to be excluded to explain the thermal emission of most of our H$_{2}$ sources. Similarly, slow C--shocks are not favored either since they are associated with rather low temperatures ($<$300 K). From Fig\ \ref{diag_h2} we see that temperatures of 1000--2500 K are estimated, inconsistent with slow C--shocks. Hence, slow J--shocks or fast C--shocks, both heating the gas to 2000--3000 K, are preferred to account for the observed thermal H$_{2}$ emission in our sources.


\subsection{Morphology and kinematics of selected YSOs}
\label{dyn_yso}

In this section we gather information on the morphology and kinematics of some YSOs in
order to study their physical association with the
\HII\ regions and to get more insight into the origin of the observed
emission lines.

We first focus on the kinematics of the YSOs themselves. The emission
lines of their spectra (\brg\ and/or H$_{2}$) have been fitted by 1D
Gaussians using QFitsView. We have first measured the full width at
half maximum (FWHM) of the sky lines to estimate the spectral
resolution of the observations. We obtain FWHM$=84 \pm 6$ \kms, thus a
resolution R$=3575 \pm 250$.  Then we have measured the central
positions and the widths of the H$_2$ and \brg\ lines (only for the
lines presenting a good signal to noise ratio). These quantities are
given in Table \ref{vel_yso}. The first and second columns identify
the YSO; the indication ``ext'' means that we have measured the
nebulous extended emission in the vicinity of the YSO.  The following
columns give, for a selection of lines, the LSR velocity ($V_{\rm
  LSR}$) and the intrinsic line width ($\Delta V$). The absence of
measurement for some lines is due to the poor quality of the data at
hand.  We estimate that the uncertainty on $V_{\rm LSR}$ is $\sim 10$
\kms. The relative uncertainty on the measured line widths if of
10$\%$, but the uncertainty on $\Delta V$ is large. Its is about 20
(40) \kms\ for wide (narrow) lines. This is mainly due to the large
uncertainty on the instrumental width. Hence, values given in Table
\ref{vel_yso} are only indicative.

\begin{table*}
   \caption{Velocity measurements. For each line, the first value is the LSR radial velocity while the second value (in parenthesis) is the line width. }\label{vel_yso}
   \begin{tabular}{l l l l l l l l l}
\hline\hline
Region & source & \brg\ & H$_2$ 1.9576 & H$_2$ 2.0338 & H$_2$ 2.1218 & H$_2$ 2.2233 & H$_2$ 2.4066 & H$_2$ 2.4237 \\
\hline
RCW~79 & 18     & -58 (196) & -- --    & -- --    & -- --    & -- --    & -- --     & -- --\\
       & 91     & -75 (202) & -- --    & -- --    & -57 (60) & -56 (--) & -- --     & -- --\\
       & 91a    & -63 (152) & -- --    & -- --    & -- --    & -- --    & -- --     & -- --\\
       & 91b    & -- --     & -- --    & -- --    & -58 (72) & -56 (80) & -- --     & -- -- \\
       & 237 ext& -57 (--)  & -- --    & -- --    & -53 (--) & -- --    & -- --     & -- --\\
       & 237    &  -- --    & -- --    & -62 (83) & -60 (78) & -62 (--) & -59 (63)  & -56 (75) \\
       & 1C4    & -49 (--)  & -- --    & -- --    & -46 (--) & -- --    & -- --     & -- -- \\
       & 2C3    &  -- --    & -- --    & -43 (--) & -45 (33) & -46 (nr) & -44 (nr)  & -45 (nr)\\
       & 3C3    & -- --     & -95 (95) & -80 (103)& -79 (88) & -76 (75) & -79 (84)  & -77 (76)\\
       & 3C3 bow& -- --     & -88 (--) & -73 (--) & -72 (72) & -71 (70) & -67 (50)  & -69 (49)\\ 
       & CHII  & -42 (--)  & -- --    & -- --    & -49 (--) & -- --    & -- --     & -- -- \\  
\hline
RCW~82 & 4      & -52 (228) & -- --    & -- --    & -- --    & -- --    & -- --     & -- --\\
       & 7 ext  & -- --     & -- --    & -- --    & -56 (43) & -- --    & -- --     & -- --\\
       & 35     & -50 (209) & -- --    & -- --    & -100 (188)& -- --   & -- --     & -- --\\
       & 51 ext &  -- --    & -- --    & -- --    & -56 (nr) & -- --    & -- --     & -- --\\
       & 51     & -74 (181) & -- --    & -- --    & -59 (66) & -- --    & -- --     & -- --\\
       & 98     & -- --     & -- --    & -57 (--) & -55 (69) & -56 (--) & -47 (45)  & -51 (nr)\\
\hline   
RCW~120& 49     & -29 (162) & -- --    & -- --    & -30 (142)& -- --    & -- --     & -- -- \\
       & 51 ext & -10 (--)  & -- --    & -- --    & -- --    & -- --    & -- --     & -- --\\
       & 51     & -5 (162)  & -- --    & -- --    & -- --    & -- --    & -- --     & -- --\\
       & 76     & -- --     & -- --    & -30 (--) & -28 (78) & -- --    & -- --     & -- --\\
       & 103    & -12 (137) & -67 (--) & -74 (--) & -79 (111)& -76 (--) & -- --     & -- --  \\
       & 107    & -30 (171) & -- --    & -- --    & -- (103) & -- --    & -- --     & -- --\\
\hline
\end{tabular}\\
\\
{Note:  All velocities are in \kms. The notation ``nr'' stands for ``not resolved'' and ``ext'' refers to extended emission.}
\end{table*}

The $V_{\rm LSR}$ velocities can be compared to the $V_{\rm LSR}$
velocities of the ionized gas and associated molecular material. These
velocities are in the ranges $-52$/$-46$ \kms\ for RCW~79
(ZA06), $-52$/$-42$ \kms\ for RCW~82 (PO09), and $-14$/$-8$ \kms\ for
RCW~120 (ZA07). The comparison shows that (and taking into account the
uncertainty on the velocity):

\begin{itemize}

\item[$\bullet$] the YSOs velocities in RCW~79 are in rather good
  agreement with those of the \HII\ region, except for 3C3. In this YSO
  the H$_2$ gas approaches us with a velocity of the order of 30
  \kms. 3C3 has a nearby extended structure (a bow-arc like structure,
  see Fig.\ \ref{source_id_1}) presenting the same velocity.

\item[$\bullet$] the YSOs in RCW~82 have velocities comparable to those of
the associated region; an exception is source 35 with very different
velocities for the H$_2$ and \brg\ lines.  The H$_2$ material
approaches us with a velocity $\sim$50 \kms\ with respect to the
material emitting the \brg\ line and the RCW~82 region.

\item[$\bullet$] in RCW~120, YSO \#51 has a velocity comparable to
  that of the \HII\ region; it is the only one (see also below). YSOs
  \#49, \#76, and \#107 are approaching us with a velocity of about 20
  \kms.  The situation is worse for YSO \#103, and resembles very much
  that of source 35 in RCW~82: the \brg\ line has the same velocity as
  the RCW~120 region, whereas all the H$_2$ lines show an approach
  velocity of some 50--60 \kms. This may be indicative of flows (see below), but may 
  also suggest that these YSOs are not associated with RCW~120. We do not 
  favor this last solution as these three YSOs are observed in the direction of 
  well defined condensations in the shell of dense neutral 
  material interacting with the ionized region. Molecular observations of this 
  shell (allowing velocity measurements) should help to clarify this point.

\end{itemize}

Table \ref{vel_yso} shows another interesting feature: the
\brg\ emission lines are wide (up to 200 \kms), and always wider than
the H$_2$ lines.  This again probably indicates that they do not
originate from the same zones of the YSOs.\\

\begin{figure}[!ht]
\centering
\includegraphics[width=9cm,angle=0]{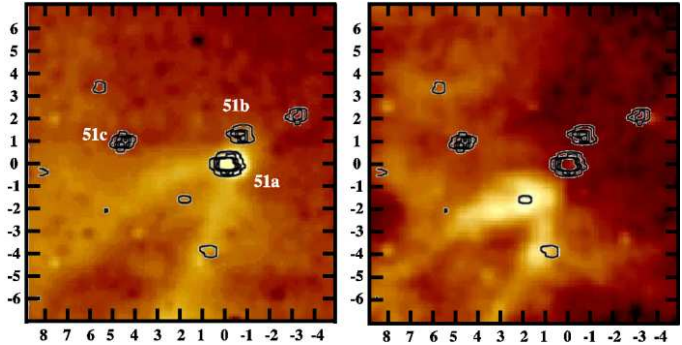}
\caption{Morphology of RCW120 51. {\it Left panel}:  \brg\ emission map. {\it Right panel}: \htwo\ emission map. The contours of the full K--band emission are overplotted.  All images have been created from the initial datacubes smoothed with a spatial Gaussian (FWHM = 3 pixels). North is up, East is to the left. The axis indicate the offsets (in arscseconds) relative to the main YSO position (coordinates given in Table \ref{tab_yso}).}\label{dyn_120_51}
\end{figure}

\begin{figure}[]
\centering
\includegraphics[width=9cm]{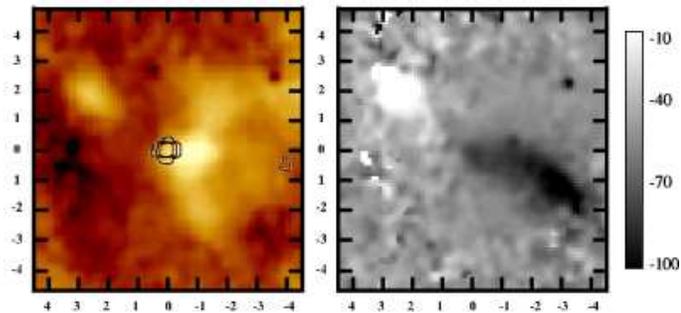}
\caption{Morphology and dynamics of RCW82 7. {\it Left panel}: \htwo\ emission map together with contours of the full K--band emission. {\it Right panel}: H$_{2}$ 2.12 \mum\ velocity map (scale in \kms). North is up, East is to the left. The axis indicate the offsets (in arscseconds) relative to the YSO position (coordinates given in Table \ref{tab_yso}).}\label{dyn_82_7}
\end{figure}

\begin{figure}[!ht]
\centering
\includegraphics[height=8cm,angle=0]{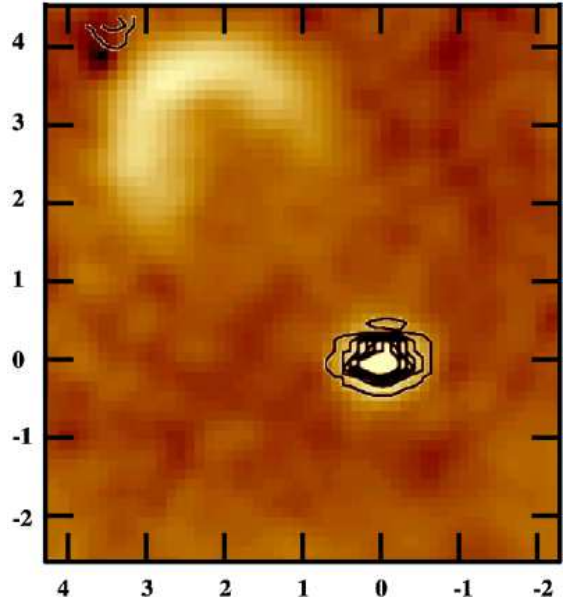}
\caption{Morphology of RCW120 C4-67. The image shows the \htwo\ emission map (created from the initial datacubes smoothed with a spatial Gaussian  -- FWHM = 3 pixels). Contours of the full K--band emission are overplotted. North is up, East is to the left. The axis indicate the offsets (in arscseconds) relative to the YSO position (coordinates given in Table \ref{tab_yso}).}\label{dyn_120_C4_67}
\end{figure}

These results indicate that the geometry of the YSO is rather complex, at least in some cases. One can gain further insight into those properties by looking at emission maps and velocity maps. Fig.\ \ref{dyn_120_51} to \ref{dyn_120_C4_67} show some illustrations. In those figures, the contours of the integrated $K$--band emission are shown on top of images centered on \brg\ and/or H$_{2}$ 2.12 \mum emission lines (and from which the continuum is subtracted). Velocity maps built from H$_{2}$ 2.12 \mum\ are also shown when they could be extracted from the data. The main conclusion regarding morphology is twofold: 1) \brg\ emission is always centered on the emission peak of the integrated $K$--band image, i.e. it coincides with the position of the YSO; 2) H$_{2}$ 2.12 \mum\ is emitted on a much wider scale, with emission peaks sometimes not corresponding to the position of the YSO. Three YSOs have been selected to illustrate these results, but they can be generalized to all YSOs (see appendix \ref{ap_yso}).  

The case of RCW120 51 is spectacular. The main $K$--band source in this region is 51a (see Fig.\ \ref{dyn_120_51}). Its spectrum  is dominated by \brg\ (Fig.\ \ref{spec_yso}) and Fig.\ \ref{dyn_120_51} confirms that is is the main emitter in this line. However, we note the presence of a \brg\ ``cone'' pointing towards 51a and, more generally, towards the center of the \HII\ region. This structure is very typical of the ionization front in the vicinity of RCW120 51. Measurements of the radial velocity along this cone ($\sim$ -6 \kms) indicates values similar to that of 51a and to that of the \HII\ region. Hence the YSO and the \brg\ cone are physically associated, and are also adjacent to the ionization front. The H$_{2}$ map of Fig.\ \ref{dyn_120_51} shows the existence of another emission cone, the orientation of which is the same as that of the \brg\ emission cone. However its top is not located on source 51a, but rather on a faint $K$--band source located south and east of it (see Fig.\ \ref{dyn_120_51}). The H$_{2}$ velocities along this cone are also consistent with those of the \HII\ region. We are thus directly probing the structure of the ionization front: seen from the \HII\ region, the line of sight crosses first a \brg\ cone tracing the ionized gas behind which is located the H$_{2}$ cone located closer to the PDR and the molecular material. Along this \htwo\ cone structure, we measure a ratio $\frac{I_{1-0s(1)}}{I_{2-1S(1)}}$ (respectively $\frac{I_{1-0s(1)}}{I_{3-2S(3)}}$) of 6.2$\pm$0.1 (respectively 12.0$\pm$0.1) intermediate between pure thermal emission and radiative excitation. The presence of non-thermal emission is consistent with the proximity to the PDR. All in all, we have thus a direct proof that the YSO RCW120 51a is located on the ionization front.

For YSO RCW82 7, Fig.\ \ref{dyn_82_7} shows again that \htwo\ is emitted over a wider region than the $K$--band continuum. The \htwo\ emission is not centered on the YSO, but is offset to the west. In addition, diffuse emission concentrations are clearly visible especially north-east of the YSO. The \htwo\ velocity map constructed by fitting a Gaussian to the 2.12 \mum\ line reveals that this blob has a radial velocity of -10 \kms\ compared to -60 \kms\ for the YSO \footnote{The velocity map also reveal a region of lower velocity south--west of the YSO, but the emission is weaker here and the result is less significant.}. This may very well be the signature of a jet centered on the YSO and interacting with molecular gas at the position of the blob. At the distance of RCW~82, such a jet would have a projected length (YSO--blob) of $\sim$0.06 pc. Besides this peculiar \htwo\ emission, there is an arc--like \htwo\ structure extending North and South of the YSO. Its velocity is comparable to that of the YSO and to that of the \HII\ region (see above). It may thus correspond to the PDR on top of which the YSO is located. A similar morphology is identified in RCW120 49 (Fig.\ \ref{dyn_120_49}) and RCW120 103 (Fig.\ \ref{dyn_120_103}), albeit with a lower significance (see Sect.\ \ref{ap_rcw120}). Another convincing evidence for the presence of a jet comes from YSO RCW120 C4-67. It features a spectacular \htwo\ map with an arc structure a few arcseconds away from the YSO (see Fig.\ \ref{dyn_120_C4_67}). Its radial velocity is comparable to that of the YSO, but the fact that the arc curvature is directed towards the YSO suggests that it might result from the interaction of a jet and molecular material. It may also be material ejected by the YSO in the past.  

From the analysis of the morphology and kinematics of a few sources, we conclude that, in some cases, we have been able to firmly establish the association of the YSO with the ionization front, strengthening the case of triggered star formation. We have also discovered the presence of a possible outflow in RCW82 7 traced by a \htwo\ velocity gradient (and possibly also in RCW120 49 and RCW120 103). The geometry of the \htwo\ emission in RCW120 C4-67 is also suggestive of the presence of a jet.

\section{Discussion}
\label{s_disc}


\subsection{Winds and bubbles}
\label{disc_wind}

The evolution of an \HII\ region is governed by the ionizing radiation
emitted by its ionizing stars -- see for example the first order
analysis of \citet{dw97} of the evolution of an \HII\ region in an
homogeneous medium. But it may also be influenced and even dominated
by the action of the wind emitted by the central massive star
\citep{freyer03}. It is often claimed that the many bubbles observed
at all scales in the Galactic plane result from the action of such
powerful winds \citep{church06}. While this is most likely the case
when Wolf--Rayet stars are involved \citep{gs95,brig97}, the dominance
of wind effects in the early evolution of interstellar bubbles remains
to be established. This is even more true since the recent downward
revision of the mass loss rates of O stars due to clumping
\citep[e.g.][]{jc05,fullerton06}.

\citet{ck01} and \citet{freyer03} have developed analytical models and
hydrodynamical simulations of \HII\ regions with winds. At the
beginning of its evolution, a young \HII\ region should not be too
much affected by the wind emitted by its ionizing star, if any. With
time, the stellar wind gets stronger and eventually leads to the
formation of a central cavity filled with very hot (T$\sim 10^{6}$ K)
low density, shocked gas.  This gas emits at X--ray wavelengths. Such
an extended X--ray emission is a strong indirect evidence for
winds. It has been observed only in the direction of three
\HII\ regions though: M17 \citep{townsley03,povich07}, the Rosette
nebula \citep{townsley03}, and the Orion nebula \citep{gudel08}.  All
regions lie relatively nearby (0.5 to 2 kpc), which might explain that
their X--ray emission is more easily detected than in distant
regions. They are excited by a cluster of several O stars and have
ages ranging from 1 (Orion) to about 3--4 Myr (M17 and Rosette). All
three clusters contain early O stars (spectral types O4--O6) with
moderately strong winds (compared to early O type supergiants which
have larger mass loss rates). This is similar to the three
\HII\ regions we have studied. We do not know if the extended X--ray
emission is a common phenomena among bubbles or if it is due to the
very massive stellar content of these regions. Unfortunately, RCW~79,
RCW~82 and RCW~120 have not been observed at X--ray wavelengths so the
effects of winds in the \HII\ regions cannot be quantified directly.

\begin{figure}[!h]
\centering
\includegraphics[width=7.3cm]{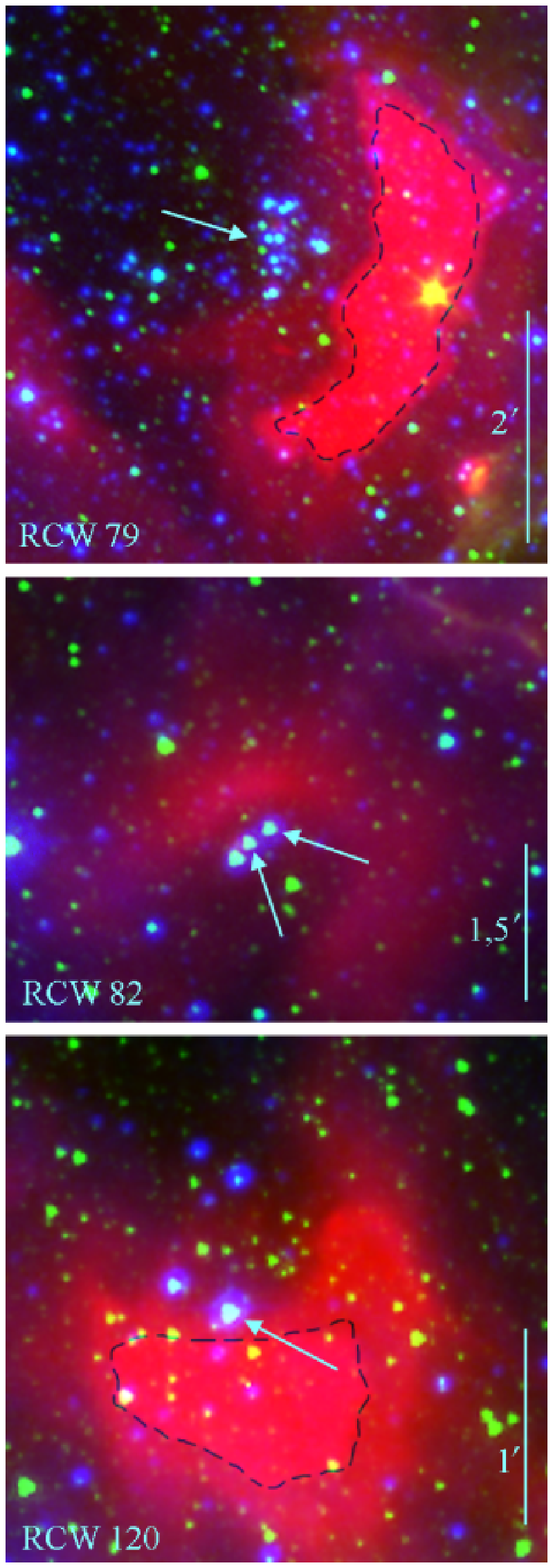}
\caption{Composite color image around the ionizing stars or cluster in the three \HII\ regions:
red is the Spitzer-MIPSGAL emission at 24~\mum, green is the Spitzer-GLIMPSE
emission at 3.6~\mum, and blue is the H$\alpha$ emission. The black dashed line encloses the zones where the 24~\mum\ emission is saturated. North is up and east is left.
The ionizing star(s) or cluster lie in the centre of the \HII\ regions, in a zone void of dust
emitting at 24~\mum. They are indicated by blue arrows. }\label{wind_24}
\end{figure}

Dust emission at 24~\mum\ may be another indicator of the effects of winds. 
Very small grains, heated by the stellar radiation and out of thermal equilibrium, radiate at 24~\mum\ \citep{jones99}.
As long as such emission is observed, it means that dust has not (yet) been swept-up by the stellar winds. The presence of a central hole in Spitzer--MIPS 24~\mum\ emission maps of some \HII\ regions might be an indication that winds are at work \citep{watson08}. 
Such a cavity could also be attributed to the radiation pressure of the central exciting stars or to dust destruction by the stars' ionizing radiation \citep{ino02,krum09}. 
However Krumholtz \& Matzner show that radiation pressure is generally unimportant for \HII\ regions driven by a handful of massive stars (thus for our three regions).
In any case if a shell is observed at 24~\mum, it means that stellar winds have not removed all the dust from the \HII\ region. Fig.\ \ref{wind_24} shows a composite image for the three \HII\ regions studied here, with red, green and blue being respectively 24~\mum, 3.6\mum\ and H$\alpha$ emission. The 24~\mum\ emission traces the hot dust. In all three regions, one clearly sees the presence of 24~\mum\ emission bows surrounding the ionizing stars. The hot dust very close to those stars is thus either blown away or destroyed by the intense UV radiation. But 5\arcsec\ to 30\arcsec\ away, i.e. well within the borders of the \HII\ regions, dust is detected. Hence, stellar winds have not (yet) removed all dust from the cavities, indicating that their effect on the dynamics of the \HII\ regions is rather limited.

To better assert the role of stellar winds and radiation in our objects, we focus on RCW~79. Since it hosts the largest number of ionizing stars of the three regions, it is the most appropriate to analyze the influence of stellar winds. 

The total number of ionizing photons emitted by the ionizing stars is $10^{49.82} $s$^{-1}$ (see Table \ref{tab_param}). This corresponds to an ionizing luminosity of $10^{39.31}$ erg s$^{-1}$. For an age of the population of 2.3 Myr, the ionizing energy released in the \ion{H}{ii} region is thus $10^{53.17}$ erg. In the previous estimate we have assumed that the ionizing luminosity was constant with time. Since the stars are rather young and close to the main sequence, this is a reasonable approximation. According to the Geneva tracks, a 40~\msun\ star has \teff\ = 42560 K and \lL\ = 5.34 on the ZAMS, and \teff\ = 39170 K and \lL\ = 5.45 after 2.3 Myr. For the corresponding radius and using the number of ionizing photons per unit area ($q_{0}$) from \citet{msh05}, the values of $Q_{0}$ (total ionizing photon flux) are $10^{49.12}$ and $10^{49.16}$ at 0 and 2.3 Myr respectively. Hence our assumption is justified.

The total wind mechanical luminosity at 2.3 Myr is $10^{36.18}$ erg s$^{-1}$ (for 12 O stars with a mass loss rate of $10^{-7}$~\myr\ and a terminal velocity of 2000 \kms). Assuming again that this value is constant between 0 and 2.3 Myr, one estimate a total wind mechanical energy release of $10^{50.04}$ erg. To check the validity of our assumption, one can use the scaling relations of \citet{vink01} (their Eq.\ 24). Using the properties of a 40 \msun\ star on the ZAMS and after 2.3 Myr of evolution \citep[using once more the tracks of ][]{mm05}, one finds a mass loss rate difference of 0.3 dex. Hence, our wind mechanical energy determination is at the very most overestimated by a factor 2. 

The 2D simulations of \citet{freyer03,freyer06} tackled the question of the effects of winds on the dynamics of \HII\ regions. They showed that the presence of the wind triggered the formation of structures (``fingers'') in the \HII\ regions, modifying the efficiencies of energy transfer (from stellar ionizing and wind mechanical energies to ionizing, kinetic and thermal energies in the \HII\ region). In their simulations, the ratio of ionizing to wind mechanical luminosities was of the order 100. They showed that in spite of this, winds could significantly affect the evolution of the \HII\ region. In our case, this ratio is at least 1000 (recall that our estimates of mass loss rates are upper limits). The main reason is that although the ionizing luminosities we find are similar to the ones used by Freyer et al., our mass loss rates are lower by nearly a factor of 10. The reason is the inclusion of clumping in our models. It is well known to lead to a downward revision of mass loss rates of O stars \citep[e.g.][]{hil91}. The values of \mdot\ used by Freyer et al.\ were derived with homogeneous models and are thus higher than our determinations. With this increase by at least a factor of 10 in the ratio of ionizing to wind mechanical luminosity, one might thus wonder whether winds still play a role. The evidence brought by the presence of 24~\mum\ dust emission argues against a strong effect of stellar winds (at least in the first 2--3 Myrs). New dedicated simulations with reduced mass loss rates are encouraged to shed more light on this issue.

\subsection{YSOs: a possible spectroscopic evolution}
\label{disc_yso}

Among the YSOs observed on the borders of the three \HII\ regions, a few have a $K$--band spectrum entirely dominated by H$_{2}$ emission. We have seen in Sect.\ \ref{yso_h2} that thermal emission due to shocks was favored to explain the different line ratios. Besides, most of these sources tend to be brighter and redder in the [24] vs $K$--[24] diagram, indicating a larger amount of circumstellar material (Sect. \ref{yso_photom}) than for the other observed YSOs. All in all, this tends to favour a picture according to which the H$_{2}$ sources are in a relatively early phase of evolution. The large 24~\mum\ emission combined to the absence of direct indicator of the presence of a disk suggests that these sources still possess large envelopes. The H$_{2}$ emission probably comes from either a jet--like structure interacting with this circumstellar material, or from an expanding shock created by stellar/disk wind and/or radiation pressure inside the envelope. One can further speculate that objects showing both H$_{2}$ lines and other features (\brg\ or CO bandheads emission) represent later stages of evolution in which the envelope partially disappeared, revealing the disk features. Their somewhat lower reddening and brightness in the [24] vs $K$--[24] diagram lend support to this hypothesis. In that case, both \brg\ and CO emission could be produced in disks irradiated by the nascent star, \brg\ being emitted on the surface and CO close to the disk midplane, in zones shielded from UV flux. This could explain that the H$_{2}$ lines are usually narrower than \brg\ in YSOs when both types of lines are detected (Table \ref{vel_yso}). The latter being emitted on the disk's surface, it would be rotationally broadened, while the former would reflect the shock conditions in the dissolving envelope. We noted in Sect.\ \ref{dyn_yso} that a couple of objects showed radial velocities different for \brg\ and H$_{2}$. In the suggested scenario, this could be explained if the H$_{2}$ emission was produced by interaction of a jet and the envelope. If this putative jet was directed towards us, one could see a blueshifted H$_{2}$ emission compared to the \brg\ line produced at the surface of the disk.

Another piece of information comes from the probable detection of jets in a few sources (Sect.\ \ref{dyn_yso}). All of these sources (RCW82 7, RCW120 49, and RCW120 C4-67) have been classified as CO emitters. It means that they possess circumstellar disks. Since bipolar outflows are predicted to develop in accreting objects \citep[e.g.][]{shu94} the detection of jets in several of our YSO confirms that we are witnessing accretion phenomena. Hence, those objects fit in the scenario presented above and correspond to evolutionary stages where the envelope has disappeared. This is consistent with the fact that all these sources are at the limit between class~I and class~II sources (see Fig.\ \ref{ccd_irac} and Table \ref{TableYSOs}). 

Obviously, the scenario presented above remains highly speculative. Future high spectral resolution spectroscopy of objects showing CO, \brg\ and H$_{2}$ emission will certainly provide more constraints on the geometry and dynamics of the line emitting regions. Nonetheless, if YSOs evolve spectroscopically according to the previous path and if they were formed simultaneously, our observations indicate that in the three \HII\ regions studied here YSOs evolve at different speed (presumably due to their different masses) since several types of objects are detected. Alternatively, if one drops the assumption that all YSOs formed at the same time, it means that the second--generation star formation event lasted less than 2 Myr. Said differently, the typical timescale for the evolution of YSOs (class 0 to class II) should be $\lesssim$ 2 Myr.


\section{Conclusion}
\label{s_conc}

We have obtained near--infrared integral field spectroscopy of three \HII\ regions (RCW~79, RCW~82, RCW~120) with SINFONI on the VLT. The main results of our study are:

\begin{itemize}

\item[$\bullet$] We have identified the ionizing sources of all three \HII\ regions. RCW~79 is powered by a cluster of O stars unraveled here for the first time. The most massive stars have a spectral type O4--6V/III. A number of later type OB stars are also detected. RCW~82 is ionized by two O9--B2V/III stars and a third star, probably of spectral type Be. Finally, a single O6--8V/III star ionizes the \HII\ region in RCW~120. 

\item[$\bullet$] The ionizing stars of each region have been analyzed with atmosphere models computed with the code CMFGEN. The derived stellar properties have been used to build the HR diagram. In the case of RCW~79, the large number of stars allowed a reliable age determination: the ionizing cluster formed 2.3$\pm$0.5 Myr ago. For the other two \HII\ regions, similar ages are suggested but are less well constrained due to the small number of sources. An upper limit of $10^{-7}$ \myr\ was derived for the mass loss rate of the ionizing stars of all three regions. 

\item[$\bullet$] The cumulative mass loss rate due to the O stars at the center of RCW~79 is $< 10^{-6}$ \myr. The resulting wind mechanical luminosity is about $10^{-3}$ the ionizing luminosity. This is ten times weaker than in the numerical simulations of \citet{freyer03}, raising the question of the quantitative role of winds on the dynamics of the three \HII\ regions studied here. The presence of hot dust emission close to the ionizing stars argues against a strong influence of stellar winds. 

\item[$\bullet$] Spectroscopy of the infrared sources on the borders of the \HII\ regions revealed typical signatures of young stellar objects. Four main categories of YSOs have been identified: (1) sources dominated by H$_{2}$ emission lines, (2) sources showing mainly \brg\ emission, (3) sources with CO bandheads in emission and (4) sources with CO bandheads in absorption. 

\item[$\bullet$] Near and mid--infrared color--color and color--magnitude diagrams indicate that all YSOs have SEDs dominated by non stellar emission. In the Spitzer/IRAC [3.6]$-$[4.5] vs [5.8]$-$[8.0] diagram, essentially all YSOs are grouped at the position between class I and class II objects. Only a couple of H$_{2}$ sources seem to be redder. The trend is more clearly seen in the [24] vs $K$--[24] diagram where the H$_{2}$ sources form clearly a distinct group with larger 24~\mum\ emission and redder $K$--[24] colors. 

\item[$\bullet$] The detailed analysis of the H$_{2}$ emitting sources through line ratios and excitation diagrams indicate that thermal emission is the dominant mechanism in the majority of sources. Slow J--shocks or fast C--shocks are the preferred explanation for this emission. Some sources show a possible weak contribution of fluorescence emission, and three are dominated by this mechanism.

\item[$\bullet$] Measurement of radial velocities and line width was performed in a few sources. In general, the derived radial velocities are consistent with that of the ionized gas in the \HII\ region, confirming that the YSOs are well associated with the regions. The exception is RCW~120, for which most sources move 20 to 50 \kms\ faster than the ionized gas. In sources showing both \brg\ and H$_{2}$ lines, the former are systematically wider (150/250 \kms vs non--resolved/100 \kms), possibly due to a different spatial origin. 

\item[$\bullet$] Morphological studies reveal that when both \brg\ and \htwo\ emission are present, \brg\ is centered on the position of the YSO while \htwo\ emission is more extended. In some cases, the \htwo\ emission peak is offset compared to that of the YSO. RCW120 51 presents a double \htwo/\brg\ cone structure following the ionization front. RCW82 7 (and possibly RCW120 49 and RCW120 C4--67) has a \htwo\ velocity structure typical of an outflow. 

\item[$\bullet$] The spectrophotometric and kinematic properties of YSOs paint the following plausible picture: H$_{2}$ dominated sources are stellar objects surrounded by an envelope in which shocks (due to jets or stellar/disk winds) produce H$_{2}$ emission; other YSOs correspond to more evolved sources in which the envelope partially or totally disappeared, revealing a disk structure from which CO and/or \brg\ emission is produced. 

\end{itemize}

Future tailored observations of a few key YSOs will provide more information on the dynamics of the circumstellar material. Additional UV and optical spectroscopy of the ionizing stars of RCW~79 will help to refine the age estimate and to reveal a possible age difference with the stars in the CHII region on the border of RCW~79. This would be a first direct proof of the existence of triggered massive star formation. Better estimates of the mass loss rates will also be possible, helping to quantify the role of stellar winds in the dynamics of \HII\ regions.

\begin{acknowledgements}
We thank the ESO staff (especially Chris Lidman and Duncan Castex) for their help and efficiency during the observations. John Hillier is thanked for making his code CMFGEN freely available and for constant help with it. We acknowledge the help of Thomas Ott with QFitsView and the advice of Frank Eisenhauer on the data reduction. We thank an anonymous referee for a careful reading of the manuscript. This work made use of the GLIMPSE and MIPSGAL Spitzer Space Telescope Legacy surveys. The Spitzer Space Telescope is operated by the Jet Propulsion Laboratory, California Institute of Technology under a contract with NASA. This publication also makes use of data products from the Two Micron All Sky Survey, which is a joint project of the University of Massachusetts and the Infrared Processing and Analysis Center/California Institute of Technology, funded by NASA. The SIMBAD database operated by CDS, Strasbourg, France was used for the completion of this study.
\end{acknowledgements}

\bibliography{biblio.bib}

\begin{appendix}

\section{Spectra}
\begin{figure}[]
\centering
\includegraphics[width=9.5cm]{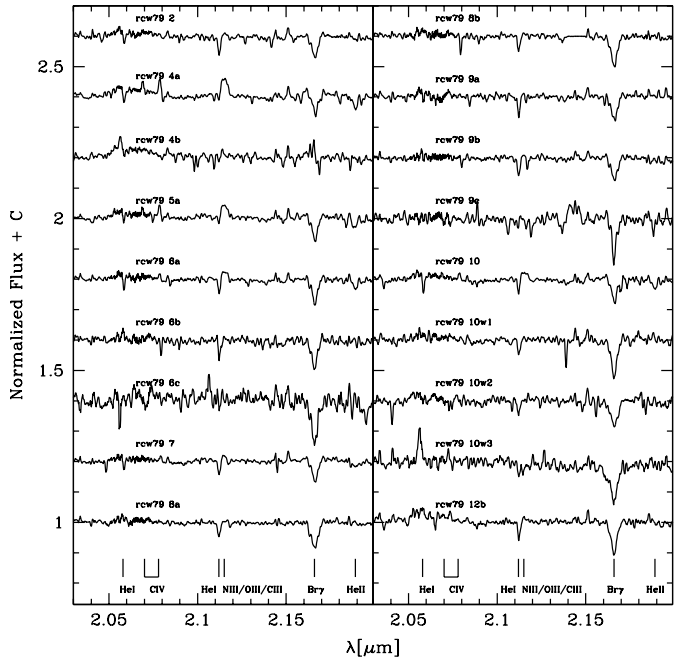}
\caption{SINFONI $K$--band spectra of the ionizing sources of RCW~79: O stars. The main stellar features are indicated.}\label{fig_rcw79_ex}
\end{figure}

\begin{figure*}[]
\begin{center}
\begin{minipage}[b]{0.4\linewidth} 
\centering
\includegraphics[width=8cm]{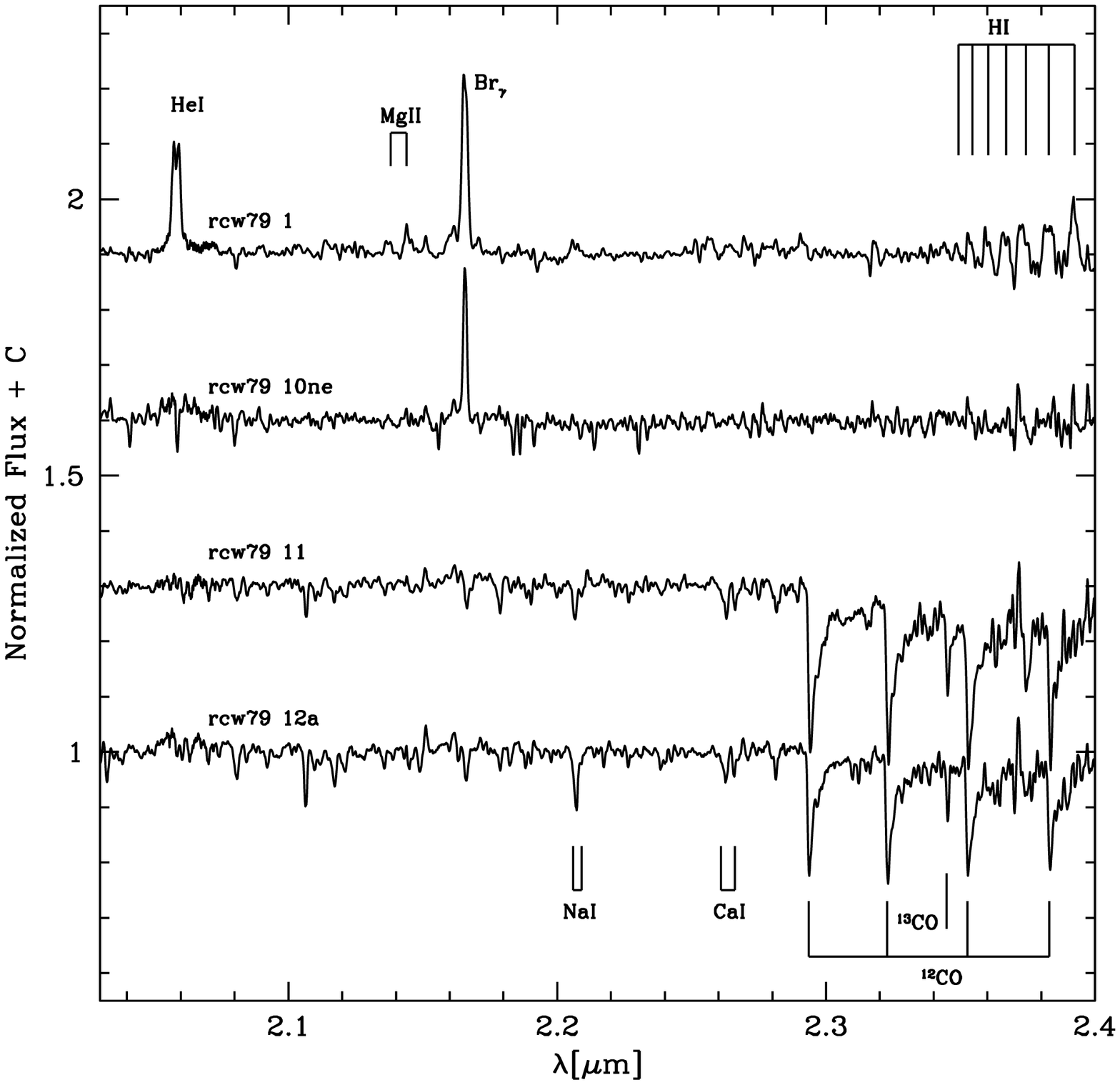}
\end{minipage}
\hspace{0.5cm} 
\begin{minipage}[b]{0.4\linewidth}
\centering
\includegraphics[width=8cm]{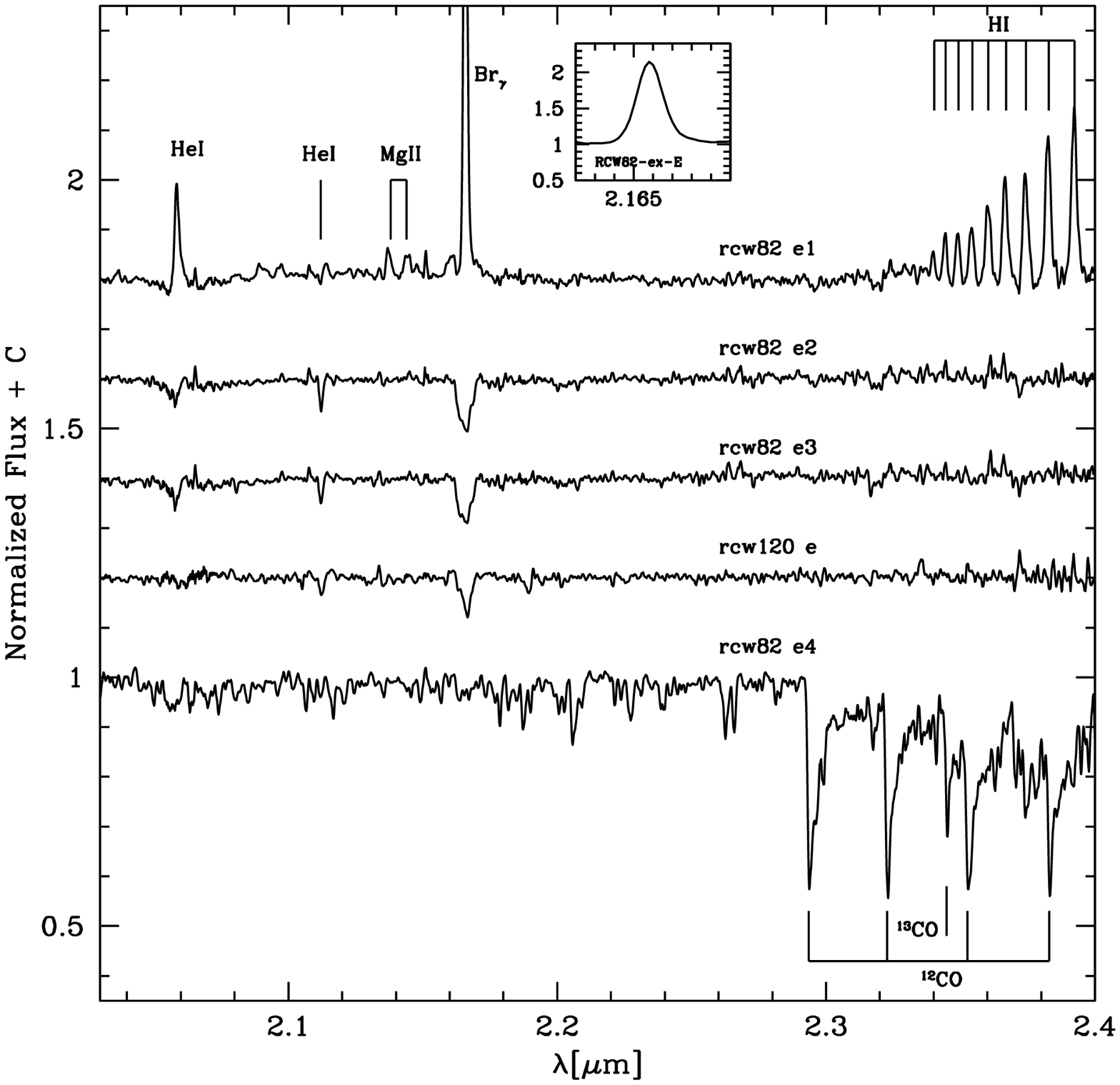}
\end{minipage}
\caption{SINFONI $K$--band spectra of central stars of \HII\ regions. {\it Left}: Be and late type stars in RCW~79. {\it Right}: ionizing stars of RCW~82 and RCW~120.} \label{fig_ex_2}
\end{center}
\end{figure*}

\begin{figure*}[]
\begin{center}
\begin{minipage}[b]{0.4\linewidth} 
\centering
\includegraphics[width=8cm]{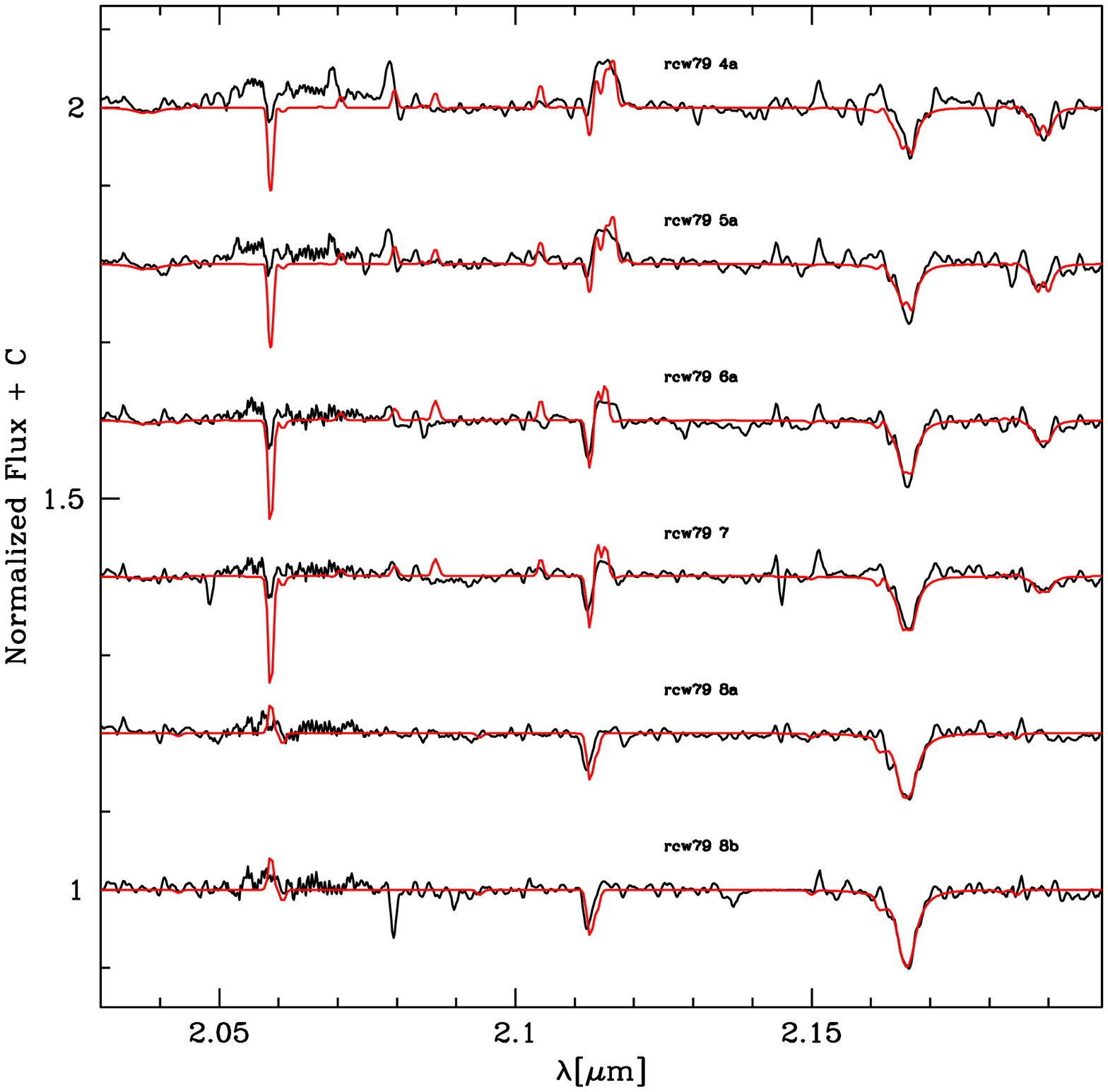}
\end{minipage}
\hspace{0.5cm} 
\begin{minipage}[b]{0.4\linewidth}
\centering
\includegraphics[width=8cm]{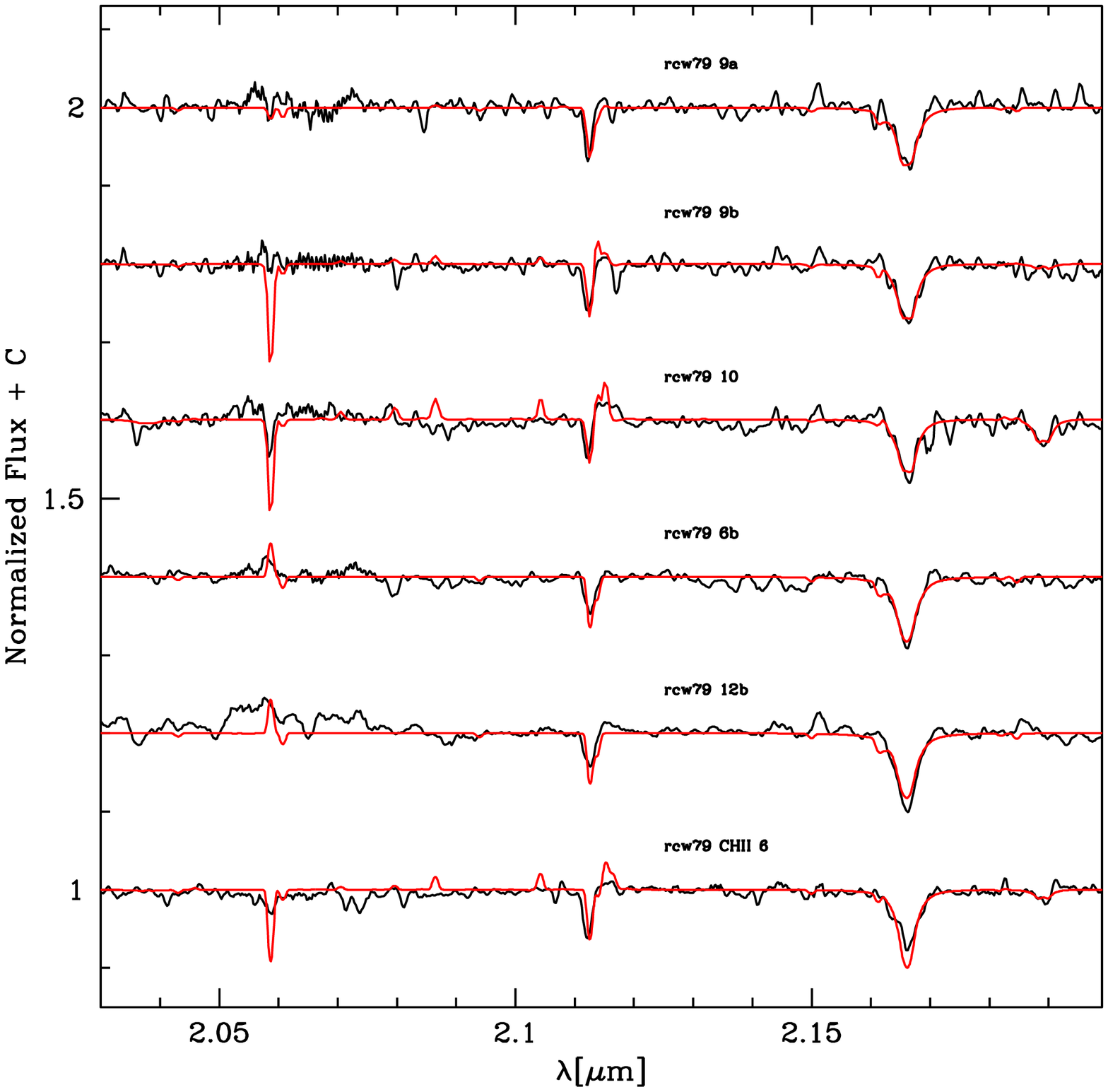}
\end{minipage}
\caption{Best CMFGEN fits (red) of RCW~79 ionizing sources spectra (black).} \label{fit_79}
\end{center}
\end{figure*}

\begin{figure}[]
\centering
\includegraphics[width=9cm]{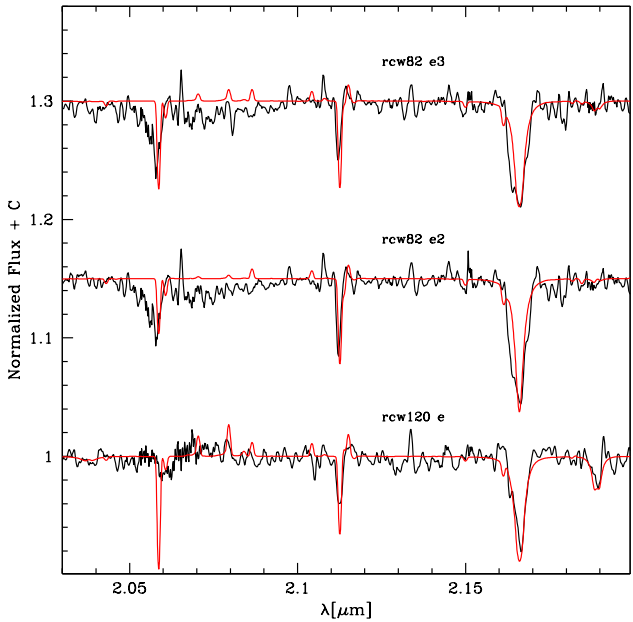}
\caption{Best CMFGEN fits (red) of RCW~82 and RCW~120 ionizing sources spectra (black).}\label{fit_82}
\end{figure}

In this Section, we present the spectra of our sources and the best fit to the ionizing sources (Figs.\ \ref{fig_rcw79_ex} to \ref{fit_82}).

\begin{figure*}
     \centering
     \subfigure[]{
          \includegraphics[width=.48\textwidth]{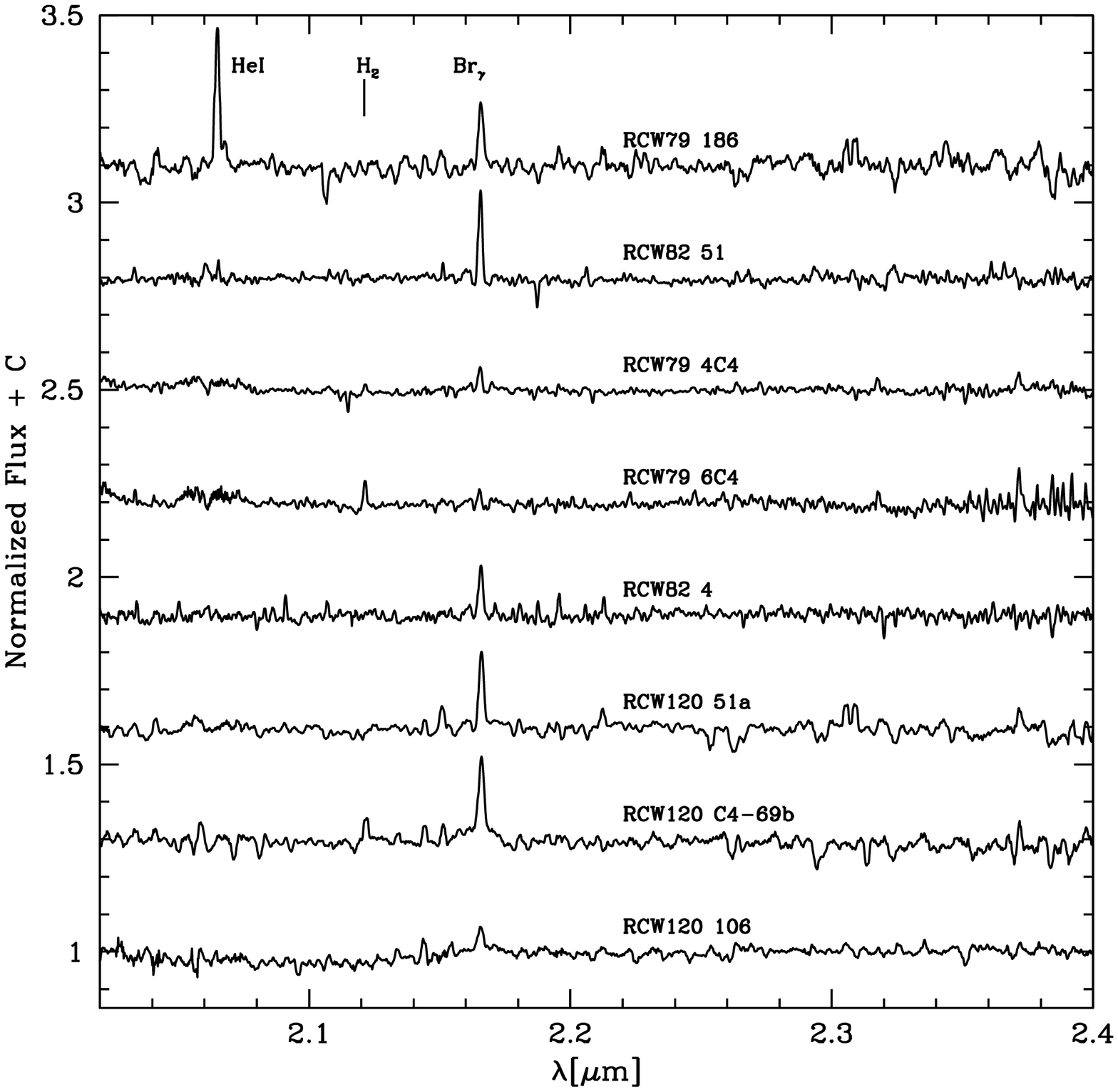}}
     \hspace{0.2cm}
     \subfigure[]{
          \includegraphics[width=.48\textwidth]{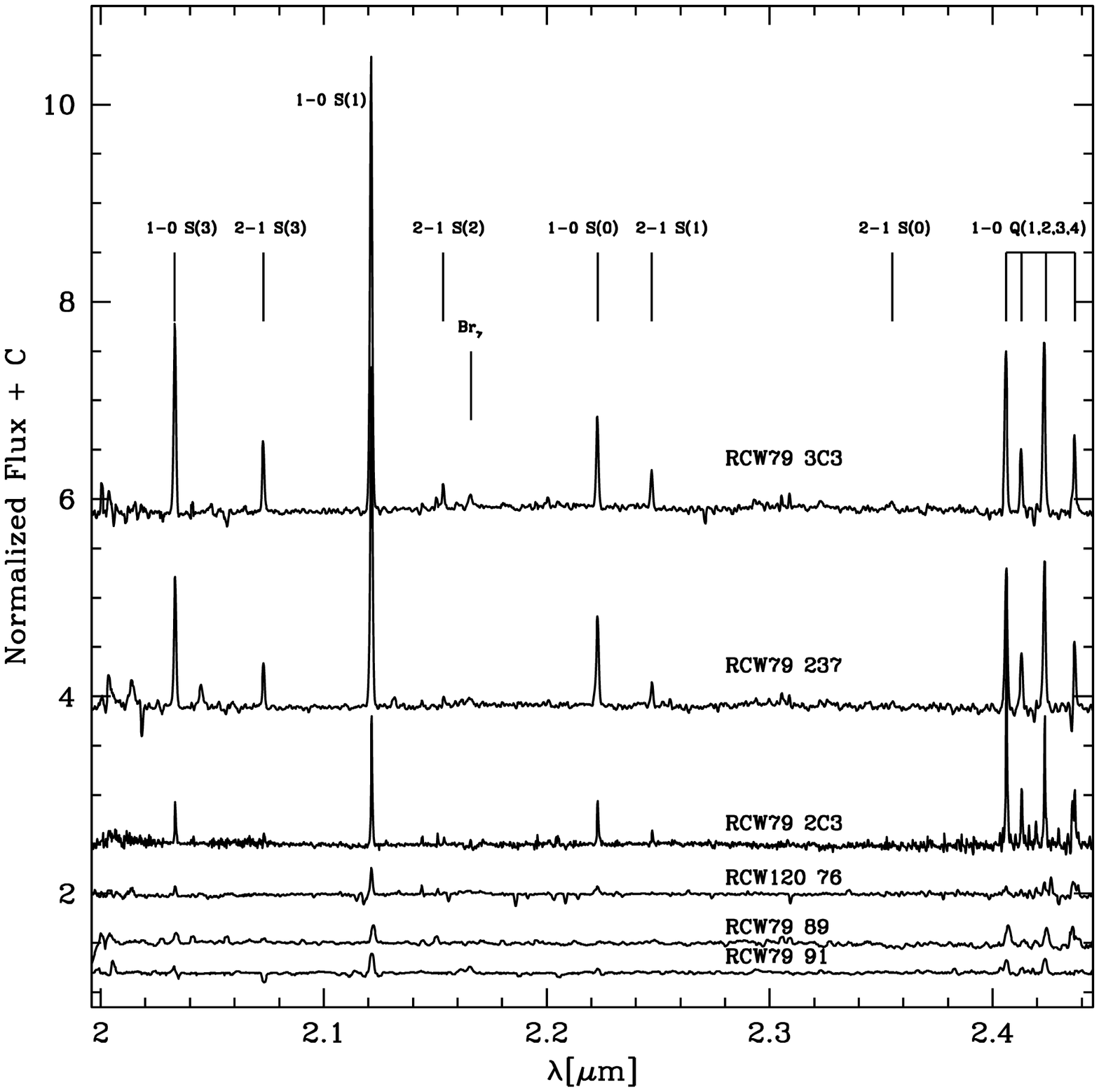}}\\
     \subfigure[]{
           \includegraphics[width=.48\textwidth]{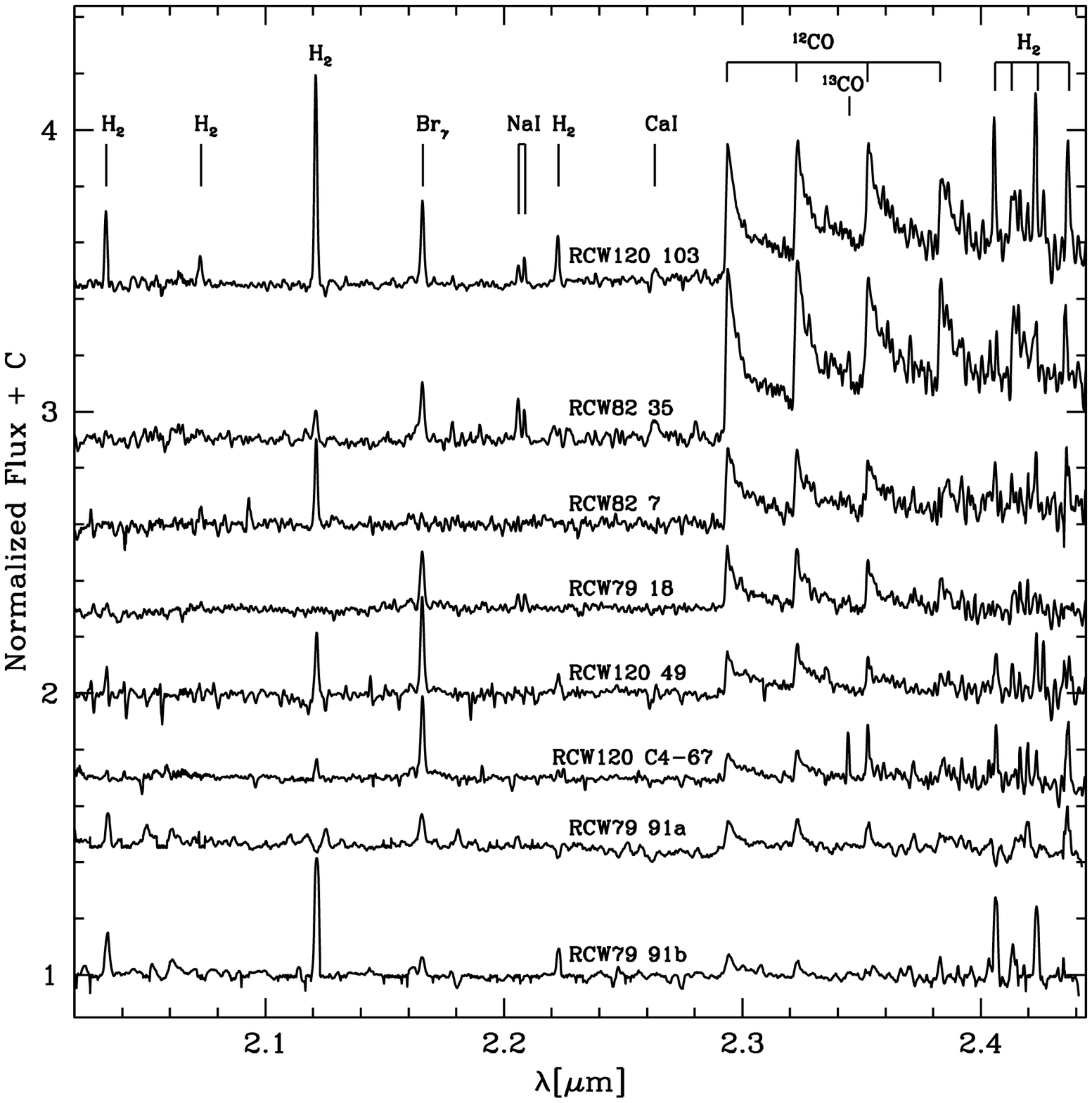}}
     \hspace{0.2cm}
     \subfigure[]{
          \includegraphics[width=.48\textwidth]{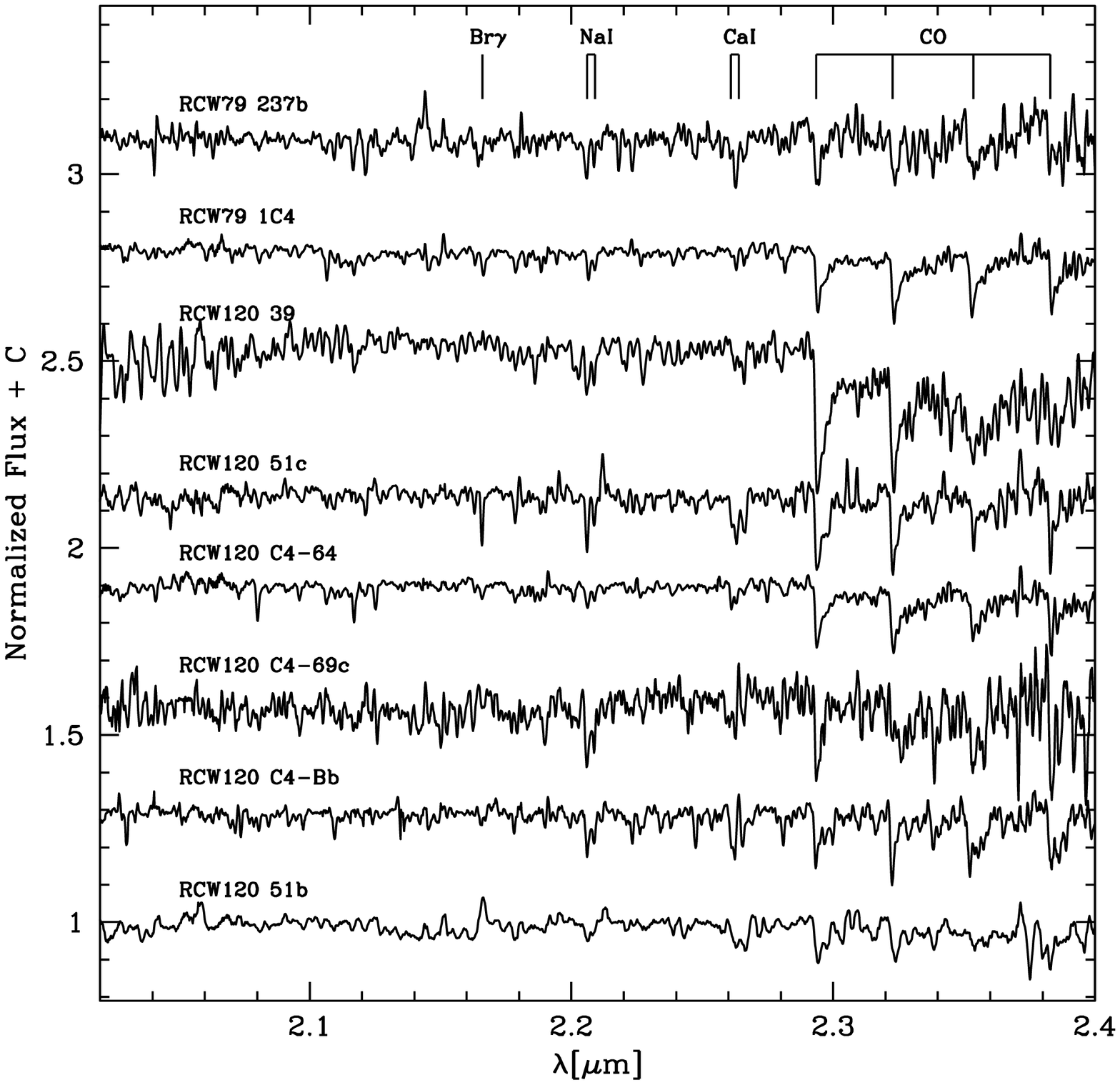}}
     \caption{Normalized SINFONI YSO spectra. Panel a, b, c and d show respectively the \brg, \htwo, CO emission and CO absorption dominated spectra.}
     \label{spec_yso}
\end{figure*}

\begin{figure}[]
\centering
\includegraphics[width=8cm]{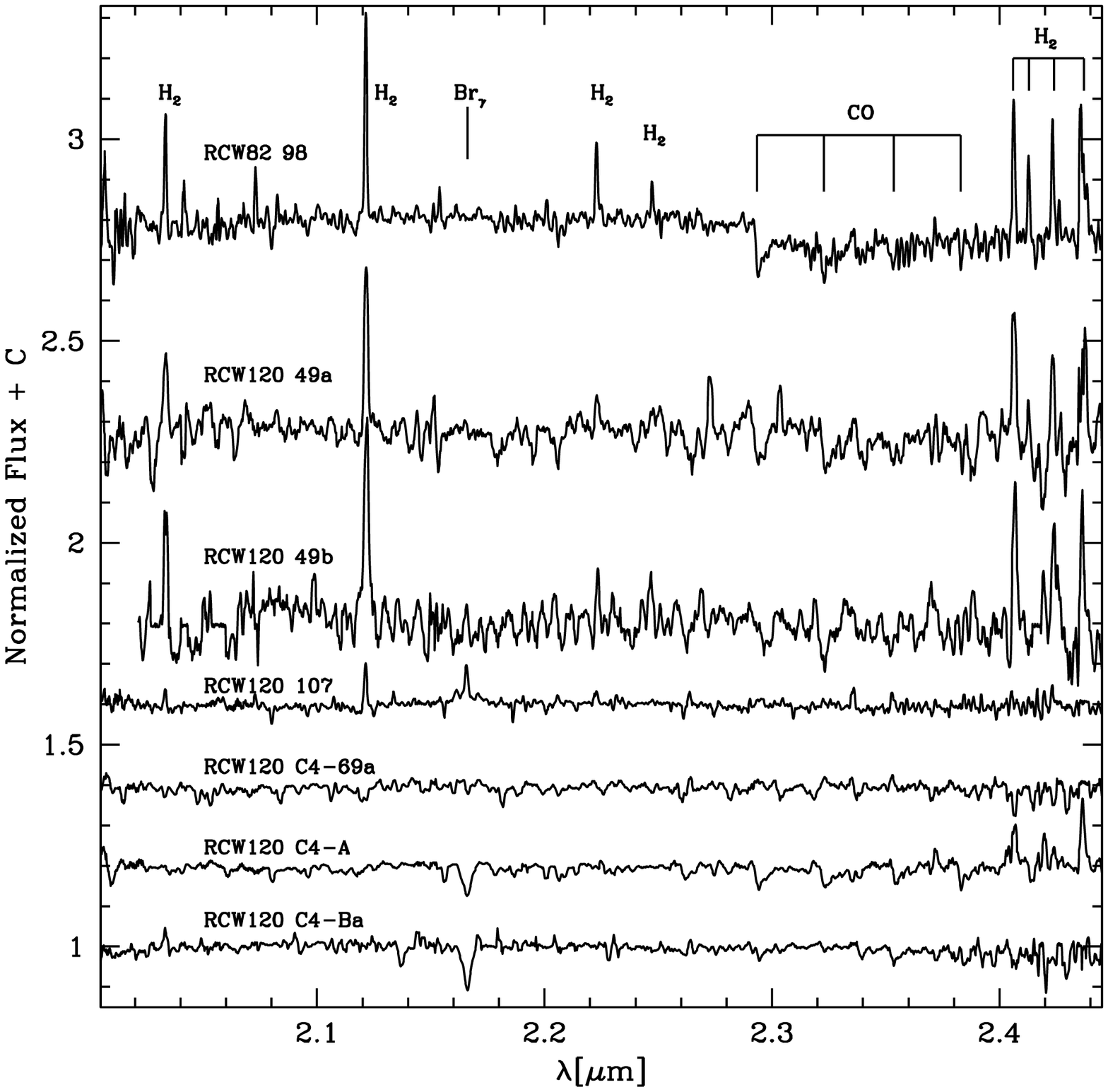}
\caption{Spectra of YSOs with composite features. }\label{spec_mixed}
\end{figure}

\section{Notes on individual objects}
\label{ap_yso}
\begin{figure*}[!h]
\centering
\includegraphics[width=16cm]{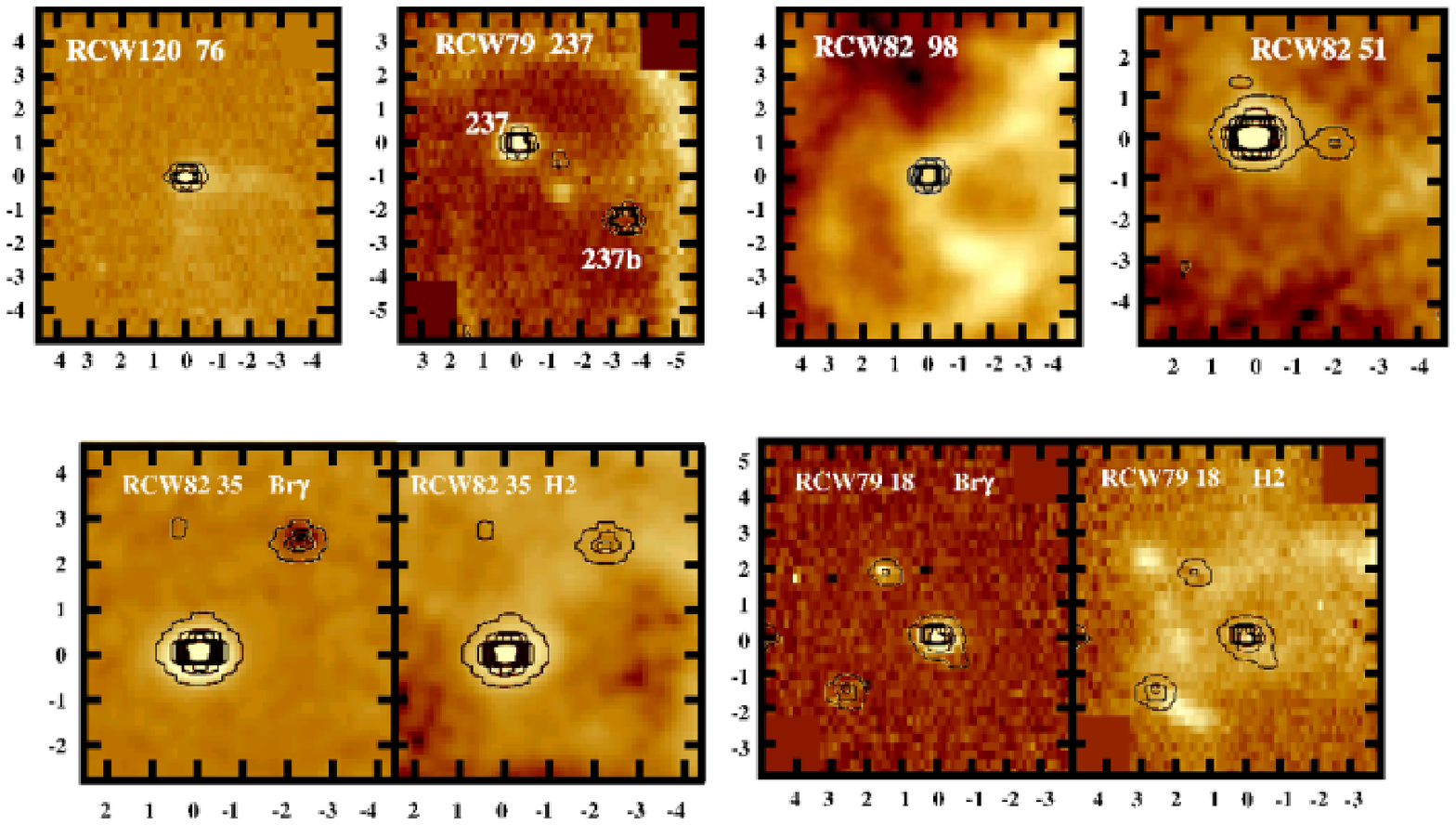}
\caption{{\it Top panels}: \htwo\ maps of YSOs with contours of the full K-band images overplotted. The \htwo\ emission is more extended than the bulk of the K-band emission. {\it Bottom panels}: \brg\ and \htwo\ maps of RCW82 35 and RCW79 18 together with contours of the full K-band emission. In all panels, North is up and East is to the left. The axis indicates the RA and DEC offsets (in arcseconds) relative to the main YSO (see Table \ref{tab_yso} for coordinates).}\label{source_id_1}
\end{figure*}

\begin{figure*}[!h]
\centering
\includegraphics[width=16cm]{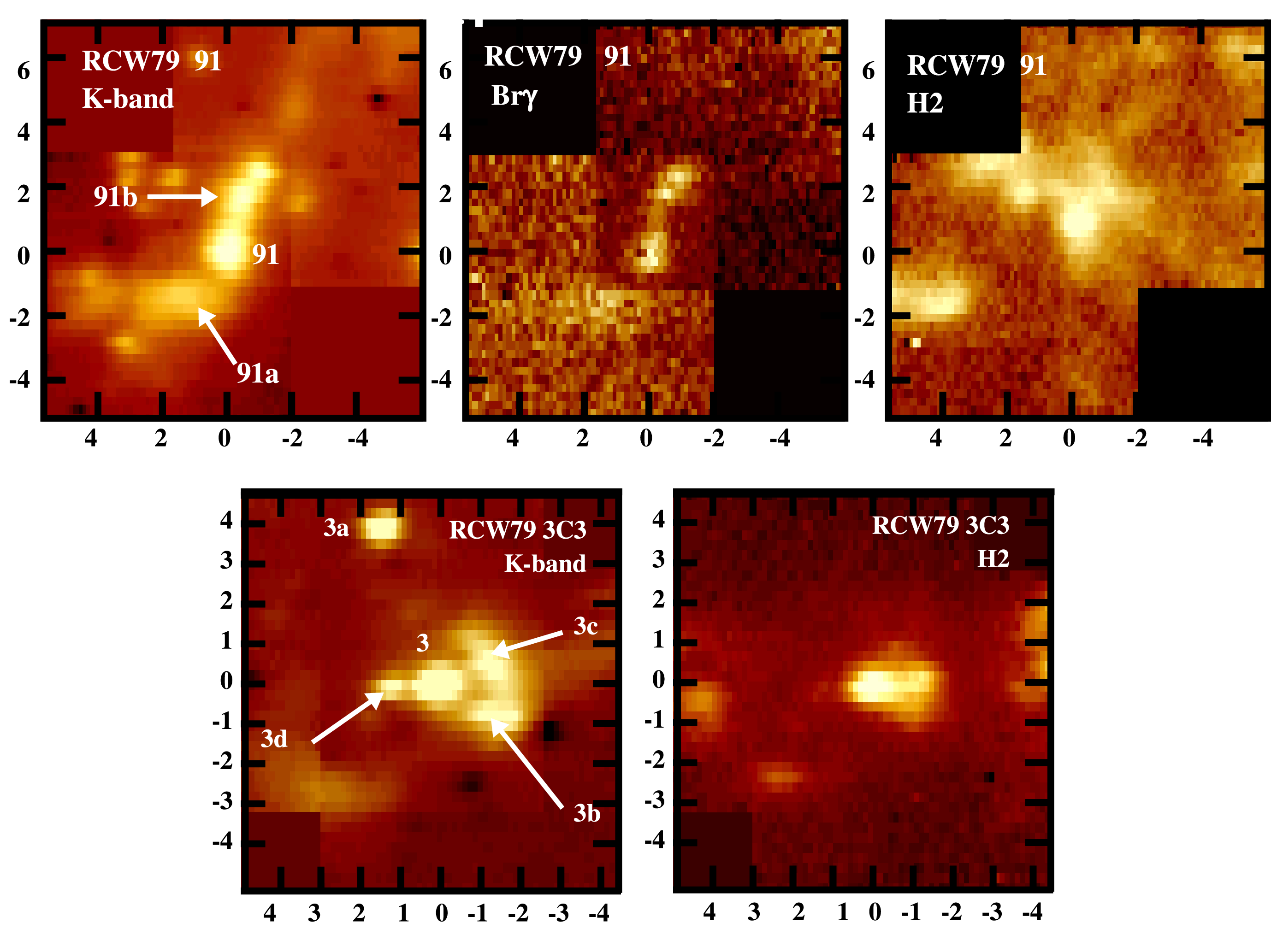}
\caption{Identification of the sub-components of YSOs RCW79 91 (top panels) and RCW79 3C3 (bottom panels) together with \brg\ and \htwo\ maps. North is up and East is to the left. The axis indicates the RA and DEC offsets (in arcseconds) relative to the main YSO (see Table \ref{tab_yso} for coordinates). }\label{source_id_2}
\end{figure*}

\begin{figure}[h]
\centering
\includegraphics[width=7cm]{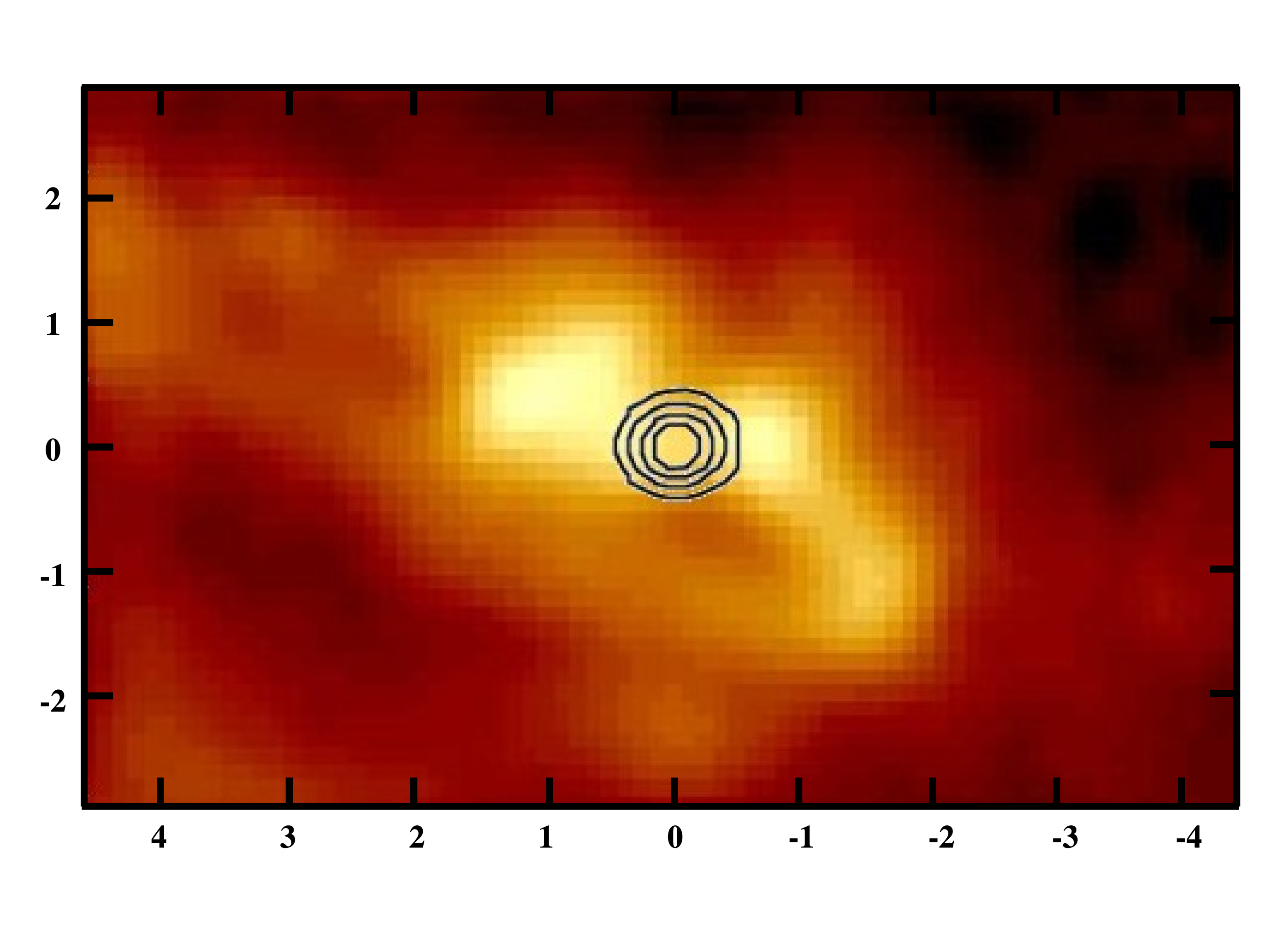}
\caption{Morphology of RCW79 2C3. The image shows the \htwo\ emission (created from the initial datacubes smoothed with a spatial Gaussian -- FWHM = 3 pixels). North is up, East is to the left. The axis indicate the offsets (in arscseconds) relative to the YSO position (coordinates given in Table \ref{tab_yso}).}\label{dyn_79_2C3}
\end{figure}

\begin{figure}[h]
\centering
\includegraphics[width=8cm]{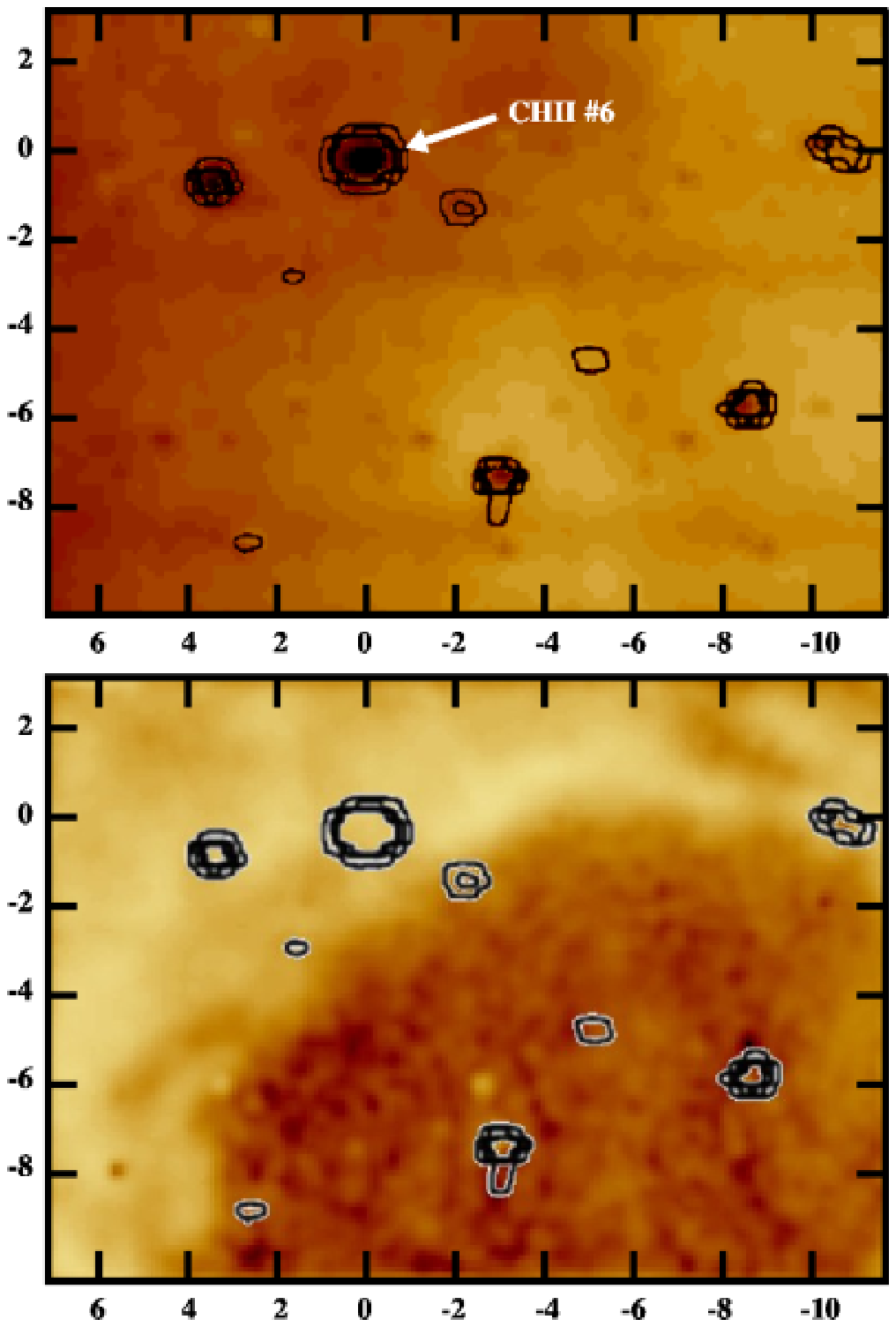}
\caption{\brg\ (top) and \htwo\ maps of the the compact \HII\ region on the border of RCW~79. The contours of the full K-band emission are shown in black. North is up and East is to the left. The axis indicates the RA and DEC offsets (in arcseconds) relative to source CHII \#6.}\label{source_id_uchii}
\end{figure}

Here, we briefly summarize the main characteristics of each region. Figs. \ref{source_id_1} to \ref{source_id_120c4} provide source identification as well as linemaps of a few YSOs.

\subsection{RCW~79}
\label{ap_rcw79}

Eight sources were observed in this region. All present spectroscopic 
features typical of YSOs. Sources 237, 91 and 3C3 are resolved in 
multiple components (respectively Figs.\ \ref{source_id_1} and \ref{source_id_2}) . \\

Sources 237-b and 1C4 show a decreasing 
continuum together with \ion{Na}{i}, \ion{Ca}{i} and CO
absorptions. Source 1C4 also shows 
Br$\gamma$ in absorption.

Sources 2C3, 3C3 and 237 are dominated by \htwo\ emission. According to 
their GLIMPSE colors, they are Class~I sources, 3C3 and 237 being the 
reddest (high $[3.6]-[4.5]$ color). This agrees with 3C3 and 237 having a rising continuum 
whereas 2C3 has a flat one. 
Fig.\ \ref{dyn_79_2C3} shows the H$_{2}$ emission of RCW79 2C3. While the $K$--band features a point source, a more elongated structure is seen in the \htwo\ map. The emission peak in the latter does not coincide with the YSO in the $K$--band: a large amount of molecular material is still present around the YSO. We could not construct the \htwo\ velocity map due to the weakness of the signal in the emitting regions. The \htwo\ emission does not come directly from the YSO, and this should be kept in mind when interpreting the mixed results of Sect.\ \ref{yso_h2} regarding the nature of the emission.

The spectrum of 237 is very similar to the one of source nr56c
\citep{leticia08}. A bright $H_2$ emission filament is observed in the
field of source \#237 (Fig.\ \ref{source_id_1}).  Its velocity is that
of the \HII\ region, and this $H2$ emission filament corresponds to
the 8~$\mu$m emission that originates from the nearby PDR.
   
Sources 18 (Fig.\ \ref{source_id_1}), 91 (Fig.\ \ref{source_id_2}),
186 and 4C4 show wide Br$\gamma$ emission.  A maser was identified by
\citet{caswell04} in the region around source 91, but its association
with any of the near-IR sources is not obvious (see ZA06).  Sources
91, 18 and 4C4 are Class~I sources, having similar colors in the
GLIMPSE color-color diagram.

The field of the compact \HII\ region shows stellar sources and
extended \brg\ and \htwo\ emission (Fig.\ \ref{source_id_uchii}). The
\brg\ emission (ionized gas) is confined within a region bounded by a
\htwo\ filament. OB stars are observed as dark spots in the
\brg\ map. Surprisingly, the two brightest ionizing stars of the
compact \HII\ region lie in the direction of the \htwo\ filament and
not in the ionized cavity; this might be a projection effect. The
extended \brg\ emission and \htwo\ filament have similar velocities,
also similar to that of the large central RCW~79 region.

Extended \brg\ and \htwo\ emission are observed in the field of source
1C4 (not shown).  The \brg\ emission is probably that of the ionized
gas in RCW~79.  The \htwo\ emission comes from the PDR. Both
emission indicate velocities similar to that of the central
\HII\ region.

Source 4C4 is a tight cluster composed of one bright objects and 6
much fainter components. They present an increasing spectrum and a
faint \brg\ emission line. No extended emission is observed in this
field.

\subsection{RCW~82}
\label{ap_rcw82}

Five candidates YSOs were observed in the direction of RCW~82, all situated on the border 
of this \HII\ region, except possibly 51. Sources 4, 7, 35, and 51 have been 
classified as Class I by PO09. Source 98 has not been classified, due to uncertain 
$J$ and $H$ measurements and to an absence of 8~$\mu$m measurement.\\

Source 4 is almost featureless, with a faint and wide \brg\ line, and a rising 
spectrum. This agrees with a Class I source. A fainter source is present in the field, 
situated south-west of 4; it shows no features, except an increasing continuum.\\

Source 7 has a rising spectrum with numerous $H_2$ lines and CO headbands in emission. 
A disk is probably present. An $H_2$ jet is observed (Fig.\ \ref{dyn_82_7}), confirming the youth 
of this object. Its evolutionary stage, I or II, is uncertain.\\

Source 35 shows all types of signatures (\brg\ and $H_2$ lines, CO
emission, and an increasing spectrum).  Its velocity field is unusual,
with \brg\ emission at the velocity of the \HII\ region, but strongly
approaching $H_2$ material. This source has probably a dominant disk
and thus is probably a Class II object; ejection is possibly present,
but needs confirmation.  An extended $H_2$ emission is observed in the
field (see Fig.\ \ref{source_id_1}), at the velocity of the
\HII\ region. It probably comes from the nearby PDR. \\

Source 51 shows a strong \brg\ emission line, an increasing spectrum,
and faint $H_2$ and CO emission. Its high emission in the 24~$\mu$m
band indicates that it is probably a rather massive YSO. It lies at
the center of a cluster (Fig.\ \ref{source_id_1}): several faint
continuum sources are present in the field (see also Fig.~17 in
PO09).\\

Source 98 (Fig.\ \ref{source_id_1}) shows faint CO features in absorption; however it is most
probably a young source (and not an evolved star) as it presents an
increasing spectra.  A bright extended $H_2$ emission is present over
the whole field (Fig.~A.8), at the velocity of the \HII\ region. Most
probably it is emitted in the vicinity of the PDR as confirmed by the
line ratios indicative of a non-thermal process ($\frac{I_{1-0s(1)}}{I_{2-1S(1)}} \sim$ 3.5 and
$\frac{I_{1-0s(1)}}{I_{3-2S(3)}} \sim$ 10).\\

\subsection{RCW~120}
\label{ap_rcw120}

Ten sources were observed in the direction of RCW~120. Eight of them 
present spectroscopic features characteristic of YSOs. Two of them are 
different. They are:

$\bullet$ RCW120 39 was classified as a Class~II in DE09. The SINFONI 
spectrum shows that it is most probably an evolved star, as it presents 
strong CO absorption features and no other signature typical of YSOs.

$\bullet$ RCW120 C4--64 was also classified as an intermediate Class~ I/II 
source in DE09. However its spectrum is slowly decreasing; there is 
no \brg\ nor H$_2$ emission lines; only the CO bands are observed in absorption.
Thus it is probably also an evolved star, located by chance on the border of 
the C4 condensation.

Note that RCW~120 lies only $\sim$12\degr\ of the Galactic center, and thus 
evolved stars of the Galactic bulge may be present in the background.\\

RCW120 76 (Fig.\ \ref{source_id_1}), 106 and 107 are rather similar,
almost featureless, with no or faint \brg\ line, no or faint H$_2$
lines and no CO emission bands, along with an increasing
continuum. They are possibly the less evolved of the YSOs surrounding
RCW~120, mostly dominated by their envelope. They were classified as
respectively Class I-II, Class I, and Class I-II in DE09, in good
agreement with the present results.\\

RCW120 49 and 103 are similar, with strong \brg\ emission, 
numerous H$_2$ lines and CO bands, superimposed on an increasing continuum. 
They are possibly more evolved than the preceding group, the presence of 
strong CO emission bands pointing to a dominant disk. They were classified as 
Class I-II and Class I in DE09.\\

\begin{figure}[h]
\centering
\includegraphics[height=12cm]{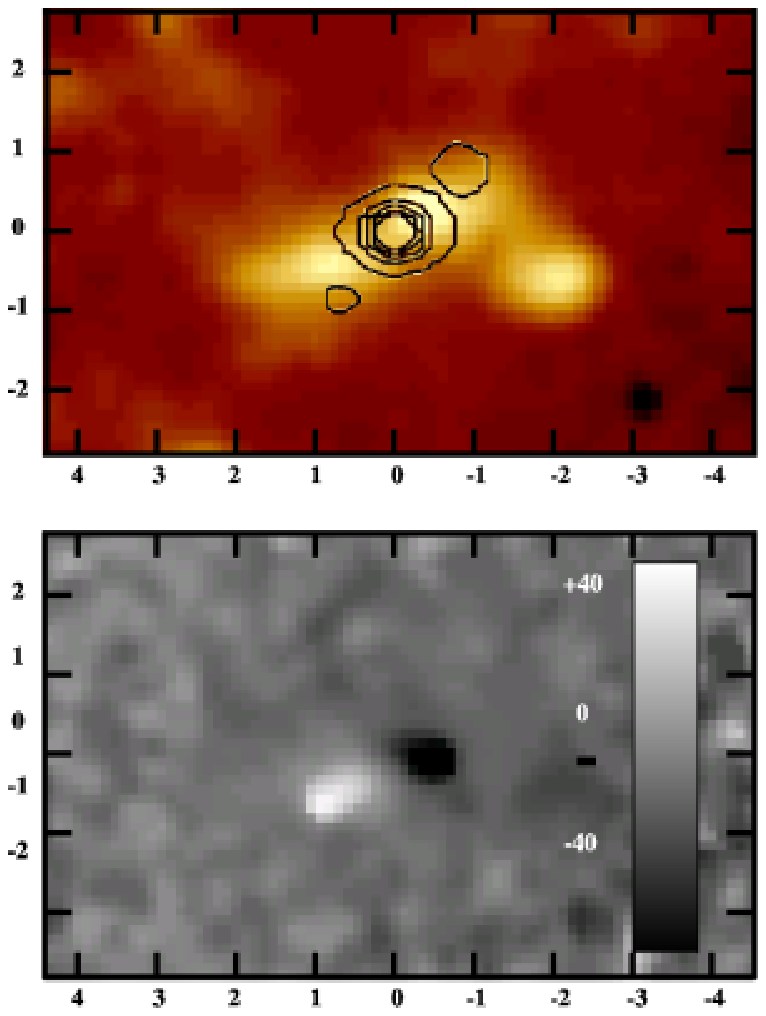}
\caption{Morphology and dynamics of RCW120 49. {\it Top}: \htwo\ emission map with contours of the full K--band emission overplotted. {\it Bottom}: \htwo\ velocity map (scale is in \kms). North is up, East is to the left. The axis indicate the offsets (in arscseconds) relative to the YSO position (coordinates given in Table \ref{tab_yso}).}\label{dyn_120_49}
\end{figure}

Fig.\ \ref{dyn_120_49} shows the \htwo\ emission and velocity maps of RCW120 49. The \htwo\ emission shows multiple peaks none of which is located at the $K$--band position of the YSO. The \htwo\ velocity map reveals that the two \htwo\ peaks on each side of the YSO have different radial velocities. We note however that the \htwo\ material at velocities of about -80 \kms\ corresponds to a weak emission region (it is slightly offset compared to the emission peak seen in the \htwo\ map). The \htwo\ profile at this position is also peculiar in the sense that it probably results from multiple components, being rather broad and featuring a ``shoulder'' on the blue side of the main emission peak. On the other hand, the bright emission peak south-east of the YSO is significantly detected at a lower velocity than the YSO itself and do not show any peculiar line profile. The analysis of the \htwo\ emission lines in Sect.\ \ref{yso_h2} indicated thermal emission with a possible small contribution from non thermal emission \footnote{Note that the analysis was performed on the spectrum extracted at the position of the \htwo\ emission peak.}. The presence of a jet would explain both this emission and the velocity map.

\begin{figure}[h]
\centering
\includegraphics[height=12cm]{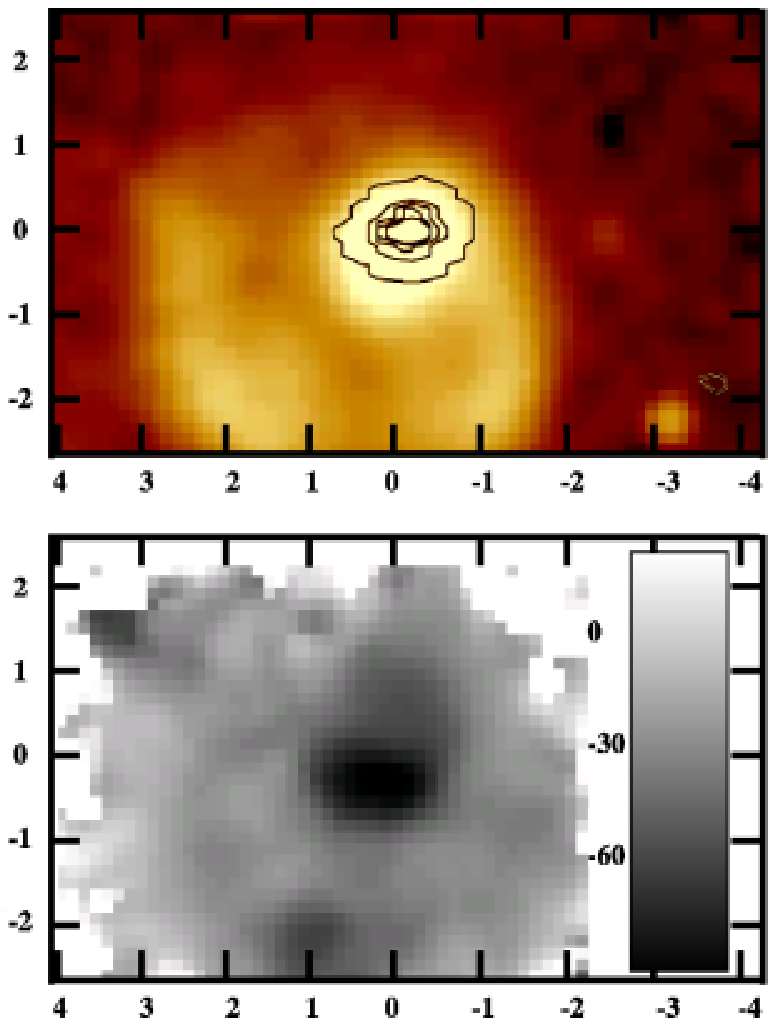}
\caption {Same as Fig.\ \ref{dyn_120_49} for YSO RCW120 103.}\label{dyn_120_103}
\end{figure}

We have seen previously that YSO RCW120 103 had peculiar kinematical properties, with \htwo\ moving much faster than \brg\ (Sect.\ \ref{dyn_yso}). Fig.\ \ref{dyn_120_103} shed more light on this problem. The \htwo\ emission peaks at the same position as the $K$--band emission. But an arc/spiral is detected south of the YSO, showing that molecular gas is present on wide scales. The velocity map confirms that a minimum (around -90 \kms) is located on the YSO. But this map tend to show that a velocity gradient exists on a direction south-east/north-west across the position of the YSO. A similar map in \brg\ did not reveal any structure. We note however that the velocity difference between both sides of the \htwo\ structure is of the order of 20 \kms, just at the limit of the resolution of our velocity measurements. Hence, we refrain from interpreting too much this possible structure and simply conclude that the geometry and dynamics of RCW120 103 is probably rather complex. This is confirmed by the velocity of the extended \htwo\ emission south of the YSO, which is intermediate between the velocity of the \HII\ region and that of the YSO. We also note that RCW82 35 which showed the same velocity discrepancy as RCW120 103 (\htwo\ velocity much larger than \brg\ velocity) did not reveal any velocity structure. 

The field of RCW120 51 contains several sources
(Fig.\ \ref{dyn_120_51}), all observed in the direction of the ionized
gas. We see on the spectra the \brg\ emission of the ionized gas. The
three brightest sources have a decreasing continuum; source 51b and
51c present CO absorption bands. These two sources are probably stars
and not YSOs. Source 51a is probably an YSO, as its spectrum show
\brg\ emission and no CO absorption. However we do not confirm the
Class I classification of DE09. YSO 51a lies at the tip of a
``finger'', a structure of the ionization front bright in the mid-IR.
The proximity of this structure has probably hampered the photometric
measurements of this object.\\

\begin{figure}[h]
\centering
\includegraphics[width=8cm]{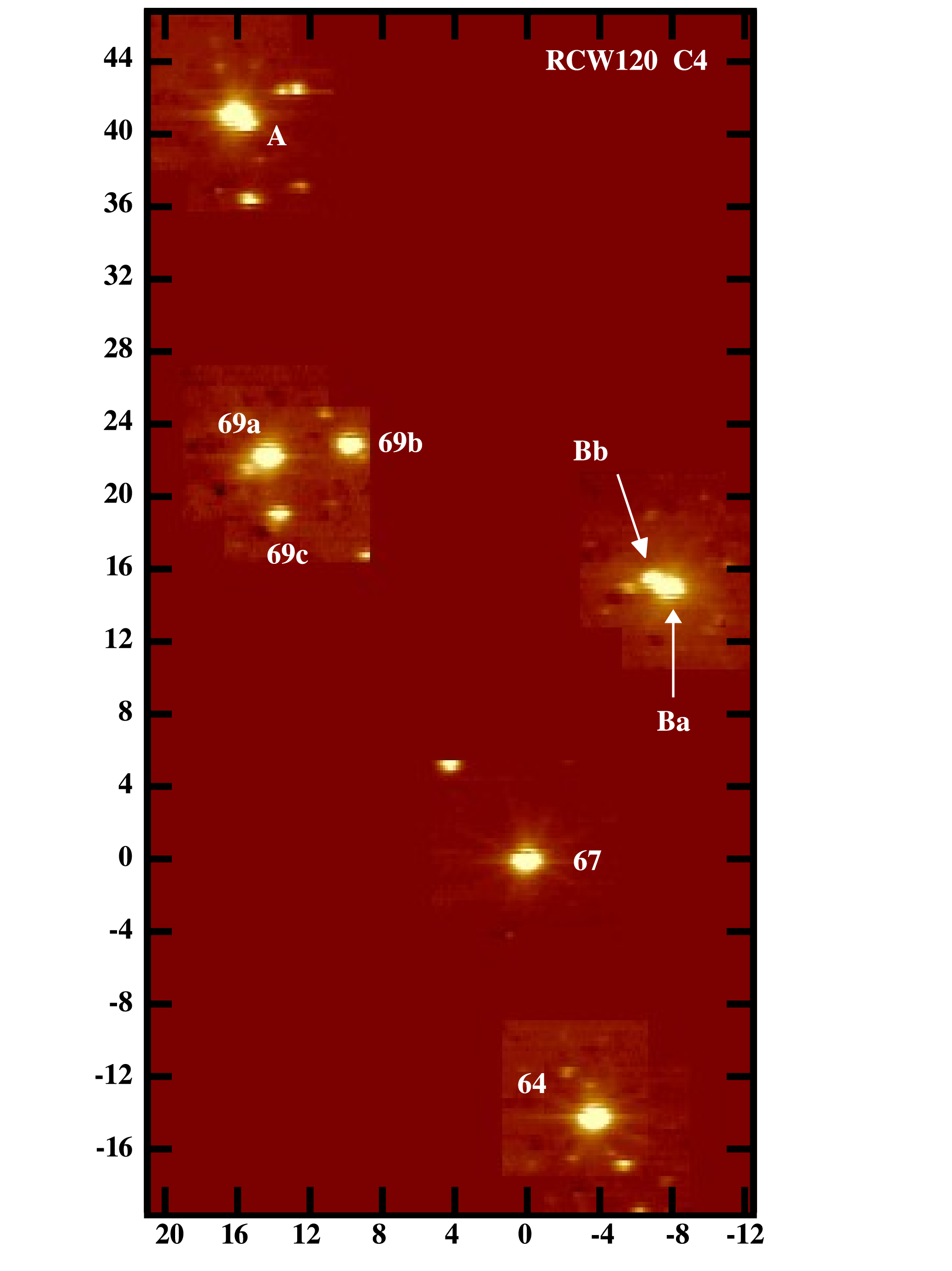}
\caption{Identification of YSOs in condensation 4 of RCW~120. North is up and East is to the left. The axis indicates the RA and DEC offsets (in arcseconds) relative to source 67.}\label{source_id_120c4}
\end{figure}

Condensation 4 contains several sources with different properties
(Fig.\ \ref{source_id_120c4}).  Sources A and Ba/Bb are respectively
at the center of zones of extended 8~$\mu$m and 24~$\mu$m emission
resembling small PDRs. The $JHK$ 2MASS photometry of the central
objects point to late B stars (DE09). The SINFONI spectra of these 3
sources show decreasing continua. Source A shows both \brg\ and CO
absorption. This might result from the blending of two sources, one
with each type of features, or be indicative of a medium type
star. Source Ba has \brg\ in absorption and is most likely a B star,
while source Bb features CO absorption bands and is either a late type
star or a YSO (see Sect.\ \ref{yso_spec}).  Three sources are observed
in the field of RCW120 69. They all have a flat continuum. RCW120
C4-69a is featureless. RCW120 C4-69c has CO absorption bands.  RCW120
C4-69b show faint CO absorption bands but \brg\ in emission. Extended
\htwo\ emission is observed south--east of the main source. These
sources are probably evolved YSOs, RCW120 C4-69b being the
youngest. RCW120 C4-69, considered as a single source, was classified
Class I by DE09. However it is not a Class I, but more probably a
Class II or III.\\

\end{appendix}

\end{document}